\documentclass[prx, reprint, amsmath,amssymb,showpacs,floatfix,longbibliography, aps, twocolumn, superscriptaddress]{revtex4-1}
\usepackage{times} 
\usepackage{graphicx}
\usepackage{dcolumn}
\usepackage{bm,bbm}
\usepackage{color}
\usepackage[dvipsnames]{xcolor}
\usepackage{hyperref}
\usepackage{enumitem}
\usepackage{cancel}
\hypersetup{colorlinks=true,citecolor=blue,linkcolor=blue, urlcolor=blue}
\hypersetup{linktocpage}
\newcolumntype{M}[1]{>{\centering\arraybackslash}m{#1}}
\newcolumntype{N}{@{}m{0pt}@{}}
\usepackage{environ}
\usepackage[caption=false]{subfig}

\usepackage{quiver} %for commutative diagrams

\NewEnviron{eqs}{%
\begin{equation}\begin{split}
    \BODY
\end{split}\end{equation}
}

\makeatletter %No subsubsections in TOC
\def\l@subsubsection#1#2{}
\makeatother

\newcommand{\ZZ}{\mathbb{Z}}
\newcommand{\ra}{\rangle}
\newcommand{\la}{\langle}
\newcommand{\chargei}{\tilde{c}_i}
\newcommand{\fluxi}{\tilde{\varphi}_i}
\newcommand{\chargej}{\tilde{c}_j}

\begin{document}

\title{Pauli stabilizer models of twisted quantum doubles} 

\date{\today}

\author{Tyler~D. Ellison}
\email[Email: ]{tyler.ellison@yale.edu}
\affiliation{Department of Physics, University of Washington, Seattle, WA 98195, USA}
\affiliation{Department of Physics, Yale University, New Haven, CT 06511, USA}

\author{Yu-An Chen}
\affiliation{Department of Physics, Condensed Matter Theory Center, and Joint Quantum Institute, University of Maryland, College Park, MD 20742, USA}
 
\author{Arpit Dua}
\affiliation{Department of Physics and Institute for Quantum Information and Matter, California Institute of Technology, Pasadena, CA 91125, USA}
 
\author{Wilbur Shirley}
\affiliation{Department of Physics and Institute for Quantum Information and Matter, California Institute of Technology, Pasadena, CA 91125, USA}
\affiliation{School of Natural Sciences, Institute for Advanced Study, Princeton, NJ 08540, USA}

\author{Nathanan Tantivasadakarn}
\affiliation{Department of Physics, Harvard University, Cambridge, MA 02138, USA}

\author{Dominic~J. Williamson}
\thanks{Current Address: Centre for Engineered Quantum Systems, School of Physics,
University of Sydney, Sydney, New South Wales 2006, Australia}
\affiliation{Stanford Institute for Theoretical Physics, Stanford University, Stanford, CA 94305, USA}

\begin{abstract}
\noindent 
We construct a Pauli stabilizer model for every two-dimensional Abelian topological order that admits a gapped boundary. 
Our primary example is a Pauli stabilizer model on four-dimensional qudits that belongs to the double semion (DS) phase of matter. 
The DS stabilizer Hamiltonian is constructed by condensing an emergent boson in a $\ZZ_4$ toric code, 
where the condensation is 
implemented at the level of the ground states by two-body measurements. 
We rigorously verify the topological order of the DS stabilizer model by identifying an explicit finite-depth quantum circuit (with ancillary qubits) that maps its ground state subspace to that of a DS string-net model.
We show that the construction of the DS stabilizer Hamiltonian generalizes to all twisted quantum doubles (TQDs) with Abelian anyons. This yields a Pauli stabilizer code on composite-dimensional qudits for each such TQD, implying that the classification of topological Pauli stabilizer codes extends well beyond stacks of toric codes -- in fact, exhausting all Abelian anyon theories that admit a gapped boundary.
We also demonstrate that symmetry-protected topological phases of matter characterized by type I and type II cocycles can be modeled by Pauli stabilizer Hamiltonians by gauging certain $1$-form symmetries of the TQD stabilizer models.
\end{abstract}

\maketitle

\renewcommand{\baselinestretch}{.96}\normalsize
\tableofcontents
\renewcommand{\baselinestretch}{1.0}\normalsize

% \tableofcontents

\section{Introduction}

Quantum error correcting codes are essential for protecting quantum information from environmental noise and faulty operations \cite{Shor1995,Aharonov2008,shor1996fault}. 
One of the most prominent classes of quantum error correcting codes are the two-dimensional {topological} quantum codes \cite{qdouble}, in which quantum information is encoded in degenerate eigenstates of topologically ordered systems \cite{wegner1971duality,Anderson1973,PhysRevLett.50.1395,PhysRevB.40.7387,WEN1990}. 
Topological quantum codes offer protection from arbitrary local errors, due to the local indistinguishability of the degenerate eigenstates \cite{dennis2002memory}, and admit fault-tolerant operations with low-overhead, derived from the properties of the anyons in the underlying topological order \cite{qdouble,Bombin2006,nayak2008non,bombin2010topological}. Despite the merits of two-dimensional topological quantum codes, it can be challenging to assess the error correcting properties of a general topologically ordered system.

A class of topological quantum codes with particularly transparent error correcting properties are the topological Pauli stabilizer codes \cite{qdouble,dennis2002memory,Terhal2015review}. These combine the robust error correction of topological quantum codes with the simple algebraic structures of Pauli stabilizer codes~\cite{gottesman1997stabilizer}. In general, topological Pauli stabilizer codes are defined by Pauli stabilizer models -- i.e., Hamiltonians whose terms are mutually commuting local products of Pauli operators \cite{qdouble}. By convention, the ground state subspace of the Pauli stabilizer model defines the logical subspace of the error correcting code. Topological Pauli stabilizer codes are exemplified by the $\ZZ_2$ toric code (TC) and the generalization to $N$-dimensional qudits referred to as the $\ZZ_N$ TC (or $\ZZ_N$ quantum double), both introduced in Ref.~\cite{qdouble}.

The characteristic properties of a topological Pauli stabilizer code can be understood in terms of the topological order of the underlying Pauli stabilizer model. The $\ZZ_N$ TC, for example, is based on the topological order of a $\ZZ_N$ gauge theory \cite{kogut1975hamiltonian}, and the properties of the gauge charges and fluxes are central to fault-tolerant quantum computations using the $\ZZ_N$ TC \cite{dennis2002memory,Fowler2012surface,Brown2017poking}. This motivates developing a classification of the topological phases of matter captured by Pauli stabilizer models. Such a classification informs us about the universal properties that can be exploited for quantum computation using a topological Pauli stabilizer code. 
This leads to the natural question: what topological phases of matter can be described by Pauli stabilizer models? 

This question was partially addressed in the works of Refs.~\cite{BDCP12, bombin2014structure,Haah2018a}. Under a technical assumption \footnote{Specifically, Refs.~\cite{BDCP12} and \cite{bombin2014structure} assumed that chiral anyon theories cannot be realized by Pauli stabilizer models.}, it was shown in Refs.~\cite{BDCP12} and \cite{bombin2014structure} that all (translation invariant) Pauli stabilizer models on qubits
% using Eq.~\eqref{eq: qudit to qubits}.} 
belong to the same phase as the $\ZZ_2$ TC or decoupled copies of the $\ZZ_2$ TC. Later, this was generalized to prime-dimensional qudits and made rigorous by Ref.~\cite{Haah2018a}. Ref.~\cite{Haah2018a} proved that, for any prime $p$, every (translation invariant) Pauli stabilizer model on $p$-dimensional qudits \footnote{Throughout, we assume the usual representations of Pauli groups, as opposed to, for example, defining the Pauli group of a four-dimensional qudit on a pair of qubits via the automorphism in Eq.~\eqref{eq: qudit to qubits}.} has the same topological order as some number of decoupled copies of the $\ZZ_p$ TC \footnote{This includes the possibility in which the model has no topological order.}. This completes the classification of (translation invariant) Pauli stabilizer models on prime-dimensional qudits, showing that the universal properties of Pauli stabilizer models on prime-dimensional qudits are fully captured by TCs. However, the works of Refs.~\cite{BDCP12, bombin2014structure,Haah2018a} leave the classification of Pauli stabilizer models on composite-dimensional qudits (i.e. products of primes) unaddressed. One can ask: are there Pauli stabilizer models on composite-dimensional qudits that capture more exotic topological phases of matter -- beyond that of decoupled layers of the $\ZZ_N$ TC?

In this work, we answer this question in the affirmative. We construct Pauli stabilizer models defined on composite-dimensional qudits that realize \textit{all Abelian anyon theories that admit a gapped boundary}. These models define topological Pauli stabilizer codes whose underlying topological order is beyond that of the $\ZZ_N$ TC or decoupled copies of the $\ZZ_N$ TC.
This represents a substantial step towards a full classification of topological Pauli stabilizer codes, which is evidently much richer than suggested by the classification for prime-dimensional qudits.
Furthermore, based on the expectation that commuting projector Hamiltonians in two spatial dimensions must have gapped boundaries \cite{kitaev2006anyons,KapustinThermal2020}, we expect that
the Pauli stabilizer models presented here yield a complete classification of topological Pauli stabilizer codes. 

We emphasize that the Pauli stabilizer models introduced in this work can be distinguished from copies of the $\ZZ_N$ TC by the properties of their anyonic excitations. 
For example, we define a Pauli stabilizer model that belongs to the double semion (DS) topological phase of matter. Similar to the $\ZZ_2$ TC, the DS topological order has four types of anyons. However, in marked contrast, the DS topological order features semionic excitations -- which produce a statistical phase of $i$ upon interchange. As there is no such excitation for a $\ZZ_2$ TC, the DS stabilizer model belongs to a distinct phase of matter. Instead, it can be interpreted as a \textit{twisted} gauge theory, more formally known as a twisted quantum double (TQD), where the `twist' manifests in the exotic exchange statistics of the gauge fluxes \cite{DW90,levin2012braiding,tqd}.
 
More generally, the Pauli stabilizer models presented here capture all TQDs with Abelian anyons, which according to Refs.~\cite{KS11} and \cite{KKOSS21}, account for every Abelian topological order with gapped boundaries \footnote{For discussion on gapped boundaries of two-dimensional topological orders see Refs.~\cite{bravyi1998quantum,beigi2011quantum,Kitaev2012}.}. Therefore, we lay the groundwork for using universal properties beyond those of the TC for quantum computing using topological Pauli stabilizer codes. We also define Pauli stabilizer models of symmetry-protected topological phases of matter. While the ground state subspaces of such models are nondegenerate on a closed manifold -- and hence, cannot be used to encode quantum information -- they are of interest as they evade the no-go theorems of Ref.~\cite{EKLH21}. 

The Pauli stabilizer models described in this work are constructed from copies of the $\ZZ_N$ TC by condensing (i.e., proliferating) certain bosonic anyons \cite{Bais2009Condensate,Kong2014condensation,Burnell2018}. This exploits the fact that all TQDs with Abelian anyons can be obtained from decoupled copies of the $\ZZ_N$ TC through condensation, as pointed out in Refs.~\cite{Bais2009Condensate} and \cite{Schuch2017holographic}. 
We show further that the condensation can be implemented within the stabilizer formalism using few-body Pauli measurements. As a result, condensation maps Pauli stabilizer models to Pauli stabilizer models. We note that this builds off of earlier works, in which boson condensation was used to construct tensor network representations of TQD ground states \cite{Schuch2017holographic,Schuch2018condensation} and symmetry-protected topological states \cite{Shenghan2017condensation}.

It is worth noting that exactly solvable models of TQDs have been introduced in previous works. In particular, commuting projector Hamiltonians of TQDs were first described in Refs.~\cite{levin2005,levin2012braiding,tqd}. These models were then leveraged to build non-Pauli stabilizer models of TQDs in Refs.~\cite{Eisert2021nonpauli,Ortiz2019semion}. While these models indeed capture the characteristic properties of TQDs, their potential for fault-tolerant quantum computation is relatively opaque (although, see Ref.~\cite{Delgado2020semionthresholds}). The TQD models presented here have the key feature that they can be fully understood within the familiar Pauli stabilizer formalism. This opens up the possibility for novel applications of TQDs to quantum computation and provides algebraically simple models for simulating topological phases of matter. 

The outline of this paper is as follows. In Section~\ref{sec: double semion stabilizer code}, we introduce the paradigmatic example of our construction -- the DS stabilizer model. 
We then derive the DS stabilizer Hamiltonian by condensing a boson in the $\ZZ_4$ TC, and establish its topological order by mapping its ground state subspace to that of the DS string-net model using a finite-depth quantum circuit (with ancilla). 
Next, to generalize the DS stabilizer model to Pauli stabilizer models of TQDs, we provide a primer on Abelian anyon theories in Section~\ref{sec: primer}. 
Subsequently, in Section~\ref{sec: TQD stabilizer codes}, we construct Pauli stabilizer models of TQDs, starting with a review of the characteristic data of TQDs with Abelian anyons. The associated Pauli stabilizer models are constructed by condensing anyons in decoupled layers of TCs. In Section~\ref{sec: SPT}, we derive Pauli stabilizer models of SPT phases by condensing gauge charges in the Pauli stabilizer models of TQDs. Appendix~\ref{app: string-net ground states} gives further details on the mapping of the DS stabilizer Hamiltonian to the DS string-net model, and Appendices~\ref{app: K matrix} and \ref{app: fusion group} provide examples of TQDs and a general formula for the group structure of the anyons in TQDs.

\section{Double semion stabilizer model} \label{sec: double semion stabilizer code}

Before describing the general construction of topological Pauli stabilizer models of twisted quantum doubles with Abelian anyons, we provide a concrete example. Our example is a Pauli stabilizer model defined on four-dimensional qudits that belongs to the double semion (DS) phase of matter. In Section~\ref{sec: DS Pauli stabilizer code}, we define the Pauli stabilizer Hamiltonian and show explicitly that its anyonic excitations exhibit the characteristic properties of the anyons in the DS topological order. In the subsequent section, Section~\ref{sec: DS construction}, we construct the DS stabilizer model from a $\ZZ_4$ toric code (TC) by condensing certain excitations. We then conclude the section with Section~\ref{sec: relation to nonPauli}, where we use a finite-depth quantum circuit to map the ground state subspace of the DS stabilizer Hamiltonian to that of the DS string-net model of Refs.~\cite{LW05} and \cite{LG12}.
 
We start by recalling the characteristic properties of the anyons in the DS phase, which we label by $\{1,s,\bar{s},s\bar{s}\}$. Similar to the anyons in the $\ZZ_2$ TC phase, these form a $\ZZ_2 \times \ZZ_2$ group under fusion, with the fusion rules given by:
\begin{eqs} \label{eq: DS fusion}
s \times s =1, \quad \bar{s} \times \bar{s} = 1, \quad s \times \bar{s} = s \bar{s}.
\end{eqs}
The exchange statistics and braiding of the anyons are determined by the function $\theta:\{1,s,\bar{s},s\bar{s}\} \to U(1)$, defined by:
\begin{align} \label{eq: DS statistics}
\theta(1)=1, \quad \theta(s)=i, \quad \theta(\bar{s}) = -i, \quad \theta(s\bar{s})=1.
\end{align}
{Physically, $\theta(a)$ is the phase obtained upon exchanging two identical $a$ anyons.
Eq.~\eqref{eq: DS statistics} tells us that $s$ is a semion, $\bar{s}$ is an anti-semion, and $s\bar{s}$ is a boson, i.e., the wave function incurs a phase of $i$, $-i$, and $1$, respectively. }
The statistics of the anyons are an invariant of topological order -- they cannot be changed by any local perturbations of the Hamiltonian that preserve the energy gap. Therefore, the statistics in Eq.~\eqref{eq: DS statistics} distinguish the DS topological order from the $\ZZ_2$ TC phase. 
The braiding relations of the anyons can then be deduced from the exchange statistics $\theta$ using the expression in Eq.~\eqref{eq: braiding M def}, introduced later in the text.
Here, we simply state that the semions and anti-semions have the following braiding relations: 
\begin{align} \label{eq: DS braiding}
B_\theta(s,s) = -1, \quad B_\theta(\bar{s},\bar{s})=-1, \quad B_\theta(s, \bar{s}) = 1,
\end{align}
where $B_\theta(a,a')$ is the phase accrued from a full braid of the anyons $a$ and $a'$. 
We show that the anyonic excitations of the DS stabilizer model, described in the next section, obey Eqs.~\eqref{eq: DS fusion} and \eqref{eq: DS statistics}, which indicates that it belongs to the DS phase.

\subsection{Definition of the Pauli stabilizer model} \label{sec: DS Pauli stabilizer code}

\begin{figure}
\centering
    \includegraphics[width=.25\textwidth]{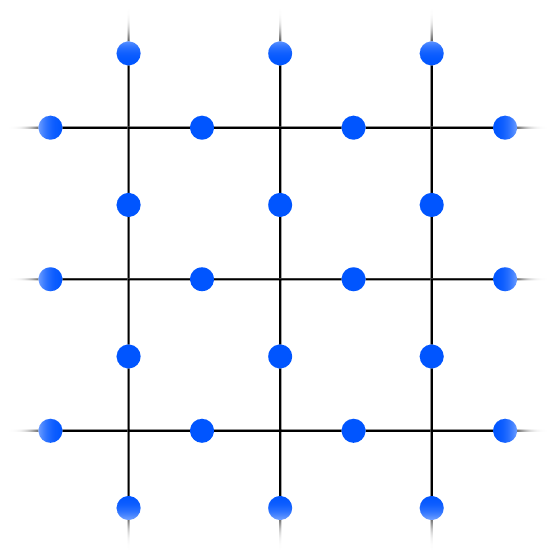}
     \caption{The DS stabilizer model is defined on a square lattice with a single four-dimensional qudit (blue) at each edge.}
     \label{fig: squareDSdof}
\end{figure}

Here, we describe a Pauli stabilizer Hamiltonian belonging to the DS phase. The Hamiltonian is defined on a square lattice with a four-dimensional qudit at each edge, as shown in Fig.~\ref{fig: squareDSdof}. The generalized Pauli X and Pauli Z operators at an edge $e$ are:
\begin{align}
X_e = \sum_{j \in \ZZ_4} |j+1\rangle \langle j|, \quad Z_e = \sum_{j \in \ZZ_4} i^j |j\rangle \langle j|,
\end{align}
where we have labeled the computational basis states by $j \in \ZZ_4$.
These operators are sometimes referred to as the shift and clock operators, respectively. They satisfy the relations:
\begin{align}
X_e^4 =1, \quad Z_e^4 =1,
\end{align}
and for any pair of edges $e$ and $e'$, they obey the commutation relations:
\begin{align} \label{eq: commutation X and Z}
Z_eX_{e'} = 
\begin{cases}
i X_{e'}Z_e & e=e' \\
X_{e'}Z_e & e \neq e'.
\end{cases}
\end{align}
% They can be represented by the matrices:
% \begin{align}
% X_e = \begin{pmatrix} 
% 0 & 0 & 0 & 1\\
% 1 & 0 & 0 & 0 \\
% 0 & 1 & 0 & 0 \\
% 0 & 0 & 1 & 0
% \end{pmatrix}, \quad 
% Z_e = \begin{pmatrix} 
% 1 & 0 & 0 & 0\\
% 0 & i & 0 & 0 \\
% 0 & 0 & -1 & 0 \\
% 0 & 0 & 0 & -i
% \end{pmatrix}.
% \end{align}

The DS stabilizer Hamiltonian then takes the form: 
\begin{align}
H_\text{DS} \equiv -\sum_v A_v - \sum_p B_p - \sum_e C_e + \text{h.c.},
\end{align}
where the sums are over all vertices $v$, plaquettes $p$, and edges $e$, respectively. We find it convenient to represent the Hamiltonian terms graphically, as shown below:
\begin{equation} \label{eq: DS terms}
\begin{gathered}
A_v \equiv \vcenter{\hbox{\includegraphics[scale=.23,trim={.5cm 0cm 1.5cm 0cm},clip]{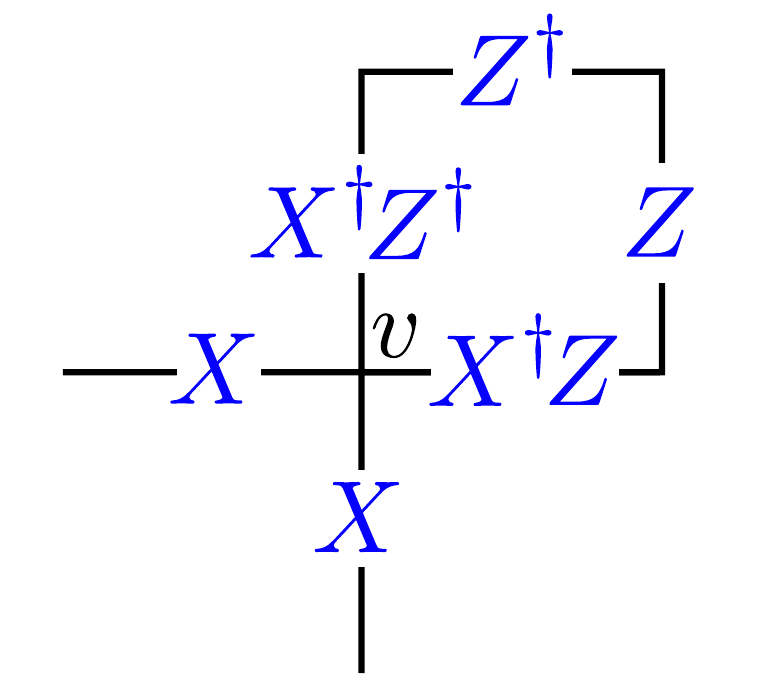}}}, \qquad B_p \equiv \vcenter{\hbox{\includegraphics[scale=.23,trim={0cm 0cm 0cm 0cm},clip]{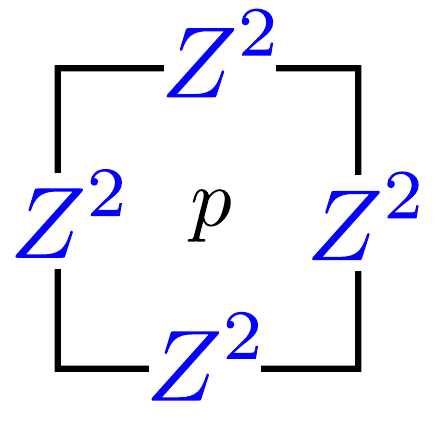}}}, \\
 C_e \equiv \vcenter{\hbox{\includegraphics[scale=.23,trim={0cm 0cm 0cm 0cm},clip]{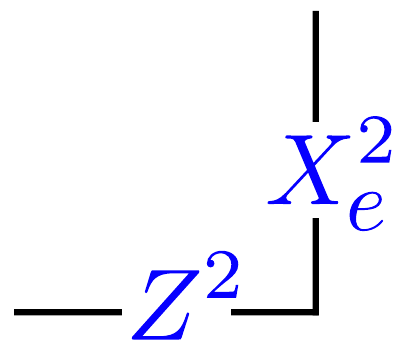}}}, \,\,\, \vcenter{\hbox{\includegraphics[scale=.23,trim={0cm 0cm 0cm 0cm},clip]{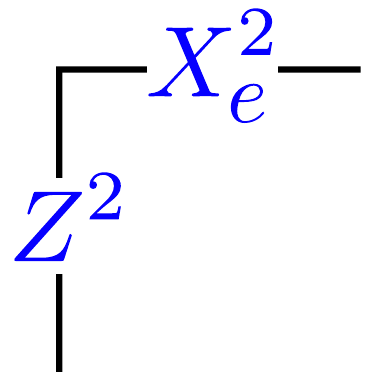}}}.
 \end{gathered}
\end{equation}
% \begin{equation} \nonumber
% A_v \equiv \vcenter{\hbox{\includegraphics[scale=.23,trim={.5cm 0cm 1.5cm 0cm},clip]{Figures/stabilizer_Av-eps-converted-to.pdf}}}, \qquad B_p \equiv \vcenter{\hbox{\includegraphics[scale=.23,trim={0cm 0cm 0cm 0cm},clip]{Figures/stabilizer_Bp-eps-converted-to.pdf}}}, 
% \end{equation}
% \begin{equation}\label{eq: DS terms}
%  C_e \equiv \vcenter{\hbox{\includegraphics[scale=.23,trim={0cm 0cm 0cm 0cm},clip]{Figures/stabilizer_Ceh-eps-converted-to.pdf}}}, \,\,\, \vcenter{\hbox{\includegraphics[scale=.23,trim={0cm 0cm 0cm 0cm},clip]{Figures/stabilizer_Cev-eps-converted-to.pdf}}}.
% \end{equation}
Note that the definition of $C_e$ depends on whether the edge $e$ is vertical or horizontal.
Using the commutation relations in Eq.~\eqref{eq: commutation X and Z}, it can be checked that the terms are mutually commuting. 

Given that the Hamiltonian terms commute with one another, they define the stabilizer group $\mathcal{S}_\text{DS}$:
\begin{align} \label{eq: DS stabilizer group}
\mathcal{S}_\text{DS} \equiv \langle \{A_v\}, \{B_p\}, \{C_e\} \rangle,
\end{align}
where the angled bracket notation denotes that $\mathcal{S}_\text{DS}$ is generated by the terms of $H_\text{DS}$.
By construction, the logical subspace of the stabilizer group $\mathcal{S}_\text{DS}$ coincides with the ground state subspace of $H_\text{DS}$. Explicitly, the logical subspace $\mathcal{H}_L$ is the mutual $+1$ eigenspace of the stabilizers:
\begin{align}
\mathcal{H}_L \equiv \{|\psi \rangle : S |\psi \rangle = |\psi \rangle, \forall S \in \mathcal{S}_\text{DS}\}.
\end{align}

On a torus, the DS Hamiltonian has a four-fold ground state degeneracy. This can be seen by counting the number of independent constraints imposed on the ground state subspace by the stabilizers. Letting $N_v$ denote the number of vertices, there are $N_v$ vertex terms, $N_v$ plaquette terms, and $2N_v$ edge terms. These constraints are not entirely independent, however. In particular, the vertex term squares to a product of edge terms and plaquette terms.
Therefore, although $A_v$ has order four, it only contributes $N_v$ order two constraints. There are also two global relations among the vertex terms and plaquette terms: 
\begin{align}
\prod_v A_v =1, \quad \prod_p B_p =1.
\end{align}
Consequently, there are only $N_v-1$ independent vertex terms and $N_v-1$ independent plaquette terms. All together, we find that there are $4N_v-2$ independent order two constraints. 
Given that there are two four-dimensional qudits per vertex, yielding a total Hilbert space of dimension $4^{2N_v}$, the ground state subspace is four-dimensional:
\begin{align}
\text{dim}(\mathcal{H}_L)=4^{2N_v} / 2^{4N_v-2} = 4.
\end{align}

\begin{figure}
\centering
    \includegraphics[width=.5\textwidth]{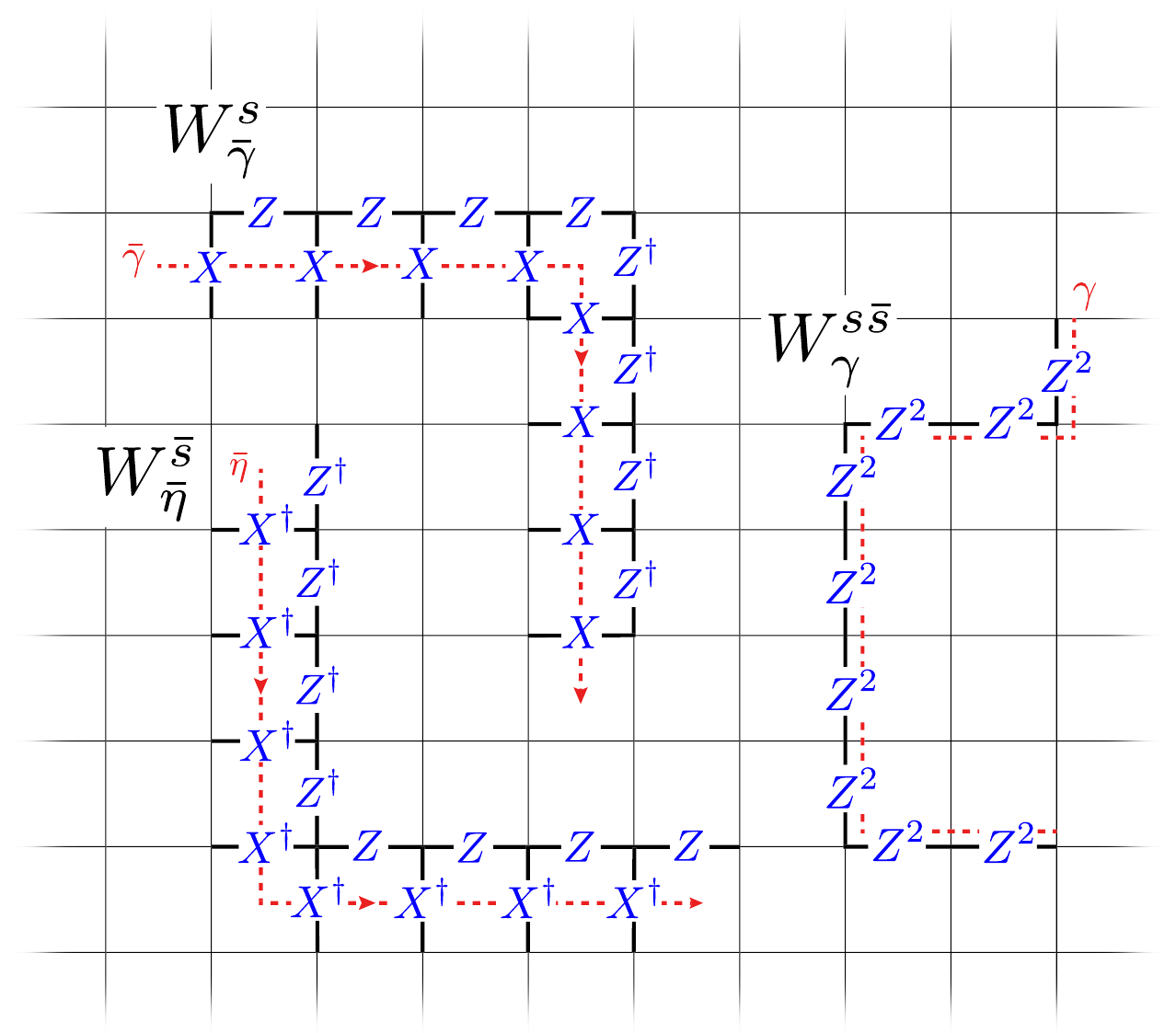}
     \caption{The string operators $W^s_{\bar{\gamma}}$ and $W^{\bar{s}}_{\bar{\eta}}$ are defined along the oriented paths (red dashed lines) $\bar{\gamma}$ and $\bar{\eta}$ in the dual lattice. The $s\bar{s}$ string operator $W^{s\bar{s}}_\gamma$ is defined along an un-oriented path $\gamma$ in the direct lattice. The DS string operators are composed of the short string operators pictured in Eq.~\eqref{eq: DS short strings}.}
     \label{fig: DSstrings}
\end{figure}

We now characterize the topological order of $H_\text{DS}$ by considering its anyonic excitations.
The anyonic excitations of $H_\text{DS}$ are created by string operators such as those depicted in Fig.~\ref{fig: DSstrings}. There are three types of string operators: $W^s_{\bar{\gamma}}$, $W^{\bar{s}}_{\bar{\gamma}}$, and $W^{s\bar{s}}_{{\gamma}}$. The first two are defined along an oriented path $\bar{\gamma}$ in the dual lattice, while the third is defined along an un-oriented path $\gamma$ in the direct lattice. To make the string operators explicit, we decompose them into products of short string operators:
\begin{align} \label{eq: DS string operators}
W^s_{\bar{\gamma}} = \prod_{e \in \bar{\gamma}} W^s_e, \quad W^{\bar{s}}_{\bar{\gamma}} = \prod_{e \in \bar{\gamma}} W^{\bar{s}}_e, \quad  W^{s\bar{s}}_{\gamma} = \prod_{e \in \gamma} W^{s\bar{s}}_e.
\end{align} 
The short string operators in Eq.~\eqref{eq: DS string operators} are represented pictorially as:
\begin{eqs} \label{eq: DS short strings}
W^{s}_e &\equiv \vcenter{\hbox{\includegraphics[scale=.22,trim={0cm 0cm 0cm 0cm},clip]{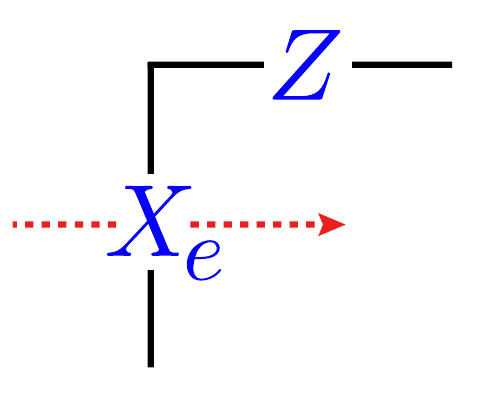}}}, \,\, \vcenter{\hbox{\includegraphics[scale=.22,trim={0cm 0cm 0cm 0cm},clip]{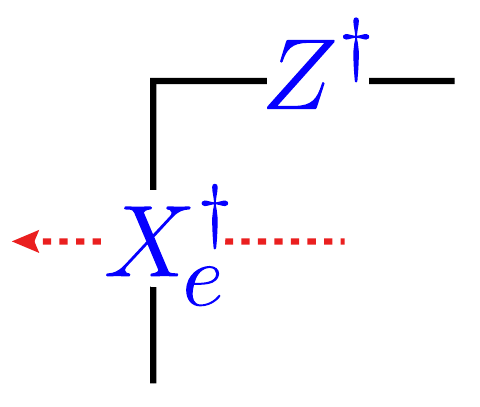}}}, \,\, \vcenter{\hbox{\includegraphics[scale=.22,trim={0cm 0cm 0cm 0cm},clip]{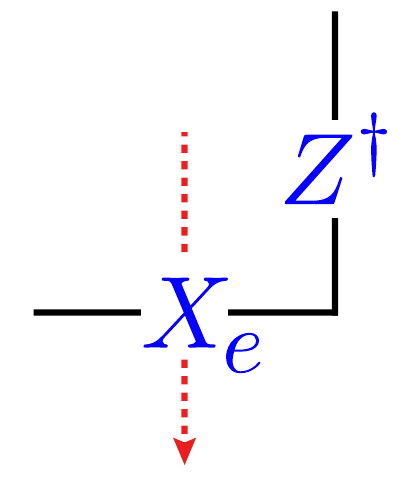}}}, \,\, \vcenter{\hbox{\includegraphics[scale=.22,trim={0cm 0cm 0cm 0cm},clip]{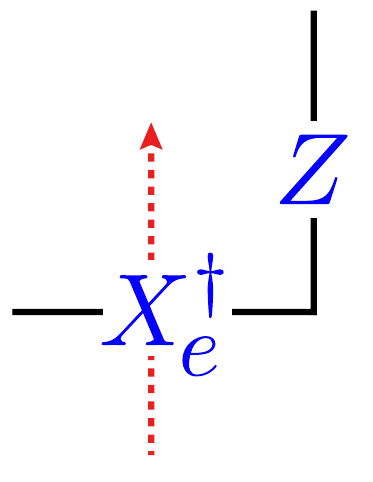}}}, \\
W^{\bar{s}}_e &\equiv \vcenter{\hbox{\includegraphics[scale=.22,trim={0cm 0cm 0cm 0cm},clip]{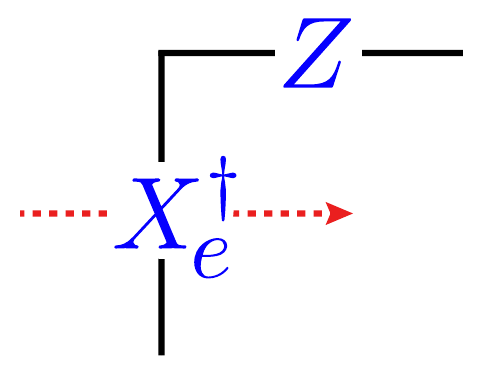}}}, \,\, \vcenter{\hbox{\includegraphics[scale=.22,trim={0cm 0cm 0cm 0cm},clip]{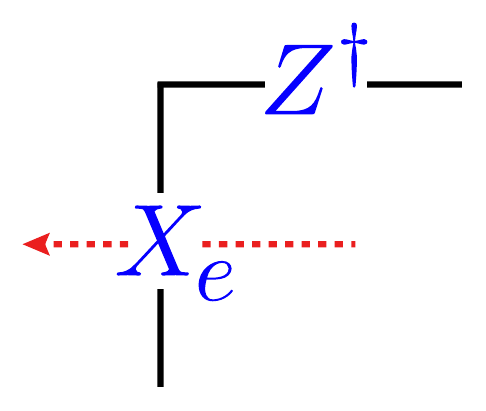}}}, \,\, \vcenter{\hbox{\includegraphics[scale=.22,trim={0cm 0cm 0cm 0cm},clip]{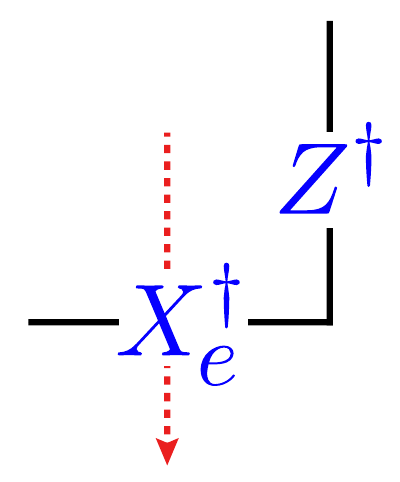}}}, \,\, \vcenter{\hbox{\includegraphics[scale=.22,trim={0cm 0cm 0cm 0cm},clip]{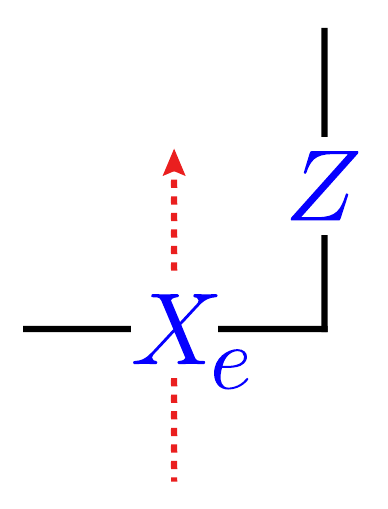}}}, \\
W^{s\bar{s}}_e &\equiv \vcenter{\hbox{\includegraphics[scale=.22,trim={0cm 0cm 0cm 0cm},clip]{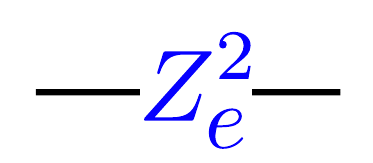}}}, \,\,\,\vcenter{\hbox{\includegraphics[scale=.22,trim={0cm 0cm 0cm 0cm},clip]{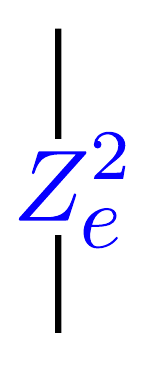}}}.
\end{eqs}
where the dashed red lines denote the orientation of the path on the dual lattice. Note that the expressions for $W^s_{\bar{\gamma}}$ and $W^{\bar{s}}_{\bar{\gamma}}$ in Eq.~\eqref{eq: DS string operators} are ambiguous up to a phase, since we have not specified an ordering of the short string operators.
Nonetheless, any choice of ordering yields a string operator. 
Importantly, the operators in Eq.~\eqref{eq: DS string operators} commute with the Hamiltonian terms along the length of the path, and only fail to commute with vertex terms and plaquette terms at the endpoints. 
Notice also that the $A_v$ and $B_p$ terms of the Hamiltonian are small loops of the string operators, i.e.:
\begin{align}
A_v \propto \prod_{e \in \bar{\gamma}_v} W^{s}_e, \quad B_p = \prod_{e \in \gamma_p} W^{s\bar{s}}_e,
\end{align}
where $\bar{\gamma}_v$ is the counter-clockwise oriented path through the edges connected to $v$ and $\gamma_p$ is the path formed by the edges bordering $p$.

We have suggestively labeled the string operators by the anyons $s$, $\bar{s}$, and $s\bar{s}$ of the DS phase. Indeed, the string operators create anyonic excitations of the DS phase, as verified below. The first property to check is the fusion rules of the excitations. The fusion rules are obtained by multiplying string operators that share the same endpoints. We find that fusing two $s$ string operators along the same path $\bar{\gamma}$ in the dual lattice gives \footnote{This holds for either orientation of $\bar{\gamma}$.}:
\begin{align}
W^s_{\bar{\gamma}} \times W^s_{\bar{\gamma}}= \vcenter{\hbox{\includegraphics[scale=.33,trim={0cm 0cm 0cm 0cm},clip]{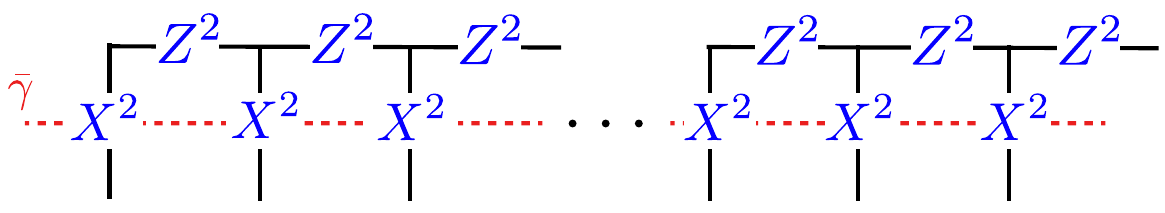}}}.
\end{align}
Up to stabilizers along $\bar{\gamma}$, this is equivalent to \footnote{$W^s_{\bar{\gamma}}$ can be modified so that it squares to stabilizers. However, we consider the string operator $W^s_{\bar{\gamma}}$ to be consistent with the general form in Eq.~\eqref{eq: DS string operators}.}:
\begin{align}
W^s_{\bar{\gamma}} \times W^s_{\bar{\gamma}}= \vcenter{\hbox{\includegraphics[scale=.32,trim={0cm 0cm 0cm 0cm},clip]{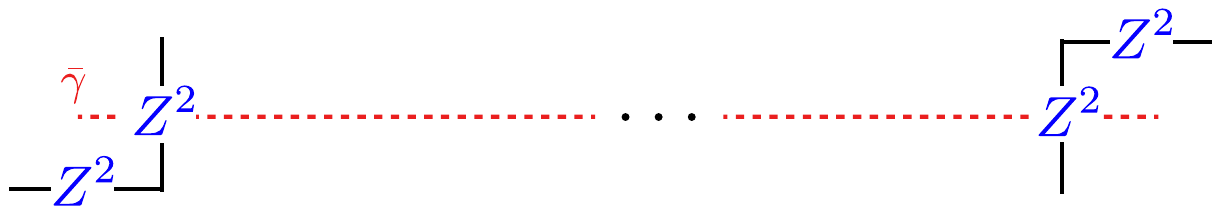}}}.
\end{align}
Therefore, the string operators labeled by $s$ multiply to local operators at the endpoints. Since the trivial anyon $1$ represents local excitations created by local operators, this means that $s \times s = 1$. Similarly, it can be checked that the string operators $W^{\bar{s}}_{\bar{\gamma}}$ and $W^{s\bar{s}}_\gamma$ satisfy the fusion rules of the anyons in the DS phase up to stabilizers and local operators at the endpoints. Thus, the anyonic excitations of $H_\text{DS}$ satisfy the fusion rules in Eq.~\eqref{eq: DS fusion}. 

\begin{figure}
\centering
    \includegraphics[width=.5\textwidth]{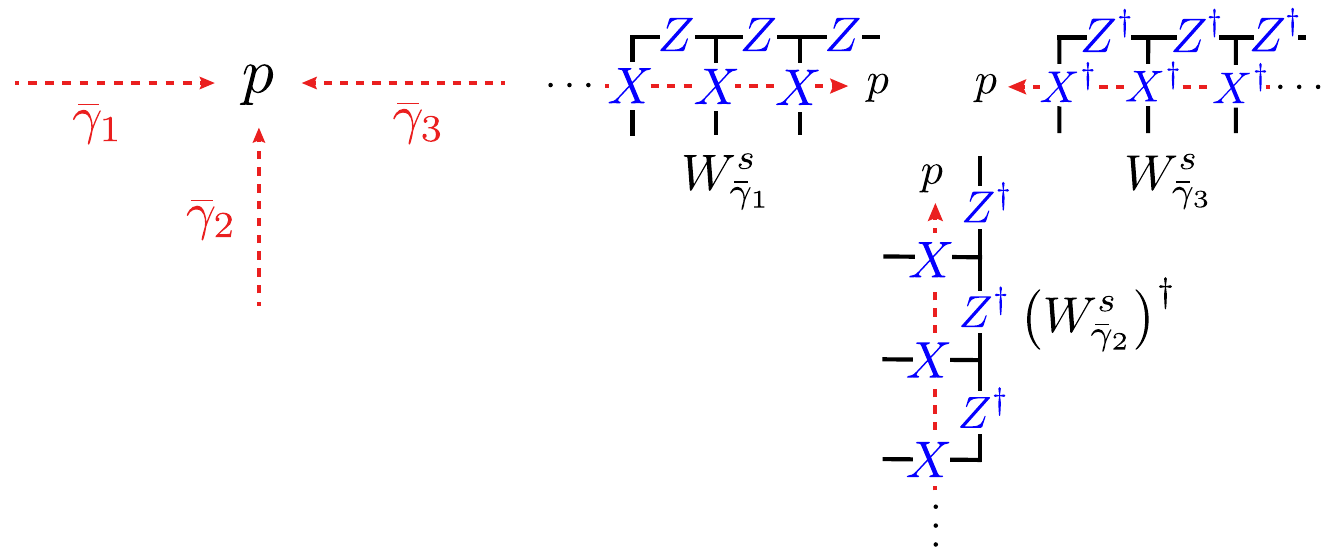}
     \caption{The exchange statistics of the anyon $s$ can be computed using the formula in Eq.~\eqref{eq: statistics formula}. $\bar{\gamma}_1$, $\bar{\gamma}_2$, and $\bar{\gamma}_3$ are oriented paths in the dual lattice incident upon the same plaquette $p$. The string operators $(W^s_{\bar{\gamma}_2})^\dagger$ and $W^s_{\bar{\gamma}_3}$ fail to commute, giving the statistics $\theta(s) =  i$.}
     \label{fig: statistics}
\end{figure}

The next property to check is that the anyonic excitations created by the string operators in Eq.~\eqref{eq: DS string operators} have the same statistics as the anyons of the DS topological order. This is accomplished by following the prescription described in Refs.~\cite{LW06,HFH18,KL20}. Let $\bar{\gamma}_1$, $\bar{\gamma}_2$, and $\bar{\gamma}_3$ be paths sharing a common endpoint $p$ and ordered counter-clockwise around $p$, as in Fig.~\ref{fig: statistics}. Then, the exchange statistics $\theta(a)$ of the anyon $a$ is computed by the expression:
\begin{align} \label{eq: statistics formula}
W^a_{\bar{\gamma}_1}(W^a_{\bar{\gamma}_2})^\dagger W^a_{\bar{\gamma}_3} = \theta(a) W^a_{\bar{\gamma}_3} (W^a_{\bar{\gamma}_2})^\dagger W^a_{\bar{\gamma}_1}.
\end{align}
The computation of $\theta(s)$ is pictured in Fig.~\ref{fig: statistics}. We find that $\theta(s)$, $\theta(\bar{s})$, and $\theta(s \bar{s})$ are:
\begin{align} 
\theta(1)=1, \quad \theta(s)=i, \quad \theta(\bar{s}) = -i, \quad \theta(s\bar{s})=1,
\end{align}
which match the statistics of the DS anyons in Eq.~\eqref{eq: DS statistics}. Therefore, the anyonic excitations of $H_\text{DS}$ have the same fusion and statistics as the anyons in the DS topological order. This implies that $H_\text{DS}$ indeed belongs to the DS phase.

\begin{figure}
\centering
    \includegraphics[width=.32\textwidth]{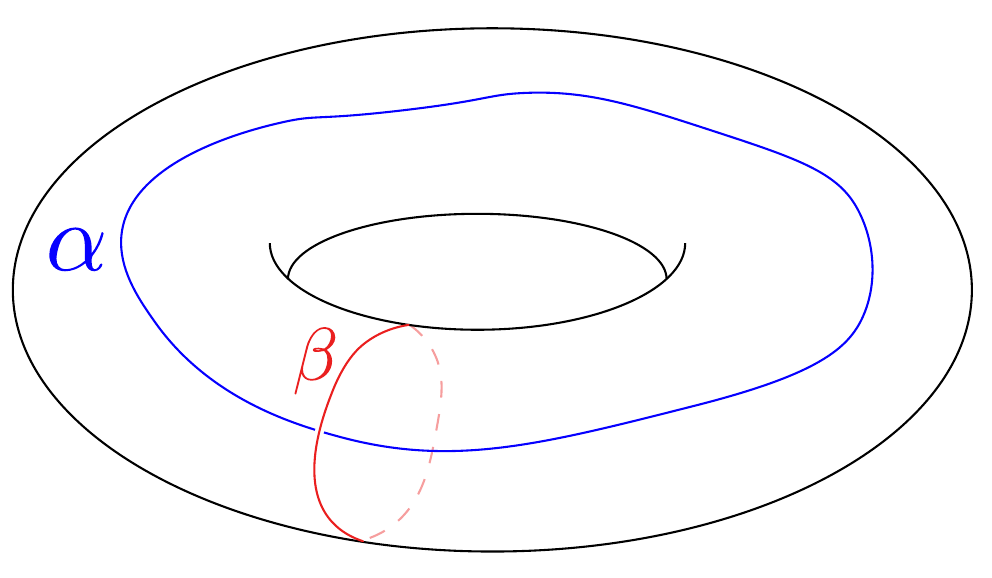}
     \caption{The logical operators of the DS stabilizer code on a torus can be represented by string operators along non-contractible loops such as $\alpha$ (blue) and $\beta$ (red).}
     \label{fig: torusstrings}
\end{figure}

The four-fold degeneracy of $H_\text{DS}$ on a torus can now be understood in terms of the DS anyons. Let $\alpha$ and $\beta$ be generators of the non-contractible cycles on the torus, pictured in Fig.~\ref{fig: torusstrings}. Then, the long string operators $W^s_{\alpha}$, $W^s_{\beta}$, $W^{\bar{s}}_{\alpha}$ and $W^{\bar{s}}_{\beta}$ form Pauli X and Pauli Z on the logical subspace $\mathcal{H}_L$. To make this explicit, we define:
\begin{align}
\bar{X}_1 \equiv W^s_{\alpha}, \quad \bar{Z}_1 \equiv W^s_{\beta}, \quad \bar{X}_2 \equiv W^{\bar{s}}_{\alpha}, \quad \bar{Z}_2 \equiv W^{\bar{s}}_{\beta}.
\end{align}
In the logical subspace, the long string operators satisfy:
\begin{align} \label{eq: DS logical relations}
\bar{X}_1^2 \sim 1, \quad \bar{Z}_1^2 \sim 1, \quad \bar{X}_2^2 \sim 1, \quad \bar{Z}_2^2 \sim 1, 
\end{align}
where $\sim$ emphasizes that the relations hold only in the logical subspace.
The only nontrivial commutation relations between the long string operators are: 
\begin{align} \label{eq: DS logical commutation}
\bar{Z}_1\bar{X}_1 = - \bar{X}_1\bar{Z}_1, \quad \bar{Z}_2\bar{X}_2 = - \bar{X}_2\bar{Z}_2.
\end{align}
Thus, they have the same relations as Pauli X and Pauli Z operators on a pair of qubits. The relations in Eq.~\eqref{eq: DS logical relations} follow from the fusion rules of $s$ and $\bar{s}$, while the commutation relations in Eq.~\eqref{eq: DS logical commutation} follow from the braiding relations in Eq.~\eqref{eq: DS braiding}. Indeed, in this model, the braiding relations of $s$ and $\bar{s}$ can be computed from the commutation relations of string operators that intersect at a single point (see Ref.~\cite{KL20} for a more general statement). 

\subsection{Construction of the Pauli stabilizer model} \label{sec: DS construction}

The DS stabilizer model can be derived from a $\ZZ_4$ TC by condensing a certain bosonic anyon. To demonstrate this, we first describe the construction of the DS stabilizer model at the level of the anyons. We then implement the construction at the lattice level to arrive at the DS stabilizer model.  

\subsubsection*{Anyon-level construction}

Recall that the anyons of the $\ZZ_4$ TC form a $\ZZ_4 \times \ZZ_4$ group, generated by the anyons $e$ and $m$. The sixteen anyons of the $\ZZ_4$ TC are shown in table below:
\begin{align} \nonumber
    \includegraphics[width=.3\textwidth]{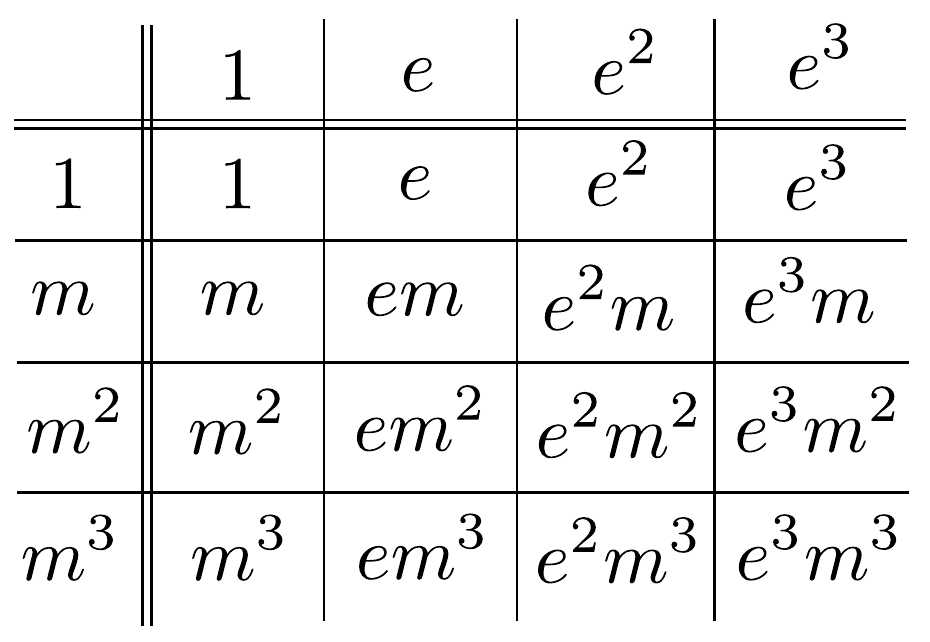}
\end{align}
The statistics of an arbitrary anyon $e^pm^q$ (with $p,q \in \ZZ_4$) is given by:
\begin{align} \label{eq: Z4 QD statistics}
\theta(e^pm^q) = i^{pq}.
\end{align}
Note that the $e$ and $m$ anyons are often interpreted as the gauge charge and gauge flux, respectively, of a $\ZZ_4$ gauge theory.
According to Eq.~\eqref{eq: Z4 QD statistics}, $e$ and $m$ are bosons and have the Aharanov-Bohm phase $i$, as expected:
\begin{align}
\theta(e) = 1, \quad \theta(m) = 1, \quad B_\theta(e,m) = i,
\end{align} 

The construction of the DS stabilizer model from the $\ZZ_4$ TC is motivated by the fact that the $\ZZ_4$ TC has  both semionic excitations and anti-semionic excitations. In particular, $em$ is a semion and $em^3$ is an anti-semion. Further, their fusion product $e^2$ is a boson, similar to the fusion of the semion $s$ and the anti-semion $\bar{s}$ of the DS anyon theory.  The only obstacle to identifying $em$ and $em^3$ with $s$ and $\bar{s}$, respectively, is that $em$ and $em^3$ have order four under fusion, while $s$ and $\bar{s}$ have order two. Indeed, both $em$ and $em^3$ square to the boson $e^2m^2$:
\begin{align} \label{eq: em and em3 fusion}
em \times em = e^2m^2, \quad em^3 \times em^3 = e^2m^2.
\end{align}
The resolution is to condense the boson $e^2m^2$. As described below, condensing $e^2m^2$ ensures that $em$ and $em^3$ represent anyons with the same fusion rules as $s$ and $\bar{s}$. Moreover, the condensation of $e^2m^2$ leaves us with precisely the anyon theory of the DS phase.

The condensation of $e^2m^2$ has two effects. The first is that anyons that braid nontrivially with $e^2m^2$ become confined. As demonstrated later in this section, after condensation, the string operators of the confined anyons create excitations along the length of the string. Therefore, the string operators create nonlocal excitations, and the confined anyons do not correspond to anyonic excitations in the condensed theory. We call the remaining anyons, i.e., those that braid trivially with $e^2m^2$, deconfined anyons. 

The second effect of condensation is that deconfined anyons related by fusion with $e^2m^2$ become identified. This means, in particular, that $e^2m^2$ is identified with the trivial anyon $1$. To emphasize the identification of anyons after condensation, we use square brackets. Thus, in the condensed theory, we have:
\begin{align}
[e^2m^2] = [1].
\end{align}
More generally, for any deconfined anyon $a$ in the $\ZZ_4$ TC, we write:
\begin{align}  \label{eq: esquared msquared identification}
[a] = [a \times e^2m^2].
\end{align} 
Note that the confined anyons are inconsistent with the identification of $e^2m^2$ with the trivial anyon, since braiding with $e^2m^2$ differs from braiding with $1$.   

\begin{table}
    \includegraphics[width=.3\textwidth]{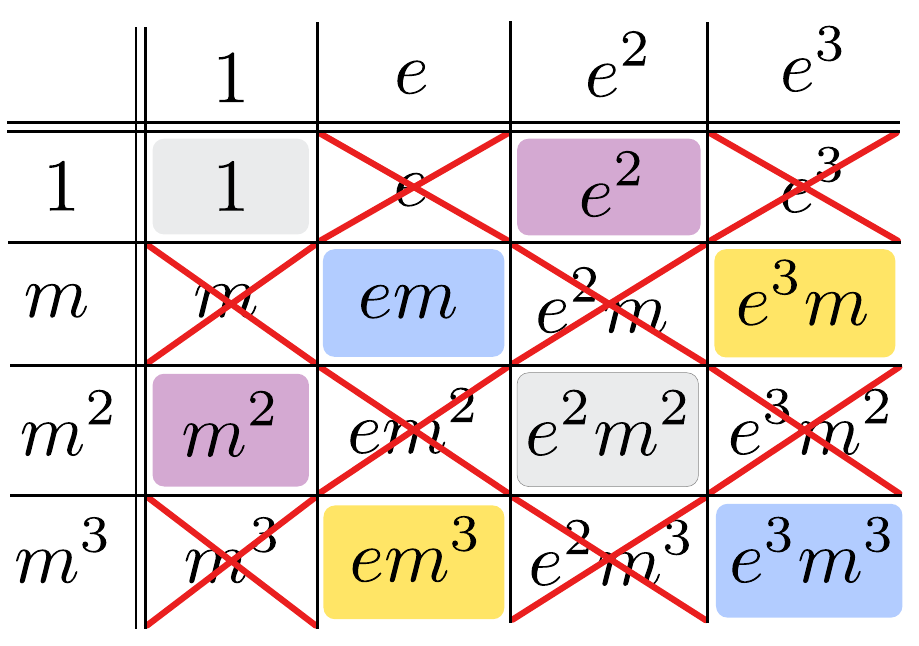}
     \caption{The condensation of $e^2m^2$ excitations in the $\ZZ_4$ TC has two effects. First, the anyons that braid nontrivially with $e^2m^2$ become confined (crossed out in red). Second, the remaining deconfined anyons become identified (shaded in matching colors) if they differ by fusion with $e^2m^2$.}
     \label{table: TCanyonscondensed}
\end{table}

The effects of condensing $e^2m^2$ are shown in Table~\ref{table: TCanyonscondensed}. The anyons that remain after condensation can be labeled by:
\begin{eqs}
\{[1],[em],[em^3],[e^2]\}.
\end{eqs}
Given Eq.~\eqref{eq: esquared msquared identification}, the anyons of the condensed theory satisfy the fusion rules:
\begin{equation}
\begin{gathered}
	{[em]} \times [em] = [1], \quad  [em^3] \times [em^3] =[1], \\ 
	[em] \times [em^3] = [e^2].
\end{gathered}
\end{equation}
Thus, in particular, $[em]$ and $[em^3]$ have order two under fusion, just like $s$ and $\bar{s}$. Furthermore, the exchange statistics of the anyons in the condensed theory are:
\begin{align}
\theta([1]) =1, \quad \theta([em])=i, \quad \theta([em^3])=-i, \quad \theta([e^2])=1.
\end{align}
The exchange statistics above are well-defined, as can be checked using Eq.~\eqref{eq: braiding M def} in Section~\ref{sec: primer} and the fact that $em$ and $em^3$ have trivial braiding relations with $e^2m^2$. We now see that the anyons $\{[1],[em],[em^3],[e^2]\}$ have the same properties as the anyons of the DS phase. Therefore, condensing $e^2m^2$ in the $\ZZ_4$ TC produces the anyon theory of the DS phase.

\subsubsection*{Lattice-level construction}

With this, we are prepared to construct the DS stabilizer model at a lattice level, starting with the lattice model for the $\ZZ_4$ TC. The $\ZZ_4$ TC is defined on a square lattice with a four-dimensional qudit at each edge (Fig.~\ref{fig: squareDSdof}). The Hamiltonian is the following sum of vertex terms and plaquette terms:
\begin{align}
H_\text{TC} = -\sum_v A^\text{TC}_v - \sum_p B^\text{TC}_p + \text{h.c.},
\end{align}
where $A^\text{TC}_v$ and $B^\text{TC}_p$ are given by:
\begin{align}
A^\text{TC}_v \equiv \vcenter{\hbox{\includegraphics[scale=.23,trim={0cm 0cm 0cm 0cm},clip]{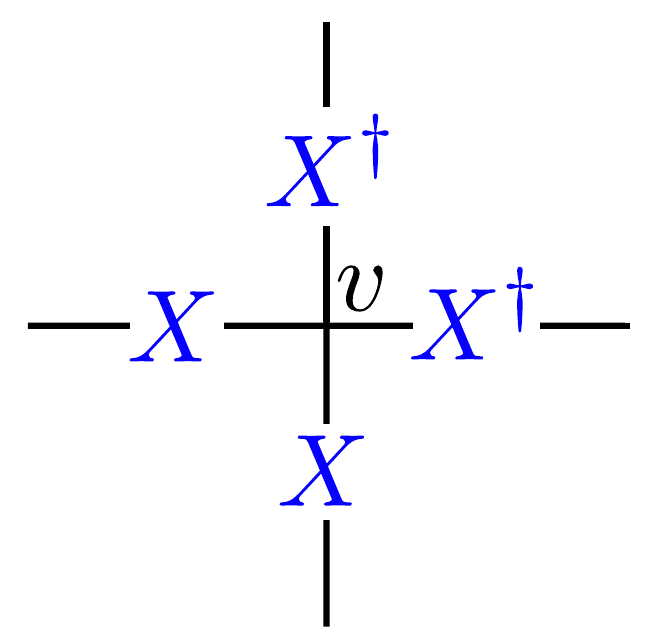}}}, \quad B^\text{TC}_p \equiv \vcenter{\hbox{\includegraphics[scale=.23,trim={0cm 0cm 0cm 0cm},clip]{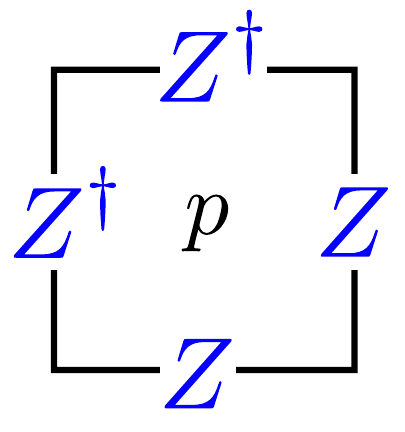}}}.
\end{align}
We define the stabilizer group $\mathcal{S}_\text{TC}$ as the group generated by the vertex terms and plaquette terms of $H_\text{TC}$:
\begin{align}
\mathcal{S}_\text{TC} \equiv \langle \{ A^\text{TC}_v \}, \{ B^\text{TC}_p \} \rangle.
\end{align}

\begin{figure}
\centering
    \includegraphics[width=.5\textwidth]{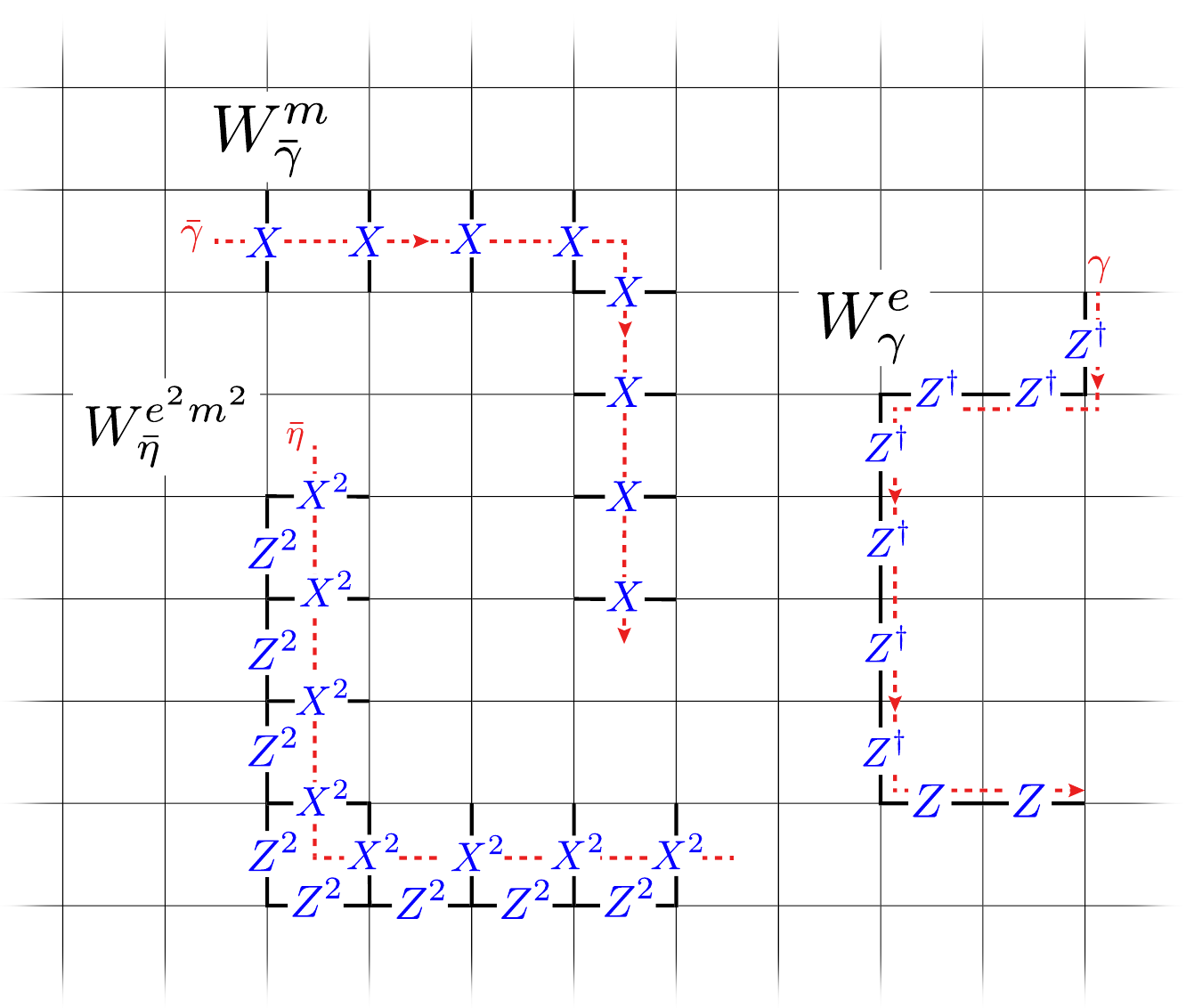}
     \caption{The $e$ and $m$ string operators generate string operators for all of the anyons in the $\ZZ_4$ TC. The $e$ string operator $W^e_\gamma$ is defined along an oriented path $\gamma$ in the direct lattice, while the $m$ string operator $W^m_{\bar{\gamma}}$ is defined along an oriented path $\bar{\gamma}$ in the dual lattice. The $e^2m^2$ string operator $W^{e^2m^2}_{\bar{\eta}}$ is defined along an un-oriented path $\bar{\eta}$ in the dual lattice.}
     \label{fig: TCstrings}
\end{figure}

The anyonic excitations of the $\ZZ_4$ TC are generated by products of the $e$ and $m$ string operators, illustrated in Fig.~\ref{fig: TCstrings}.
Fig.~\ref{fig: TCstrings} also shows a string operator for a pair of $e^2m^2$ excitations.
The $e^2m^2$ string operators can be generated by the short string operators:
\begin{align} \label{eq: Z4 QD short strings}
W^{e^2m^2}_e = \vcenter{\hbox{\includegraphics[scale=.23,trim={0cm 0cm 0cm 0cm},clip]{Figures/stabilizer_Ceh-eps-converted-to.pdf}}}, \,\,\, \vcenter{\hbox{\includegraphics[scale=.23,trim={0cm 0cm 0cm 0cm},clip]{Figures/stabilizer_Cev-eps-converted-to.pdf}}}.
\end{align}
Note that there is ambiguity in how the $e^2$ and $m^2$ excitations are bound together to form $e^2m^2$. In Eq.~\eqref{eq: Z4 QD short strings}, we have made an arbitrary choice for the short string operators. This choice ultimately determines the form of the vertex terms and the string operators of the DS stabilizer model. 
We point out that the short string operators $W^{e^2m^2}_e$ chosen in Eq.~\eqref{eq: Z4 QD short strings} are  precisely the $C_e$ terms from Eq.~\eqref{eq: DS terms}, and throughout the rest of this section, we use the notation $C_e$ to agree with Section~\ref{sec: DS Pauli stabilizer code}.

The next step in the construction of the DS stabilizer model is to condense the $e^2m^2$ excitations. 
To this end, we define $\mathcal{S}_C$ to be the group of $e^2m^2$ string operators. This is generated by the set of short $e^2m^2$ string operators $\{C_e\}$:
\begin{align}
\mathcal{S}_C \equiv \langle \{C_e\} \rangle.
\end{align}
In particular, $\mathcal{S}_C$ includes open string operators, which create, annihilate, and more generally proliferate the $e^2m^2$ excitations. Therefore, after condensing $e^2m^2$, the states $|\psi \rangle$ in the ground state subspace should satisfy:
\begin{align} \label{eq: condensed theory logical requirement}
C |\psi \rangle = |\psi \rangle, \quad \forall C \in \mathcal{S}_C.
\end{align}
In other words, the stabilizer group for the condensed theory should contain $\mathcal{S}_C$ as a subgroup. This is certainly the case for the DS stabilizer group in Eq.~\eqref{eq: DS stabilizer group}.

Only certain elements of the stabilizer group $\mathcal{S}_\text{TC}$ are compatible with the condition in Eq.~\eqref{eq: condensed theory logical requirement}. These are, in particular, the elements of $\mathcal{S}_\text{TC}$ that commute with every element of $\mathcal{S}_C$. We define $\mathcal{S}_\text{TC}^C$ to be the subgroup formed by the stabilizers of $\mathcal{S}_\text{TC}$ that commute with the elements of $\mathcal{S}_C$:
\begin{align}
\mathcal{S}_\text{TC}^C \equiv \{S \in \mathcal{S}_\text{TC} : S C  = C S, \, \forall C \in \mathcal{S}_C\}.
\end{align}
To understand the elements of $\mathcal{S}_\text{TC}^C$, it is useful to note that the stabilizer group $\mathcal{S}_\text{TC}$ is generated by small loops of $e$ and $m$ string operators, corresponding to $B^\text{TC}_p$ and $A^\text{TC}_v$, respectively. This means that the elements of $\mathcal{S}_\text{TC}$ are, in general, products of closed string operators of the $\ZZ_4$ TC anyons, such as loops of $em$ string operators or $e^2$ string operators. Therefore, the elements of $\mathcal{S}_\text{TC}^C$ are loops of string operators that commute with the open $e^2m^2$ string operators. These correspond to closed string operators of anyons that braid trivially with $e^2m^2$. As a result, $\mathcal{S}^C_\text{TC}$ can be generated by small loops of $em$ string operators and small loops of $e^2$ string operators. These are exactly the $A_v$ and $B_p$ terms of the DS Hamiltonian in Eq.~\eqref{eq: DS terms}:
\begin{align}
\mathcal{S}^C_\text{TC} = \langle \{A_v\}, \{B_p\} \rangle.
\end{align}
$\mathcal{S}^C_\text{TC}$ is designed to exclude the closed string operators of confined anyons. 

Finally, the stabilizer group of the condensed theory is generated by $\mathcal{S}_\text{TC}^C$ and $\mathcal{S}_C$, i.e., it is given by:
\begin{align} \label{eq: DS stabilizer group derived}
\langle \mathcal{S}_\text{TC}^C, \mathcal{S}_C \rangle = \langle  \{A_v\}, \{B_p\}, \{C_e\} \rangle.
\end{align}
The justification for taking $\langle \mathcal{S}_\text{TC}^C, \mathcal{S}_C \rangle$ to be the stabilizer group of the condensed theory comes from considering the corresponding Hamiltonian, which is precisely the DS stabilizer Hamiltonian $H_\text{DS}$. We show that $H_\text{DS}$ exhibits the two effects of condensation described below Eq.~\eqref{eq: em and em3 fusion}. Namely, the string operators of the confined $\ZZ_4$ TC anyons create confined excitations, and the string operators of deconfined $\ZZ_4$ TC anyons create identical anyonic excitations, if they differ by an $e^2m^2$ string operator.

To give an example of a confined excitation in the DS stabilizer model, let us consider the $e$ string operators of the $\ZZ_4$ TC.  
An open $e$ string operator $W_\gamma^e$, such as the one depicted in Fig.~\ref{fig: TCstrings}, fails to commute with the vertex terms $A_v$ of the DS stabilizer Hamiltonian at the endpoints of the path $\gamma$. It also fails to commute with the edge terms $C_e$ along the length of $\gamma$. Therefore, separating the vertex excitations at the endpoints of $\gamma$ comes with an energetic penalty that grows linearly with the length of the string operator. This is indicative of confinement. On an infinite plane, it is not possible to create a single vertex excitation (i.e., change the eigenvalue of $A_v$ by $i$) without creating either an extensive number of edge excitations or a second vertex excitation. We note that confined excitations, such as those created by the $e$ string operator in the DS stabilizer model, are only possible in Pauli stabilizer models on composite-dimensional qudits. For translationally invariant Pauli stabilizer models built from prime-dimensional qudits on an infinite plane, it is always possible to choose the local stabilizer terms of the Hamiltonian so that one can violate any given single stabilizer term using some (possibly nonlocal) product of Pauli operators \cite{Haah2018a}. 

In addition to confinement, we see that anyons created by string operators of the deconfined $\ZZ_4$ TC anyons become identified, if the corresponding $\ZZ_4$ TC anyons differ by $e^2m^2$.
As a concrete example, the string operators for $em$ and $e^3m^3$ differ by an $e^2m^2$ string operator. Since the $e^2m^2$ string operators are stabilizers in the DS stabilizer model, the $em$ and $e^3m^3$ string operators create the same anyons when applied to a ground state of $H_\text{DS}$. Therefore, the deconfined anyons $em$ and $e^3m^3$ have become identified, and $H_\text{DS}$ indeed describes the theory obtained after condensing $e^2m^2$ in a $\ZZ_4$ TC.

Interestingly, the construction of the DS stabilizer code described above can be understood in terms of the effects of Pauli measurements on stabilizer groups. Recall that, given a stabilizer group $\mathcal{S}$, the measurement of a product of Pauli operators $P$ results in a modified stabilizer group. The modified stabilizer group is generated by $\pm P$ (depending on the measurement outcome \footnote{Here, for simplicity, we have assumed that $P$ has $\pm 1$ eigenvalues, which is the case for the $C_e$ operators.}) and the elements of $\mathcal{S}$ that commute with $P$. Starting with the stabilizer group $\mathcal{S}_\text{TC}$, measurements of the short string operators $\{C_e\}$ produce the stabilizer group:
\begin{align}
\langle \mathcal{S}^C_\text{TC}, \{\pm C_e\} \rangle = \langle  \{A_v\}, \{B_p\}, \{\pm C_e\} \rangle.
\end{align}
By post-selecting for $+1$ measurement outcomes or by error correction, we obtain the stabilizer group $\mathcal{S}_\text{DS}$.
Hence, the condensation of $e^2m^2$ can be implemented by simply measuring the set of operators $\{C_e\}$. This yields an efficient construction of the DS stabilizer code from a $\ZZ_4$ TC, requiring only two-body Pauli measurements.

\subsection{Relation to the string-net model} \label{sec: relation to nonPauli}

\begin{figure*}
\subfloat[\label{fig: squaretotridof}]{ \includegraphics[width=.35\textwidth]{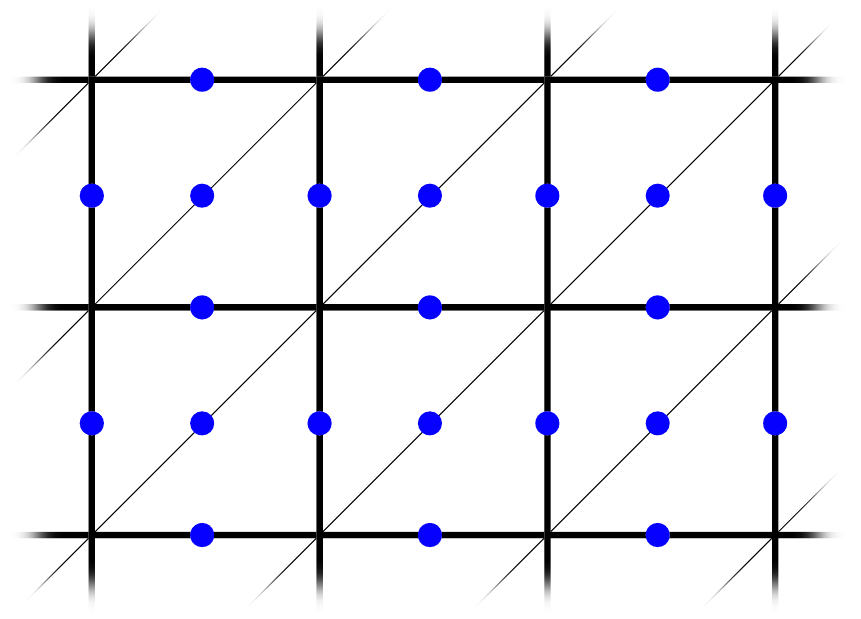}}
\quad \,\,\,\,\,
\subfloat[\label{fig: triDSdof}]{\includegraphics[width=.40\textwidth]{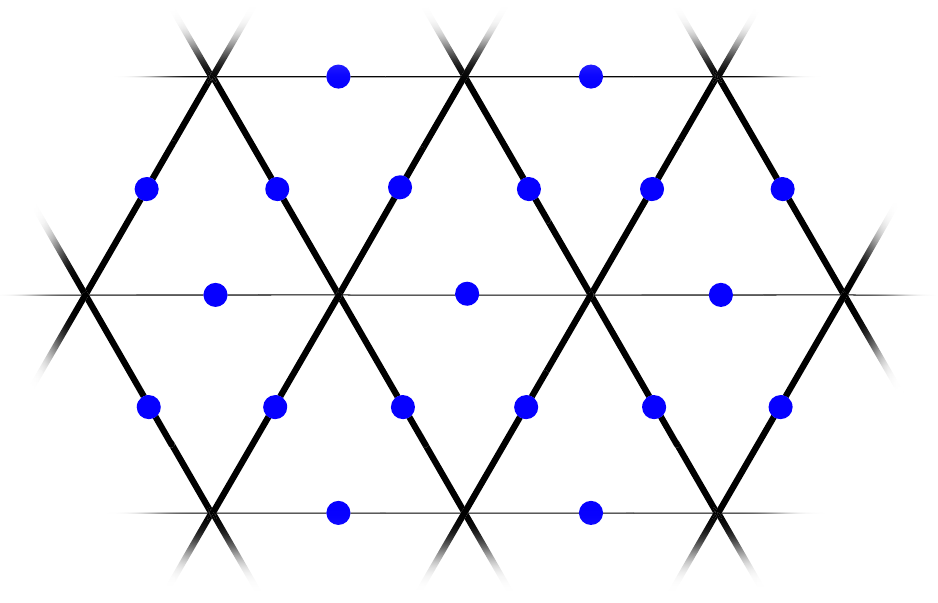}} \!\!\!\!\!\!\!\!\!\!\!\!\!\!\!\!\!\!\!\!\!\!\!\!\! \\
\subfloat[\label{fig: triDSqubits}]{\includegraphics[width=.44\textwidth, trim={.37cm 0cm 0cm 0cm},clip]{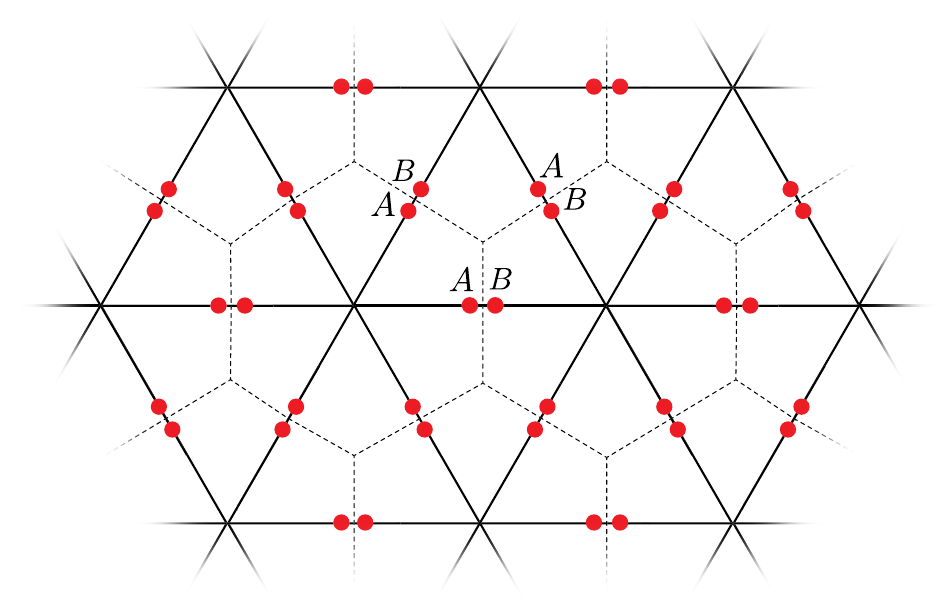}}
\quad \,
\subfloat[\label{fig: snDSdof}]{ \includegraphics[width=.33\textwidth]{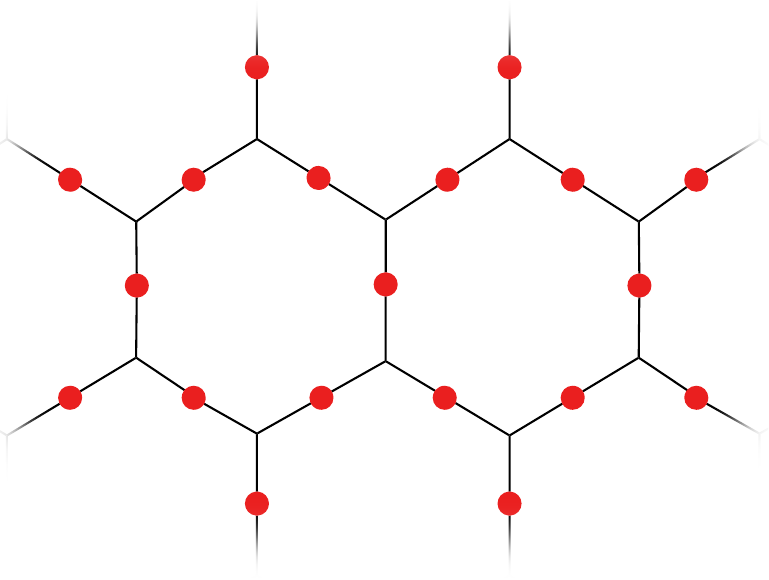}}
\caption{(a) We add an ancillary qudit to the center of each square plaquette. (b) We rotate (and shear) the lattice so that the qudits now lie on the edges of regular triangular lattice. (c) We add a pair of ancillary qubits (red) to each edge, and label them by $A$ and $B$. The transformation in Eq.~\eqref{eq: qudit to qubits} is then used to map the Hamiltonian on qudits to a Hamiltonian on qubits. Subsequently, the ancillary qudits can be removed (blue dots suppressed). (d) The finite-depth quantum circuit $\mathcal{U}_{AB}$ disentangles the $B$ qubits from the system, giving us a single qubit on each edge of the hexagonal dual lattice. This is the lattice on which the DS string-net model is defined.}
\end{figure*}

We now identify a finite-depth quantum circuit $\mathcal{U}$ (with ancillary degrees of freedom) that maps the ground state subspace of the DS stabilizer Hamiltonian $H_\text{DS}$ to the ground state subspace of the DS string-net model $H_\text{DS}^\text{s-n}$. This establishes that $H_\text{DS}$ belongs to the same topological phase of matter as $H_\text{DS}^\text{s-n}$, by the arguments of Ref.~\cite{chen2010local}. Our strategy is to map $H_\text{DS}$ to $H_\text{DS}^\text{s-n}$ using the following operations: 
\begin{itemize}
\item adding and removing ancillary degrees of freedom,
\item conjugating $H_\text{DS}$ by layers of $\mathcal{U}$, 
\item making ground space-preserving changes to the Hamiltonian.
\end{itemize}
This transformation shows that the ground state subspace of the Hamiltonian $\mathcal{U}H_\text{DS}\mathcal{U}^\dagger$ is equivalent to that of $H_\text{DS}^\text{s-n}$ (up to adding ancillary degrees of freedom to $H_\text{DS}$ and $H_\text{DS}^\text{s-n}$). 
We refer to Appendix~\ref{app: string-net ground states} for a more explicit mapping of the ground states using notation from simplicial cohomology.

Let us start by briefly recalling the DS string-net model of Refs.~\cite{LW05} and \cite{LG12}. The DS string-net model is defined on a hexagonal lattice with a qubit at each edge (see Fig.~\ref{fig: snDSdof}). The Hamiltonian $H_\text{DS}^\text{s-n}$ is a sum of two types of terms and is given by:
\begin{align} \label{eq: snDS def}
H_\text{DS}^\text{s-n} \equiv -\sum_v A_v^\text{s-n} - \sum_p B_p^\text{s-n} + \text{h.c.}.
\end{align}
The vertex terms $A_v^\text{s-n}$ and plaquette terms $B_p^\text{s-n}$ are represented as:
\begin{eqs} \label{eq: snDS terms}
A_v^\text{s-n} &\equiv \vcenter{\hbox{\includegraphics[scale=.23,trim={0cm 0cm 0cm 0cm},clip]{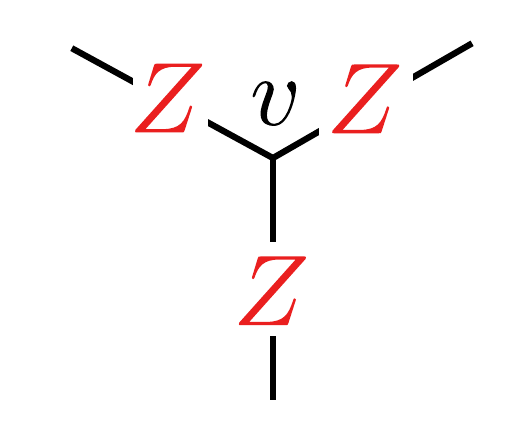}}}\!\!\!, \,\,\, \,\vcenter{\hbox{\includegraphics[scale=.23,trim={0cm 0cm 0cm 0cm},clip]{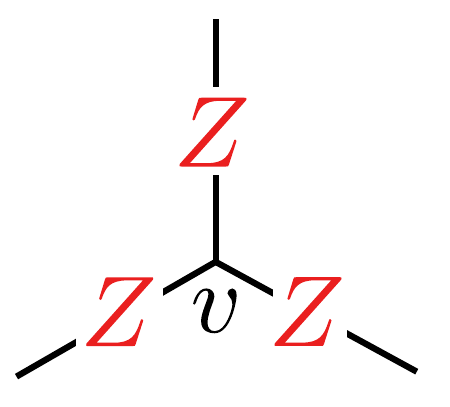}}}, \\
B_p^\text{s-n} &\equiv - \vcenter{\hbox{\includegraphics[scale=.23,trim={0cm 0cm 0cm 0cm},clip]{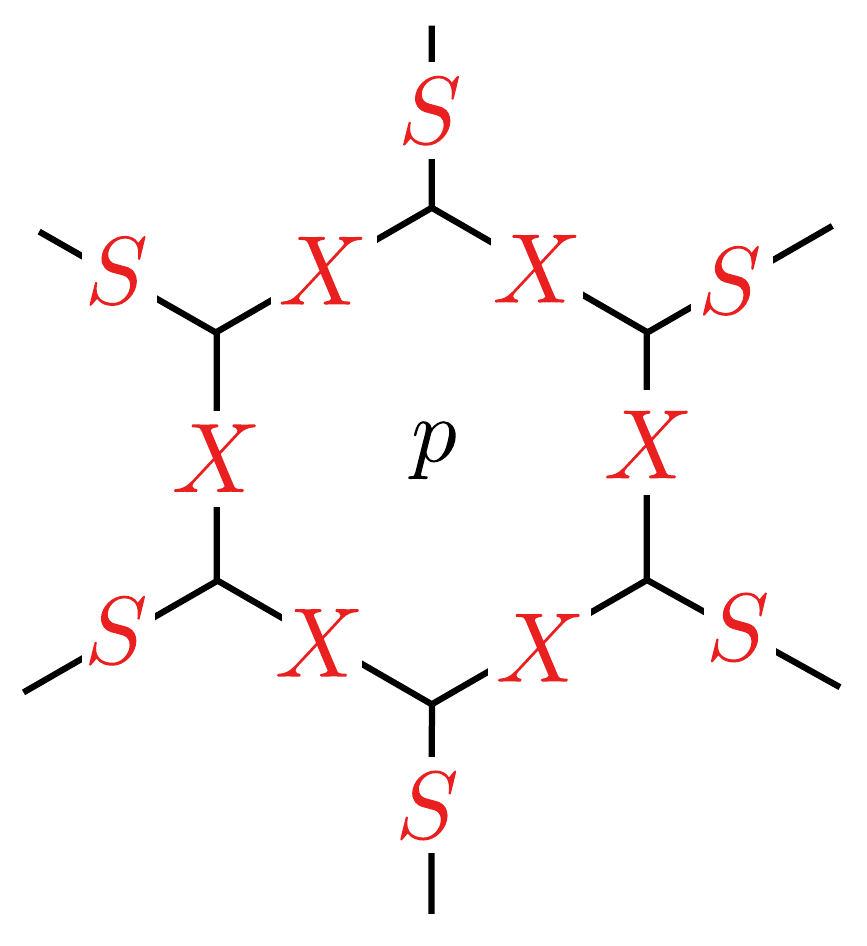}}},
\end{eqs}
where the Pauli X and Pauli Z operators in Eq.~\eqref{eq: snDS terms} are defined on qubits and $S$ is the phase gate with diagonal elements $\text{diag}(1,i)$. $H_\text{DS}^\text{s-n}$ is not a stabilizer model, since the $B_p^\text{s-n}$ terms only commute with one another up to products of $A_v^\text{s-n}$ terms. Nonetheless, the ground state subspace is the mutual $+1$ eigenspace of the Hamiltonian terms.

To relate the DS stabilizer Hamiltonian to $H_\text{DS}^\text{s-n}$, the first step is to map $H_\text{DS}$ to a DS stabilizer model on a triangular lattice. The motivation is that the triangular lattice is dual to the hexagonal lattice, on which the DS string-net model is defined. This mapping is accomplished by introducing an ancillary four-dimensional qudit to the center of each square plaquette (see Fig.~\ref{fig: squaretotridof}). For convenience, we rotate the lattice by $45^\circ$, so that the ancillary qudit lies on the horizontal edge of a regular triangular lattice, as shown in Fig.~\ref{fig: triDSdof}. We label the vertices, edges, and faces of the triangular lattice by $v$, $e$, and $f$, respectively. We take the qudits on the horizontal edges $e$ to be stabilized by the group $\langle X_{e}^2,Z^2_{e}\rangle$. The Hamiltonian on the system with ancillary qudits is then:
\begin{align} \label{eq: squareDS with ancilla}
H_\text{DS} - \sum_{e \in E_h} X_{e}^2 - \sum_{e \in E_h}Z_{e}^2,
\end{align}
where the sums are over the set of horizontal edges $E_h$.

We now couple $H_\text{DS}$ to the ancillary qudits to create a DS stabilizer Hamiltonian on the triangular lattice. We do so by conjugating $H_\text{DS}$ by a finite-depth quantum circuit $\mathcal{U}_{CX}$ composed of control-$X$ gates. We denote a control-$X$ gate with control qudit at edge $e$ and target qudit at edge $e'$ by $CX_{ee'}$. The gate $CX_{ee'}$ is defined by the mapping of operators:
\begin{equation}
\begin{aligned}
X_e &\longleftrightarrow X_eX_{e'}, & X_{e'}  &\longleftrightarrow X_{e'}, \\
Z_e &\longleftrightarrow Z_{e}, & Z_{e'} &\longleftrightarrow Z_e^\dagger Z_{e'}. 
\end{aligned}
\end{equation}
To specify the finite-depth quantum circuit $\mathcal{U}_{CX}$, we label the vertices 
of the upward pointing triangles according to:
\begin{align} \label{eq: upwards tri ordering}
\vcenter{\hbox{\includegraphics[width=.12\textwidth]{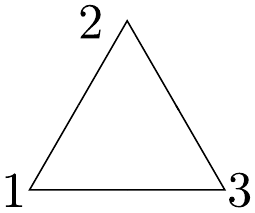}}}.
\end{align}
With this, $\mathcal{U}_{CX}$ is the product of control-$X$ gates:
\begin{align} \label{eq: UCX def}
\mathcal{U}_{CX} \equiv \prod_{f \in F_\text{up}} CX_{ 12 ,13 } CX_{ 23, 13 },
\end{align}
where the product is over the set of all upwards pointing triangles $F_\text{up}$, and the control and target qudits are specified by pairs of vertices according to the labeling in  Eq.~\eqref{eq: upwards tri ordering}. $\mathcal{U}_{CX}$ forms the first two layers of $\mathcal{U}$ \footnote{$\mathcal{U}_{CX}$ is two layers because the gates $CX_{12,13}$ and $CX_{23,13}$ have overlapping supports.}.

We construct the DS stabilizer Hamiltonian $H^I_\text{DS}$ on a triangular lattice by conjugating the Hamiltonian in Eq.~\eqref{eq: squareDS with ancilla} by $\mathcal{U}_{CX}$. Here, we have started to label the intermediate Hamiltonians with Roman numerals. Up to redefining the generators of the stabilizer group (which preserves the ground state subspace), conjugation by $\mathcal{U}_{CX}$ yields the Hamiltonian:
\begin{align} \label{eq: triDS}
H^I_\text{DS} \equiv -\sum_v A^I_v - \sum_f B^I_f -\sum_e C^I_e + \text{h.c.}.
\end{align}
The terms $A^I_v$, $B^I_f$, and $C^I_e$ are given graphically as:
\begin{eqs} \nonumber
A^I_v &= \vcenter{\hbox{\includegraphics[scale=.23,trim={0cm 0cm 0cm 0cm},clip]{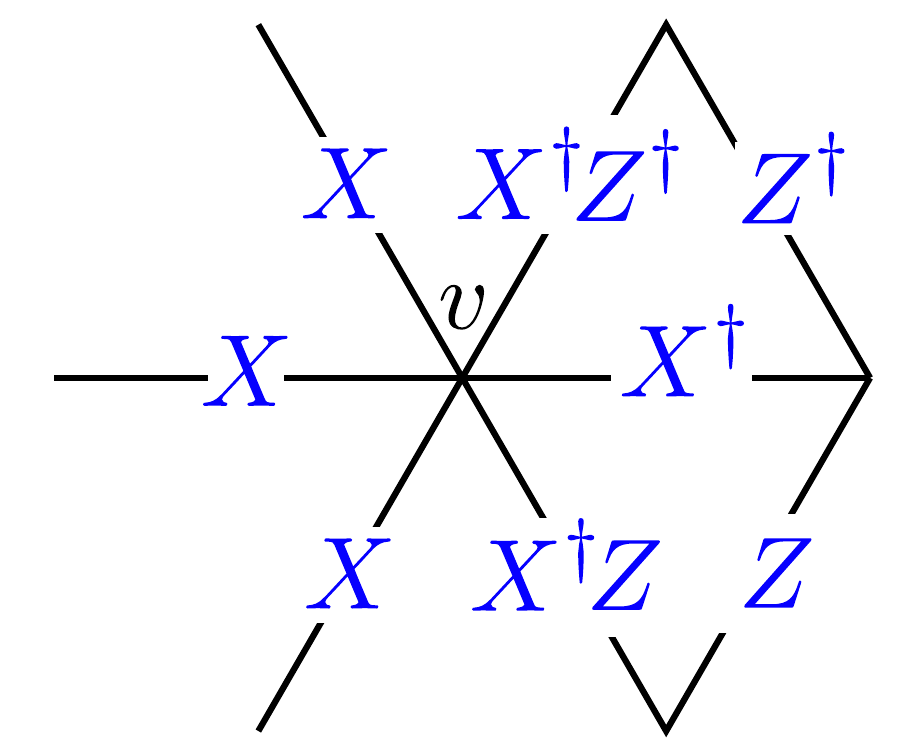}}},
\end{eqs}
\begin{eqs} \label{eq: triDS terms}
B^I_f &= \vcenter{\hbox{\includegraphics[scale=.23,trim={0cm 0cm 0cm 0cm},clip]{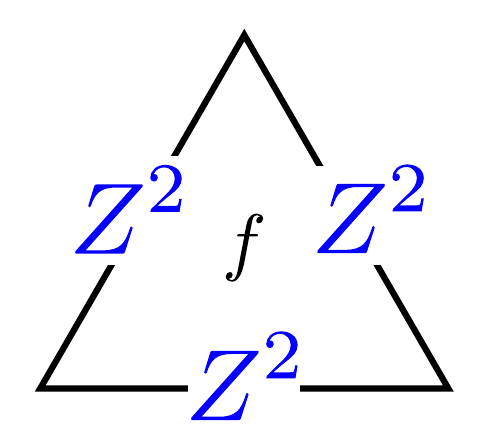}}}, \, \,  \vcenter{\hbox{\includegraphics[scale=.23,trim={0cm 0cm 0cm 0cm},clip]{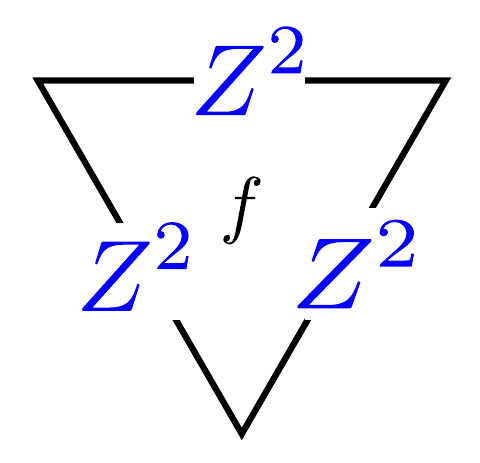}}},\\
C^I_e &= \vcenter{\hbox{\includegraphics[scale=.23,trim={0cm 0cm 0cm 0cm},clip]{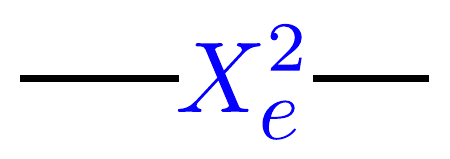}}}, \,\,\, \vcenter{\hbox{\includegraphics[scale=.23,trim={0cm 0cm 0cm 0cm},clip]{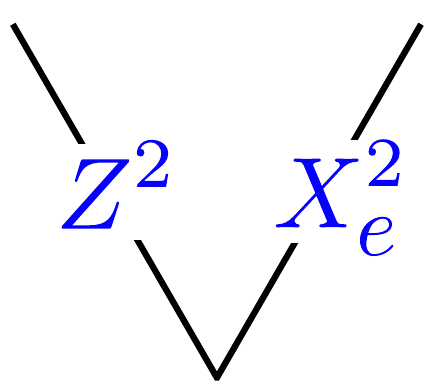}}}, \,\,\, \vcenter{\hbox{\includegraphics[scale=.23,trim={0cm 0cm 0cm 0cm},clip]{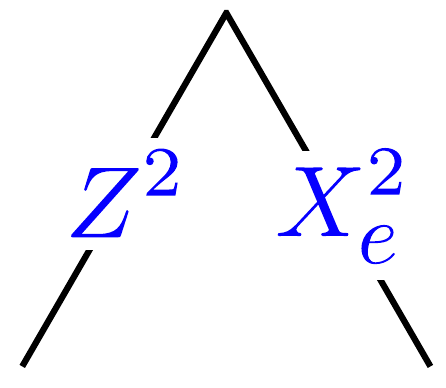}}}.
\end{eqs}
Notice that the $X_e^2$ terms in Eq.~\eqref{eq: squareDS with ancilla} have been absorbed into the definition of $C^I_e$, and the face term $B^I_f$ associated to the upwards pointing triangle comes from conjugating the $Z_e^2$ horizontal edge term. 
We point out that the construction of the DS stabilizer model in Section~\ref{sec: DS construction} carries over to the triangular lattice and produces precisely the Hamiltonian in Eq.~\eqref{eq: triDS}. 

Since the DS string-net model is defined on a system of qubits, the next step is to map each four-dimensional qudit of the DS stabilizer Hamiltonian to a pair of qubits. This can be accomplished by introducing a pair of qubits to each edge. To make this explicit, we label a pair of qubits at an edge $e$ by $A$ and $B$ (see Fig~\ref{fig: triDSqubits}) and denote the Pauli X and Pauli Z operators by $X^A_e, \, Z^A_e, \, X^B_e, \, Z^B_e$. 
The system on four-dimensional qudits can then be mapped to qubits by conjugating with a finite-depth quantum circuit $\mathcal{U}_{2,4}$. The unitary operator $\mathcal{U}_{2,4}$ is defined by the mapping of operators \footnote{We remark that the mapping in Eq.~\eqref{eq: qudit to qubits} differs from the mapping in \cite{SSC20} by conjugation with a Hadamard.}:
\begin{align}\label{eq: qudit to qubits}
Z_e \longleftrightarrow S_e^AZ^B_e, \quad X_e \longleftrightarrow X^A_e  CX_e^{AB}.
\end{align}
The operators on the left-hand sides of Eq.~\eqref{eq: qudit to qubits} are defined on the four-dimensional qudit, while the operators on the right-hand sides act on the pair of qubits. Here, $CX_e^{AB}$ is the control-$X^B$ gate with the $A$ qubit as the control qubit. The mapping of operators in Eq.~\eqref{eq: qudit to qubits} maps the Pauli stabilizer model $H^I_\text{DS}$ to a non-Pauli stabilizer model $H^{II}_\text{DS}$. Note that the qudit degrees of freedom become decoupled from the system, and may be disregarded \footnote{Indeed, the transformation in Eq.~\eqref{eq: qudit to qubits} implies:
\begin{align}
Z^2_e \longleftrightarrow Z^A_e, \quad \left(\frac{1+i}{2} Z_e + \frac{1-i}{2}Z^\dagger_e \right) \longleftrightarrow Z^B_e.
\end{align}
Therefore, the ancillary product state on qubits is mapped to a product state on qudits.}.

For simplicity, let us focus on the effects of the transformation in Eq.~\eqref{eq: qudit to qubits} on the $C_e^I$ terms. In terms of the operator algebra on the $A$ and $B$ qubits, $C_e^{II} = \mathcal{U}_{2,4} C_e^I \mathcal{U}_{2,4}^\dagger$ is:
\begin{align} \label{eq: Cetriqubits}
C_e^{II} = \vcenter{\hbox{\includegraphics[scale=.23,trim={0cm 0cm 0cm 0cm},clip]{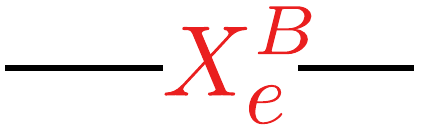}}}, \,\,\, \vcenter{\hbox{\includegraphics[scale=.23,trim={0cm 0cm 0cm 0cm},clip]{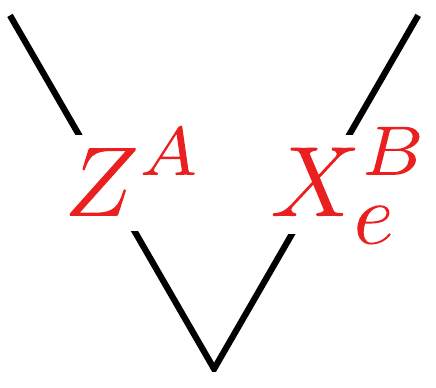}}}, \,\,\, \vcenter{\hbox{\includegraphics[scale=.23,trim={0cm 0cm 0cm 0cm},clip]{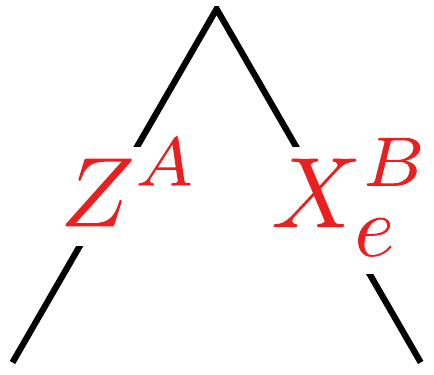}}}.
\end{align}
The form of $C_e^{II}$ is noteworthy, because, as argued below, the $B$ qubits can be disentangled from the ground states by applying control-$Z$ gates. That is, the control-$Z$ gates are designed to map each $C_e^{II}$ in Eq.~\eqref{eq: Cetriqubits} to a single site $X^B_e$ operator.

The ground state subspace of the DS string-net model can now be prepared from the ground state subspace of ${H}^{II}_\text{DS}$ by applying a finite-depth quantum circuit $\mathcal{U}_{CZ}$ composed of control-$Z$ gates. $\mathcal{U}_{CZ}$ consists of two layers and takes the form:
\begin{align}
\mathcal{U}_{CZ} \equiv \mathcal{U}_{AA} \mathcal{U}_{AB}.
\end{align}
The first layer $\mathcal{U}_{AB}$ is needed to decouple the $B$ qubits, leaving us with one qubit per edge. The second layer $\mathcal{U}_{AA}$ ensures that the ground state wave functions have the correct amplitudes, i.e., the amplitudes match those of the ground states of $H_\text{DS}^\text{s-n}$ defined on the hexagonal dual lattice. 

Let us define the layers of $\mathcal{U}_{CZ}$ more carefully, starting with $\mathcal{U}_{AB}$. 
$\mathcal{U}_{AB}$ is the product of control-$Z$ gates: 
\begin{align} \label{eq: UAB def}
\mathcal{U}_{AB} \equiv \prod_{\langle 123 \rangle} CZ^{AB}_{ 12 , 23 }.
\end{align}
The product above is over all faces $\la 123 \ra$, and $CZ^{AB}_{12 , 23 }$ is the control-$Z$ gate between the $A$ site on the edge $\langle 12 \rangle$ and the $B$ site on the edge $\langle 23 \rangle$ with the ordering of vertices:
\begin{eqs} \label{eq: updownordering}
\vcenter{\hbox{\includegraphics[width=.12\textwidth]{Figures/stabilizer_uptriordering-eps-converted-to.pdf}}}\!\!\!\!,  \vcenter{\hbox{\includegraphics[width=.12\textwidth]{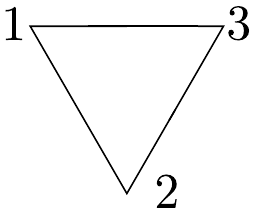}}}.
\end{eqs}
Given an arbitrary face $\langle 123 \rangle$, conjugation by $\mathcal{U}_{AB}$ maps the Pauli X operators $X^A_{ 12 }$ and $X^B_{ 23 }$ according to:
\begin{align} \label{eq: AB operator mapping}
X^A_{ 12 } \longleftrightarrow X^A_{ 12 }Z^B_{ 23 }, \quad X^B_{ 23 } \longleftrightarrow Z^A_{ 12 }X^B_{23 },
\end{align}
where $X_{vv'}^B$ denotes the $B$ site Pauli X operator on the edge $\langle vv' \rangle$.
Importantly, this maps each edge operator $C^{II}_e$ to a single site operator $X_e^B$:
\begin{align} \label{eq: Ceconjugated}
    \mathcal{U}_{AB}C^{II}_e\mathcal{U}_{AB}^\dagger = \, \vcenter{\hbox{\includegraphics[scale=.23,trim={0cm 0cm 0cm 0cm},clip]{Figures/stabilizer_Cetri3-eps-converted-to.pdf}}}, \,\,\,\,\, \vcenter{\hbox{\includegraphics[scale=.23,trim={0cm 0cm 0cm 0cm},clip]{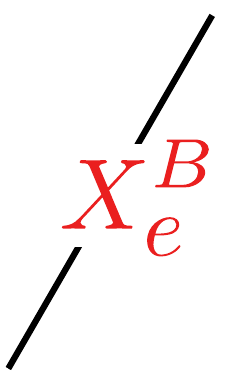}}}, \,\,\,\,\,\vcenter{\hbox{\includegraphics[scale=.23,trim={0cm 0cm 0cm 0cm},clip]{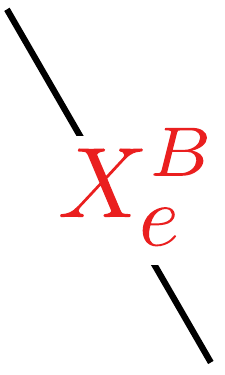}}}.
\end{align}

\setlength{\arrayrulewidth}{0.2mm}
\setlength{\tabcolsep}{4.5pt}
\renewcommand{\arraystretch}{1.5}

\begin{table}[t]
    \small
    \centering
    \begin{tabular}{ |c | c | c | c | c |}
    \hline
    $a_{12}$ & $a_{13}$ & $a_{23}$ & $CZ_{12,23} |a_{12} a_{13} a_{23} \rangle$ &  $S^\dagger_{12}S_{13}S^\dagger_{23}  |a_{12} a_{13} a_{23} \rangle$ \\
    \hline
    0 & 0 & 0 & 1 & 1   \\
    \hline
    0 & 1 & 1 & 1 & 1   \\
    \hline
    1 & 0 & 1 & $-1$ & $-1$   \\
    \hline
    1 & 1 & 0 & 1 & 1   \\
    \hline
    \end{tabular}
    \caption{\label{tab: checking da and aUa} The identity in Eq.~\eqref{eq: CZ to S identity} can be checked explicitly on the computational basis states that satisfy $B^{III}_f=1$. We show the calculation for an upwards pointing triangle $\langle 123 \rangle$, whose vertices are ordered according to Eq.~\eqref{eq: updownordering}. $a_{vv'}$ denotes the $\{0,1\}$ value at the edge $\langle vv' \rangle$. $CZ_{12,23}$ and $S^\dagger_{12} S_{13} S^\dagger_{23}$ are diagonal in the computational basis and their eigenvalues (tabulated above) agree on the computational basis states satisfying $B^{III}_f=1$.}
\end{table}

The next layer, $\mathcal{U}_{AA}$, is a product of control-$Z$ gates between the $A$ sites. We let $CZ^{AA}_{12 , 23 }$ denote the control-$Z$ operator between the $A$ site on edge $\langle 12 \rangle$ and the $A$ site on edge $\langle 23 \rangle$ using the ordering of vertices in Eq.~\eqref{eq: updownordering}.
 $\mathcal{U}_{AA}$ is then defined to be the finite-depth quantum circuit:
\begin{align} \label{eq: UAA def}
\mathcal{U}_{AA} = \prod_{f \in F_\text{up}} CZ^{AA}_{ 12 , 23 },
\end{align}
where the product is over the set of all upward pointing triangles $F_\text{up}$. The Pauli X operators $X^A_{ 12 }$ and $X^A_{ 23 }$ defined on the edges of the upwards pointing triangle $\la 123 \ra$ are mapped as:
\begin{align} \label{eq: AA operator mapping}
X^A_{ 12 } \longleftrightarrow X^A_{ 12 }Z^A_{ 23 }, \quad X^A_{ 23 } \longleftrightarrow  Z^A_{ 12 } X^A_{ 23 }.
\end{align}
Notice that conjugation by $\mathcal{U}_{AA}$ does not affect the edge terms in Eq.~\eqref{eq: Ceconjugated}.

After conjugating $H^{II}_\text{DS}$ by $\mathcal{U}_{CZ}$, the ground states are in a product state on the $B$ sites. This follows from the fact that the ground states are $+1$ eigenstates of the $\mathcal{U}_{CZ} C^{II}_e \mathcal{U}_{CZ}^\dagger = X^B_e$ terms. This means that, without affecting the ground state subspace, we can freely replace all $B$ site operators with the identity. In what follows, we suppress the the $A$ site label and consider systems with a single qubit at each edge. 

Composing the transformations in Eqs.~\eqref{eq: UCX def}, \eqref{eq: qudit to qubits}, \eqref{eq: UAB def}, and \eqref{eq: UAA def}, we find that ${H}_\text{DS}$ is mapped by the finite-depth quantum circuit:
\begin{align}
\mathcal{U}\equiv\mathcal{U}_{CZ} \mathcal{U}_{2,4} \mathcal{U}_{CX},
\end{align}
to the Hamiltonian:
\begin{eqs} \label{eq: Hprime def}
H^{III}_\text{DS} \equiv \mathcal{U}{H}_\text{DS}\mathcal{U}^\dagger = -\sum_v A^{III}_v -\sum_f B^{III}_f,
\end{eqs}
where we have suppressed the ancillary degrees of freedom.
The vertex terms $A^{III}_v$ and face terms $B^{III}_f$ are pictured below: 
\begin{eqs}
A^{III}_v &\equiv -\vcenter{\hbox{\includegraphics[scale=.23,trim={0cm 0cm 0cm 0cm},clip]{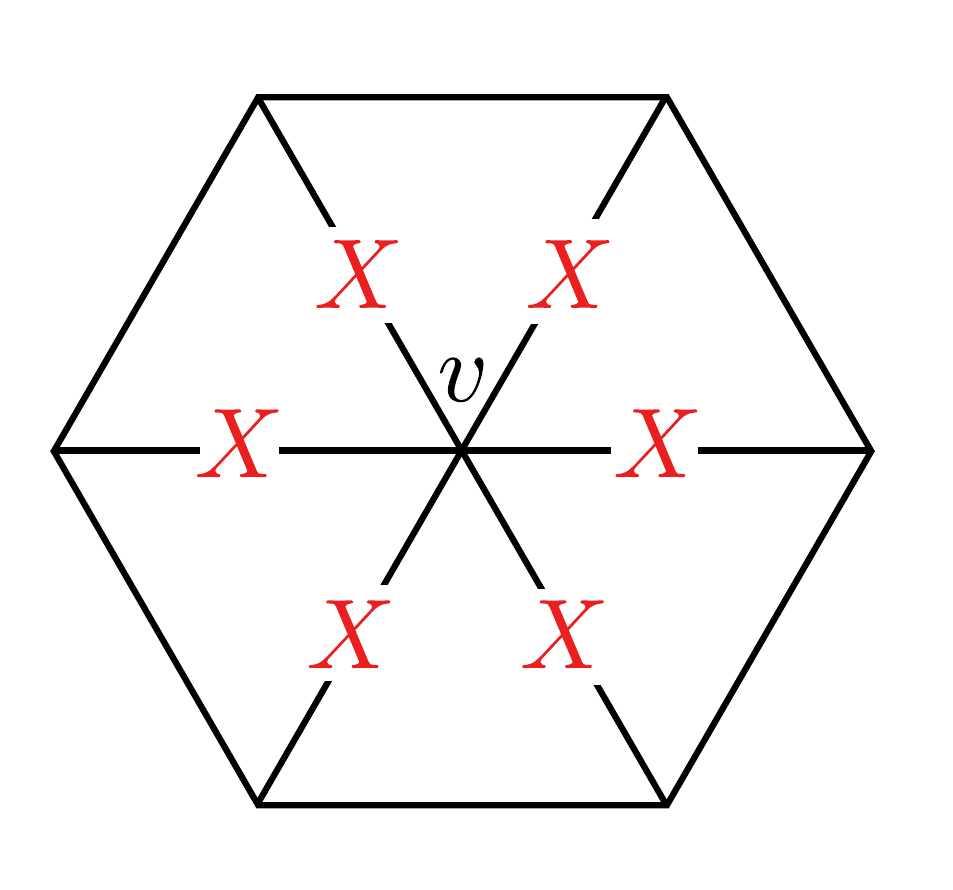}}} \!\!\!\! \times \vcenter{\hbox{\includegraphics[scale=.23,trim={0cm 0cm 0cm 0cm},clip]{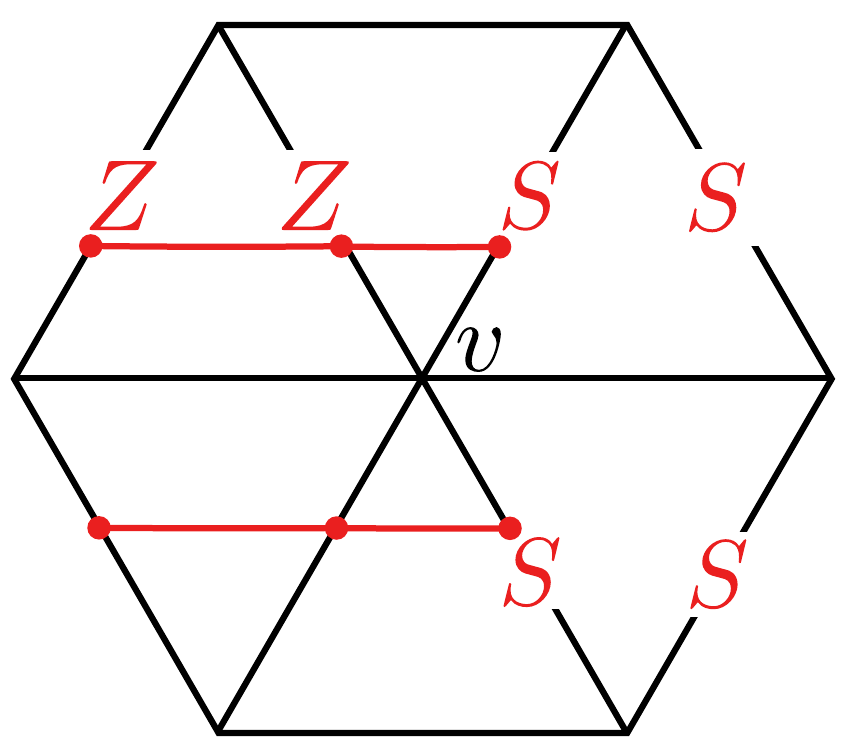}}} \\
B^{III}_f &\equiv  \!\!\! \vcenter{\hbox{\includegraphics[scale=.23,trim={0cm 0cm 0cm 0cm},clip]{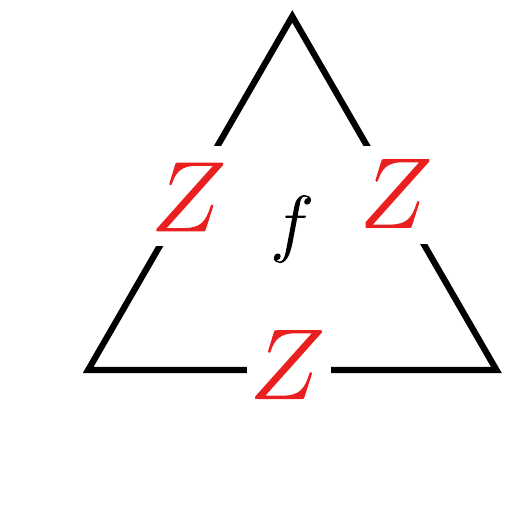}}}, \,\, \vcenter{\hbox{\includegraphics[scale=.23,trim={0cm 0cm 0cm 0cm},clip]{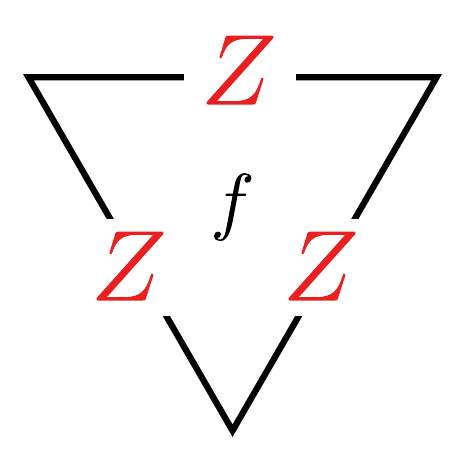}}},
\end{eqs}
% \begin{align} \nonumber
% A^{III}_v &\equiv -\vcenter{\hbox{\includegraphics[scale=.23,trim={0cm 0cm 0cm 0cm},clip]{Figures/stabilizer_Avprime1-eps-converted-to.pdf}}} \!\!\!\! \times \vcenter{\hbox{\includegraphics[scale=.23,trim={0cm 0cm 0cm 0cm},clip]{Figures/stabilizer_Avprime2-eps-converted-to.pdf}}}
% \end{align}
% \begin{align}
% B^{III}_f &\equiv  \!\!\! \vcenter{\hbox{\includegraphics[scale=.23,trim={0cm 0cm 0cm 0cm},clip]{Figures/stabilizer_HdoubleprimeBf1-eps-converted-to.pdf}}}, \,\, \vcenter{\hbox{\includegraphics[scale=.23,trim={0cm 0cm 0cm 0cm},clip]{Figures/stabilizer_HdoubleprimeBf2-eps-converted-to.pdf}}},
% \end{align}
with red edges in the definition of $A^{III}_v$ representing control-$Z$ operators. For each vertex term, there are four control-$Z$ gates.

To see that the Hamiltonian in Eq.~\eqref{eq: Hprime def} has the same ground state subspace as the string-net model (up to ancillary product states), we map $H^{III}_\text{DS}$ to ${H}_\text{DS}^\text{s-n}$ by making changes that preserve the ground state subspace. We note that the terms of  $H^{III}_\text{DS}$ are unfrustrated, meaning that the ground states are in the $+1$ eigenspace of the $A^{III}_v$ and $B^{III}_f$ operators. Consequently, we can freely multiply the $A^{III}_v$ terms by $B^{III}_f$ terms without affecting the ground states. We can also employ the identities:
\begin{eqs} \nonumber
\vcenter{\hbox{\includegraphics[width=.12\textwidth]{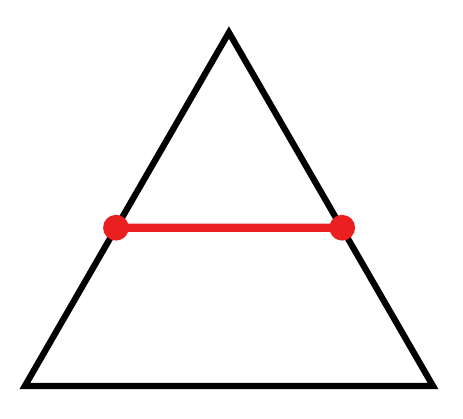}}} &=  \vcenter{\hbox{\includegraphics[width=.12\textwidth]{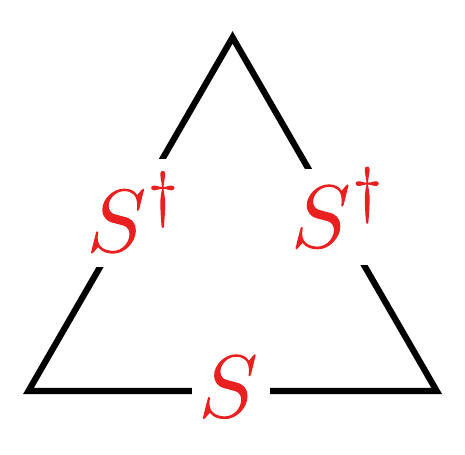}}},  
\end{eqs}
\begin{eqs} \label{eq: CZ to S identity}
\vcenter{\hbox{\includegraphics[width=.12\textwidth]{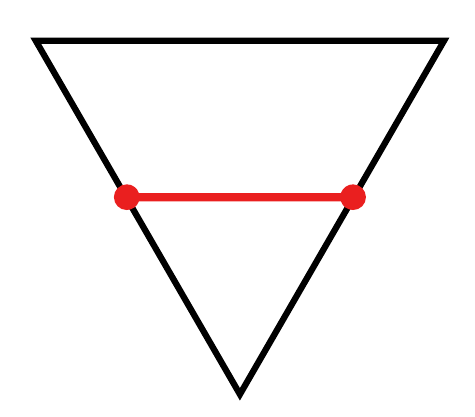}}} &= \vcenter{\hbox{\includegraphics[width=.12\textwidth]{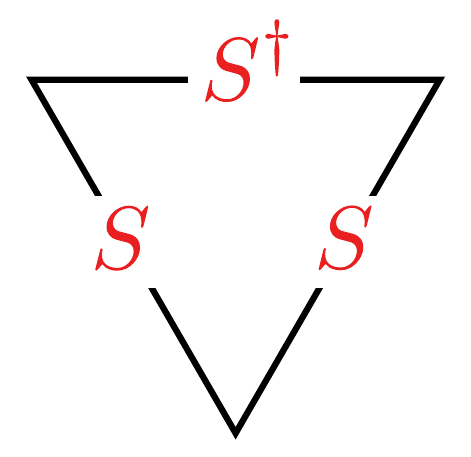}}}\!\!\!,
\end{eqs} 
% \begin{eqs} \label{eq: CZ to S identity}
% \vcenter{\hbox{\includegraphics[width=.12\textwidth]{Figures/stabilizer_CZidentity1-eps-converted-to.pdf}}} &=  \vcenter{\hbox{\includegraphics[width=.12\textwidth]{Figures/stabilizer_CZidentity2-eps-converted-to.pdf}}},  \\ 
% \vcenter{\hbox{\includegraphics[width=.12\textwidth]{Figures/stabilizer_CZidentity3-eps-converted-to.pdf}}} &= \vcenter{\hbox{\includegraphics[width=.12\textwidth]{Figures/stabilizer_CZidentity4-eps-converted-to.pdf}}}\!\!\!,
% \end{eqs} 
which holds in the $B^{III}_f = 1$ subspace, as shown in Table~\ref{tab: checking da and aUa} \footnote{Note that, although the identity for the upwards pointing triangle has two factors of $S^\dagger$, compared to one factor for the downwards pointing triangle, this can be made more symmetric by acting with $B^{III}_f = 1$ on the upwards pointing triangle.}. Note that after replacing all of the control-$Z$ gates using the identities above, the Hamiltonian terms are no longer mutually commuting. Nevertheless, they are unfrustrated and the ground state subspace is the mutual $+1$ eigenspace of the Hamiltonian terms.

Using these transformations, $H^{III}_\text{DS}$ can be mapped to:
\begin{align}
H^{IV}_\text{DS} \equiv -\sum_v A^{IV}_v - \sum_f B^{IV}_f,
\end{align}
where $A^{IV}_v$ and $B^{IV}_f$ are graphically represented by:
\begin{eqs}
A^{IV}_v &= - \vcenter{\hbox{\includegraphics[scale=.23,trim={0cm 0cm 0cm 0cm},clip]{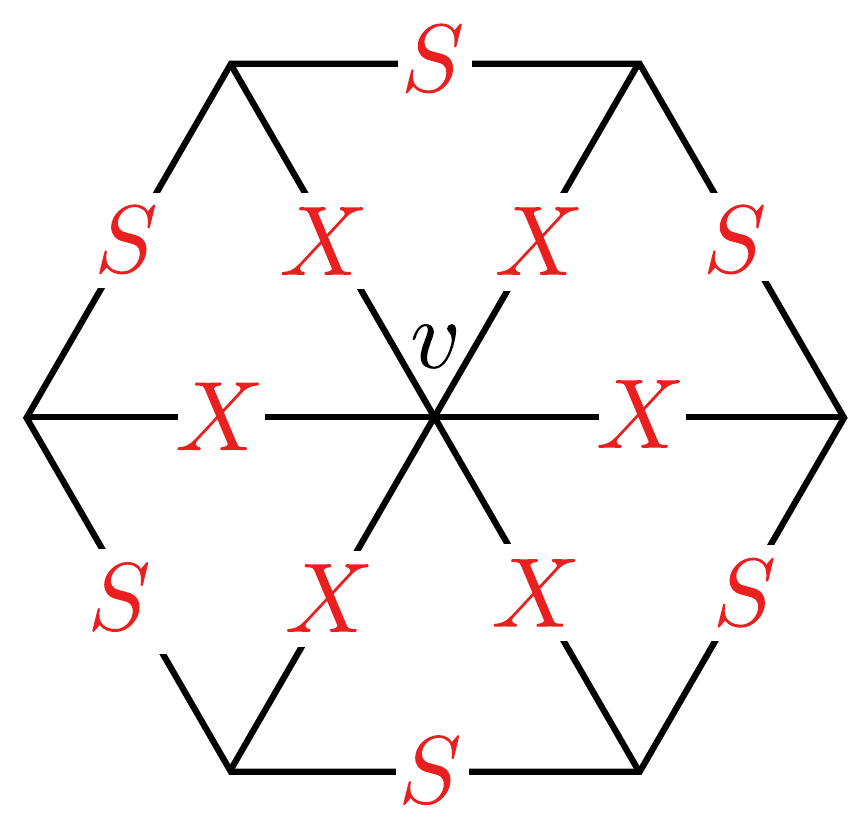}}}, \\
B^{IV}_f &= \!\!\! \vcenter{\hbox{\includegraphics[scale=.23,trim={0cm 0cm 0cm 0cm},clip]{Figures/stabilizer_HdoubleprimeBf1-eps-converted-to.pdf}}}, \,\, \vcenter{\hbox{\includegraphics[scale=.23,trim={0cm 0cm 0cm 0cm},clip]{Figures/stabilizer_HdoubleprimeBf2-eps-converted-to.pdf}}}.
\end{eqs} 
When viewed as a model on the hexagonal dual lattice, $H^{IV}_\text{DS}$ is precisely the DS string-net model in Eq.~\eqref{eq: snDS def}. By construction, $H^{IV}_\text{DS}$ has the same ground state subspace as $\mathcal{U}H_\text{DS}\mathcal{U}^\dagger$, up to adding ancillary degrees of freedom to $H_\text{DS}$ and $H^{IV}_\text{DS}$. Therefore, we have found a finite-depth quantum circuit with ancillary degrees of freedom that maps the ground state subspace of $H_\text{DS}$ to the ground state subspace of the DS string-net model.

\section{Primer on Abelian anyon theories} \label{sec: primer}

In this section, we review the characteristic data of Abelian anyon theories with the aim of generalizing the DS stabilizer model to a wider class of Abelian topological orders. We also introduce notation for boson condensation, as it is essential to the construction of the topological stabilizer models defined in the next section. Further details on Abelian anyon theories can be found in Ref.~\cite{WW20}, and we refer to Ref.~\cite{B17} for background on boson condensation. 

An Abelian anyon theory $\mathcal{A}$ is specified by a pair $(A, \theta)$, where $A$ is a finite Abelian group and $\theta$ is a function from $A$ to $U(1)$:
\begin{align} {\theta:A \to U(1)},\end{align}
that satisfies certain constraints. Intuitively, the group $A$ corresponds to the Abelian group formed by the anyons under fusion, and $\theta$ encodes the statistics and braiding of the anyons. Given this interpretation, we refer to the elements of $A$ as anyons. The DS anyon theory, for example, is specified by $A= \ZZ_2 \times \ZZ_2$ with $\theta$ given by:
\begin{align}
\theta(1)=1, \quad \theta(s)=i, \quad \theta(\bar{s})=-i, \quad \theta(s\bar{s}) = 1,
\end{align}
where the elements of $\ZZ_2 \times \ZZ_2$ have been labeled by the anyons $\{1,s,\bar{s},s \bar{s}\}$. {We call an anyon $b$ a bosonic anyon if it has trivial exchange statistics: $\theta(b)=1$.}
 
\begin{figure*}
\subfloat[\label{fig: thetasquare}]{ \includegraphics[width=.38\textwidth]{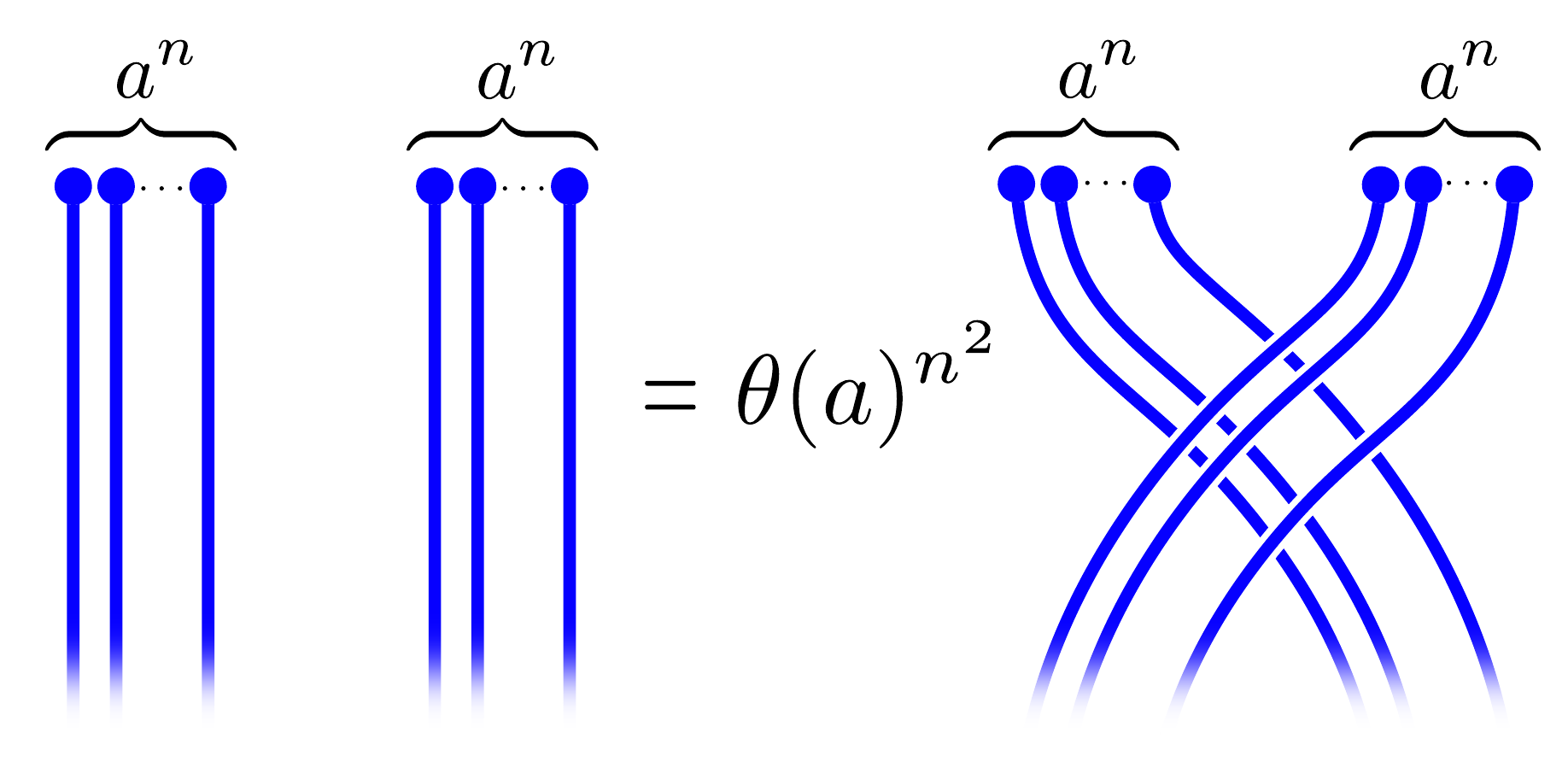}}
\qquad \qquad \quad 
\subfloat[\label{fig: braiding}]{\includegraphics[width=.5\textwidth]{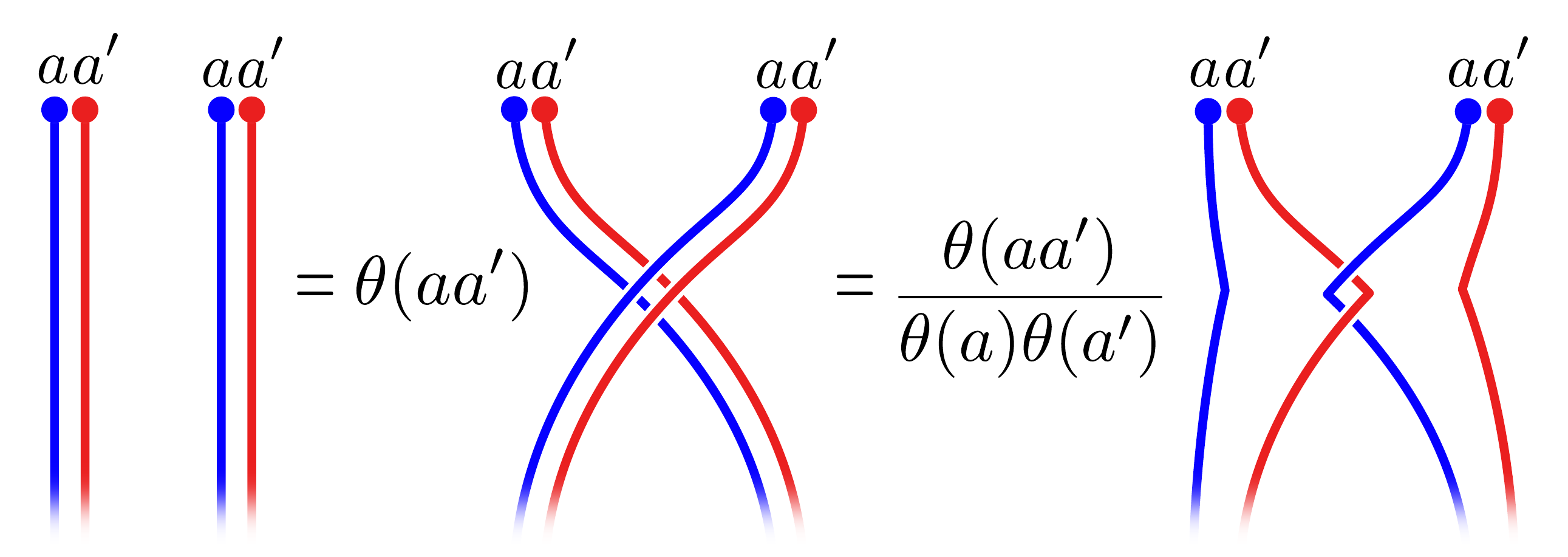}}
\caption{(a) The constraint in Eq.~\eqref{eq: theta quadratic} can be derived by considering the exchange of two $a^n$ anyons. There are $n^2$ exchanges of $a$ anyons in total, giving the phase $\theta(a)^{n^2}$. (b) The expression in Eq.~\eqref{eq: braiding M def} for the braiding relations of anyons $a$ and $a'$ can be obtained by exchanging two $aa'$ anyons. The figure in the far right shows a full braid of $a$ with $a'$, yielding the phase $B_\theta(a,a')$.}
\end{figure*}

Based on physical arguments, only certain choices of the function $\theta$ define a consistent anyon theory. The first requirement is that $\theta$ satisfies:
\begin{align} \label{eq: theta quadratic}
\theta(a^n) = \theta(a)^{n^2},
\end{align}
for any anyon $a \in A$ and integer $n$.
In words, this means that exchanging two $a^n$ anyons is equivalent to exchanging $n^2$ pairs of $a$ anyons. This is shown graphically in Fig.~\ref{fig: thetasquare}. The second constraint on $\theta$ is most naturally stated in terms of the function:
\begin{align}
B_\theta : A \times A \to U(1),
\end{align}
defined by \cite{kitaev2006anyons}: 
\begin{align} \label{eq: braiding M def}
B_\theta(a,a') \equiv \frac{\theta(aa')}{\theta(a)\theta(a')}.
\end{align}
Physically, $B_\theta(a,a')$ captures the braiding relations between the anyons $a$ and $a'$ in $A$. This can be seen by considering the exchange statistics of $aa'$, as depicted in Fig.~\ref{fig: braiding}. The second constraint on $\theta$ is then given by the fact that braiding $a^n$ around $a'$ is (tautologically) the same as braiding $n$ copies of $a$ around $a'$. That is, the function $B_\theta$ satisfies:
\begin{eqs} \label{eq: bilinear braiding}
B_\theta(a^n,a') &= B_\theta(a,a')^n, \\
B_\theta(a,{a}'^n) &= B_\theta(a,a')^n,
\end{eqs}
for all $a,a' \in A$ and any integer $n$. 

\begin{figure}
\centering
    \includegraphics[width=.25\textwidth]{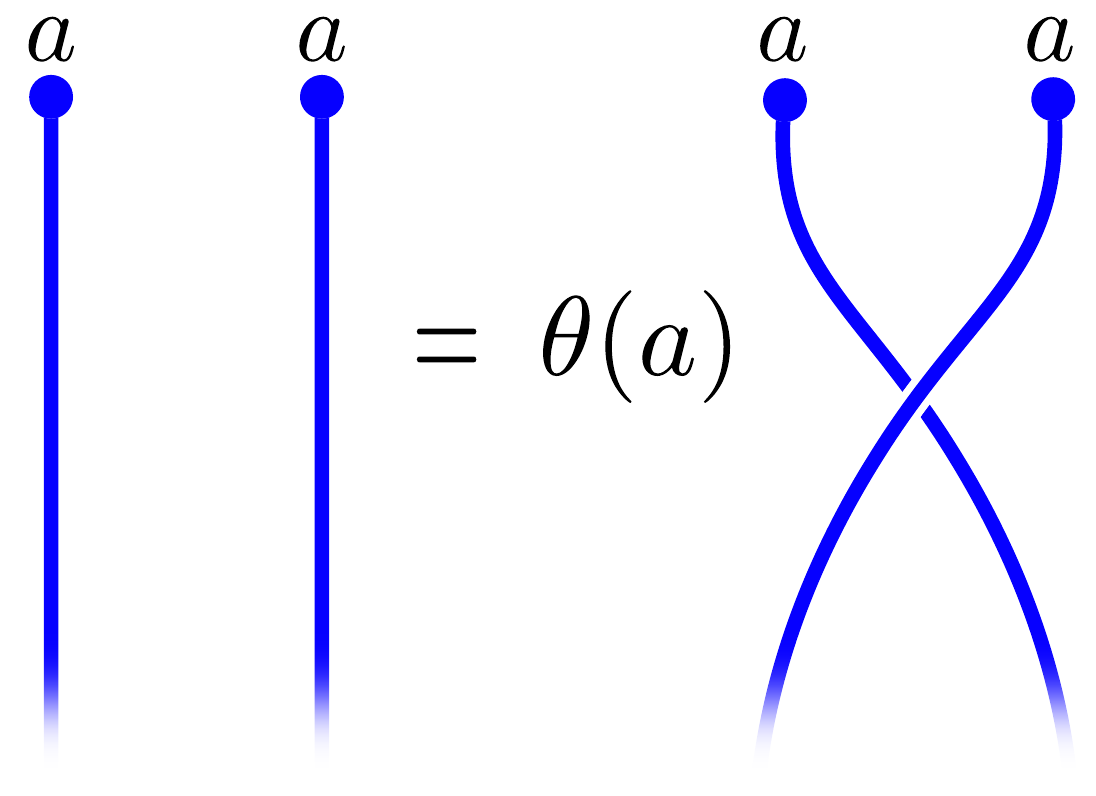}
     \caption{$\theta(a)$ captures the phase accrued from exchanging $a$ anyons.}
\end{figure}

In more formal language, the function $\theta$ corresponds to a quadratic form.
Specifically, one can define the function $q:A \to [0,1)$ by the expression:
\begin{align}
\theta(a) = e^{2\pi i q(a)}.
\end{align}
According to Eq.~\eqref{eq: theta quadratic}, we have:
\begin{align} \label{eq: quadratic form 1}
q(a^n) = n^2 q(a),
\end{align}
and by Eq.~\eqref{eq: bilinear braiding}, the function:
\begin{align} \label{eq: quadratic form 2}
b_q(a,a') \equiv q(aa')-q(a)-q(a')
\end{align}
is bilinear. Eqs.~\eqref{eq: quadratic form 1} and \eqref{eq: quadratic form 2} are the defining properties of a quadratic form. Consequently, Abelian anyon theories are classified by quadratic forms \cite{WW20}.
Furthermore, in this work, we consider anyon theories with the property that every anyon braids nontrivially with at least one anyon in $A$. That is, for every $a\in A$, there exists an $a' \in A$ such that $B_\theta(a,a') \neq 1$.
Such anyon theories are referred to as modular anyon theories and are classified, in particular, by nondegenerate quadratic forms. 
We note that, more generally, anyon theories are specified by the following three pieces of data (i) a set of anyon labels, (ii) the fusion rules of the anyons, and (iii) the so-called $F$- and $R$-symbols. In Ref.~\cite{Q99}, it is shown that the pair $(A,\theta)$ determines the data (i)-(iii) for Abelian anyon theories.

For later purposes, let us introduce the concept of stacking Abelian anyon theories. Given the anyon theories $\mathcal{A}_1 = (A_1,\theta_1)$ and $\mathcal{A}_2= (A_2, \theta_2)$, we can form an anyon theory $\mathcal{A}_1 \boxtimes \mathcal{A}_2$, which we refer to as a stack of $\mathcal{A}_1$ and $\mathcal{A}_2$. Physically, $\mathcal{A}_1 \boxtimes \mathcal{A}_2$ is the anyon theory of two decoupled topological orders that realize the anyon theories $\mathcal{A}_1$ and $\mathcal{A}_2$, respectively. The stack of $\mathcal{A}_1$ and $\mathcal{A}_2$ is defined by the pair $(A_1 \times A_2, \theta_{1,2})$, where $\theta_{1,2}$ is given by:
\begin{align}
\theta_{1,2}(a_1a_2) = \theta_1(a_1) \theta(a_2), 
\end{align}
for any anyons $a_1 \in A_1$ and $a_2 \in A_2$. Using this language, Refs.~\cite{BDCP12} and \cite{Haah2018a} showed that the anyon theories of translationally-invariant Pauli stabilizer models on prime-dimensional qudits are necessarily stacks of toric codes. 

An important tool for the construction of the Pauli stabilizer models in the next section is boson condensation. Thus, we next discuss the effects of boson condensation in Abelian anyon theories. At a physical level, the condensation of a boson $b$ amounts to proliferating the $b$ quasiparticles. As a consequence, pairs of bosons can be freely created and destroyed in the ground state. To make the effects on the anyon theory explicit, we let $\{b_i\}$ be a set of bosons with trivial mutual braiding relations. The two important effects of condensation are then as follows.
\begin{enumerate}[label=(\roman*)]
\item The first effect is that the anyons that braid nontrivially with at least one $b_i$ in $\{b_i\}$ become confined. A single confined anyon is not a local excitation in the condensed theory, since the energetic cost of separating a pair of confined anyons grows linearly with the separation (see Section~\ref{sec: DS construction}).  
The remaining anyons are referred to as deconfined anyons. A deconfined anyon $a$ satisfies:
\begin{align} \label{eq: deconfined def}
B_\theta(a,b_i) = 1, \quad \forall i.
\end{align}
It is important that the bosons $\{b_i\}$ have trivial mutual braiding relations, so that they do not confine one another upon condensation. 

\item The second effect is that anyons that are related by products of the condensed bosons become identified. More precisely, after condensation, the anyons can be organized into equivalence classes, where two anyons $a$ and $a'$ are equivalent if they satisfy:
\begin{eqs}
a=a' \times \prod_{i=1}^M b_i^{p_i},
\end{eqs}
for some set of integers $\{p_i\}$. We denote the equivalence class containing the anyon $a$ by $[a]$. In particular, the bosons $b_i$ are identified with the trivial anyon, i.e., $[b_i] = [1]$, for all $i$.
\end{enumerate}
We note that the statistics and braiding of the deconfined anyons are unaffected by the condensation. For any $[a],[a']$, we have:
\begin{align}
\theta([a]) = \theta(a), \quad B_\theta([a],[a']) = B_\theta(a,a').
\end{align} 
These are well-defined given the condition in Eq.~\eqref{eq: deconfined def}.

Lastly, we define the notion of a Lagrangian subgroup. A Lagrangian subgroup  $\mathcal{L}$ is a subgroup of $A$ composed of bosonic anyons with the following two properties:
\begin{enumerate}[label=(\roman*)]
\item The elements of $\mathcal{L}$ have trivial braiding relations with each other, i.e., for every $b,b' \in \mathcal{L}$, $B_\theta(b,b')=1$.
\item For every anyon $a \in A -\mathcal{L}$, there exists a $b \in \mathcal{L}$ such  that $a$ braids nontrivially with $b$, i.e., $B_\theta(a,b) \neq 1$.
\end{enumerate}
For example, the $s\bar{s}$ anyon of the DS anyon theory generates a Lagrangian subgroup. It trivially satisfies condition (i), and satisfies condition (ii) because it braids nontrivially with $s$ and $\bar{s}$.
The existence of a Lagrangian subgroup signals the potential for a gapped boundary in  a topologically ordered system. Specifically, a Hamiltonian whose excitations are described by the anyon theory $\mathcal{A}$ admits a gapped boundary if and only if $\mathcal{A}$ has a Lagrangian subgroup \cite{Levin13,KS11}. 

Moreover, Refs.~\cite{KS11} and \cite{KKOSS21} proved that every Abelian topological order that admits a gapped boundary can be described by a twisted quantum double (TQD). Here, we use TQD to refer to any model obtained by gauging the symmetries of a Hamiltonian belonging to a symmetry-protected topological (SPT) phase. The argument is that, if the bosons of the Lagrangian subgroup are condensed, then all of the anyons become confined (or identified with the trivial anyon). This is due to condition (ii) of a Lagrangian subgroup. Furthermore, the condensation can be implemented by gauging the $1$-form symmetries associated with the closed string operators of the bosons in $\mathcal{L}$. Since all of the anyons become confined, gauging the $1$-form symmetry produces a model for an SPT phase protected by a $0$-form symmetry. By subsequently gauging the $0$-form symmetry of the SPT model, we recover the initial anyon theory. Thus, TQDs, including the Pauli stabilizer models constructed in the next section, must exhaust all Abelian anyon theories that can be realized in systems with gapped boundaries.

\section{Twisted quantum double stabilizer models} \label{sec: TQD stabilizer codes}

We now construct Pauli stabilizer models corresponding to twisted quantum doubles (TQDs).
The construction is a generalization of the construction of the DS stabilizer model in Section~\ref{sec: double semion stabilizer code}. In particular, we start with a toric code (TC) (or stack of TCs) defined on composite-dimensional qudits and condense certain bosonic anyons so that the remaining deconfined excitations match those of a TQD with Abelian anyons. We refer to the resulting Pauli stabilizer models as TQD stabilizer models. Before describing the construction in detail, we recall some of the characteristic properties of TQDs. Note that TQD Hamiltonians were first presented in Refs.~\cite{levin2012braiding} and \cite{HW13} based on the spacetime formulation in Ref.~\cite{DW90}. These models exhibit the characteristic properties of TQDs described below, although the details of the models are not essential to follow the discussion.

\subsection{Anyon theories of twisted quantum doubles} \label{sec: TQD anyons}
 
In this work, ``TQD'' refers to any two-dimensional topologically ordered system that can be obtained by gauging the symmetry of a model with symmetry-protected topological (SPT) order. Given the classification of two-dimensional SPT phases \cite{chen2012symmetry}, TQDs are characterized by a group $G$ and a group cocycle $\omega$ belonging to $H^3[G,U(1)]$ -- the same data that specifies the SPT phase. Moving forwards, we assume familiarity with group cohomology and refer to Refs.~\cite{chen2012symmetry} and \cite{propitius} for the necessary background. 
For certain choices of $G$ and $\omega$, the corresponding TQDs host non-Abelian anyons -- see, for example, the models of Refs.~\cite{propitius,HW13,NBVdN15}. Stabilizer models, however, are unable to model topological orders with non-Abelian anyons \cite{PV16}. Therefore, we restrict our attention to TQDs whose anyons are Abelian. We refer to such TQDs as Abelian TQDs. 

A TQD is an Abelian TQD if and only if $G$ is a finite Abelian group and $\omega \in H^3[G,U(1)]$ is cohomologous to a product of type I and type II cocycles (defined below) \cite{wang2014braiding} \footnote{In other words, TQDs corresponding to type III cocycles host non-Abelian anyons \cite{propitius}.}. To make the form for $\omega$ explicit, we write $G$ as a general finite Abelian group:
\begin{align}
G=\prod_{i=1}^M \ZZ_{N_i},
\end{align}
and introduce the set of integers $\mathcal{I}$:
\begin{align} \label{eq: Abelian TQD parameters}
\mathcal{I} \equiv \{ n_i \}_{i=1}^M \cup \{ n_{ij} \}_{i,j=1}^M.
\end{align}
Here, $n_i$ is an element of $\ZZ_{N_i}$, and $n_{ij}$ belongs to $\ZZ_{N_{ij}}$, where $N_{ij}$ denotes the greatest common divisor of $N_i$ and $N_j$. Furthermore, $n_{ij}$ satisfies $n_{ij}=n_{ji}$ and $n_{ii}=2n_i$. The set of integers $\mathcal{I}$ then specifies the following group cocycle $\omega_{\mathcal{I}}\in H^3[G,U(1)]$ of an Abelian TQD \cite{propitius}:
% \begin{align} \label{eq: type I type II product}
%     \omega_{\mathcal{I}}(g,h,k) = \prod_{i=1}^M \omega_i^{n_i}(g,h,k) \prod_{j>i} \omega_{ij}^{n_{ij}}(g,h,k) \times \delta \beta(g,h,k).
% \end{align}
\begin{align} \label{eq: type I type II product}
    \omega_{\mathcal{I}}(g,h,k) = \delta \beta(g,h,k) \prod_{i=1}^M \omega_i^{n_i}(g,h,k) \!\prod_{i,j\,:\,j>i} \omega_{ij}^{n_{ij}}(g,h,k).
\end{align}
Here, $g,h,k$ are elements of $G$, 
% the second product is over all pairs $i,j$ with $1\leq i<j\leq M$, 
$\beta$ is an arbitrary $2$-cochain, and $\omega_i$ and $\omega_{ij}$ are so-called type I and type II cocycles. Explicitly, $\omega_i$ and $\omega_{ij}$ are given by \cite{propitius}:
\begin{align}
    \omega_i(g,h,k) &= e^{{\frac{2 \pi i}{N_i^2}g_i(h_i+k_i-[h_i+k_i]_{N_i})}} \\
    \omega_{ij}(g,h,k) &= e^{\frac{2 \pi i}{N_{i}N_j}g_i(h_j+k_j-[h_j+k_j]_{N_j})},
\end{align}
where $g_i,h_i,k_i \in \ZZ_{N_i}$ are the $i^\text{th}$ components of $g,h,k$, and $[\cdots]_{N_i}$ denotes addition modulo $N_i$. 
In words, Eq.~\eqref{eq: type I type II product} tells us that $\omega_\mathcal{I}$ is cohomologous to a product of type I and type II cocycles determined by the integers $\mathcal{I}$. 
Abelian TQDs are, therefore, conveniently parameterized by an Abelian group $G$ and the set of integers $\mathcal{I}$ in Eq.~\eqref{eq: Abelian TQD parameters}. 

We now describe the characteristic properties of the anyonic excitation of Abelian TQDs. 
We rely on the fact that Abelian TQDs are gauge theories, derived by gauging the symmetry of SPT models. As such, a subgroup of the anyons in an Abelian TQD can be labeled as gauge charges \cite{kogut1975hamiltonian,wang2015topological}. The gauge charges are uniquely determined by the gauging procedure and reproduce the group $G= \prod_{i=1}^M \ZZ_{N_i}$ under fusion. We use $c_i$ to denote the unit of gauge charge that generates the subgroup $\ZZ_{N_i}$. The gauge charges are always bosons with trivial mutual braiding, i.e.:
\begin{align} \label{eq: gauge charge properties}
    \theta(c_i) = 1 , \quad B_\theta(c_i,c_j) = 1, \quad \forall i,j.
\end{align}

All other anyonic excitations of Abelian TQDs carry nonzero gauge flux, implying that they braid nontrivially with at least one of the gauge charges. 
We call an anyon $\varphi_i$ an elementary flux if it carries a single unit of gauge flux, i.e., its braiding with gauge charges satisfies: 
\begin{align} \label{eq: aharonov-bohm}
     B_\theta(\varphi_i,c_j) = \begin{cases} 
      e^{2 \pi i / N_i} & \text{if } i=j \\
      1 & \text{if } i\neq j.
   \end{cases}
\end{align}
Note that there are $|G|$ possible choices for the elementary flux $\varphi_i$, differing by gauge charges. After choosing an elementary flux $\varphi_i$, for each $i$, the gauge charges and the elementary fluxes generate all of the anyons of the Abelian TQD. The group formed by the anyons has order $|G|^2$, although the fusion rules of the elementary fluxes may be nontrivial, as demonstrated by Eq.~\eqref{eq: flux charge relation} below.

Unlike the gauge charges, the elementary fluxes can have exotic exchange statistics and nontrivial mutual braiding relations.
Ref.~\cite{wang2015topological} showed that TQDs are characterized by the statistics and braiding of the elementary fluxes. For an Abelian TQD corresponding to a group $G= \prod_{i=1}^M \ZZ_{N_i}$ and a choice of elementary flux $\varphi_i$ for each $i$, one can define the quantities:
\begin{align} \label{eq: Thetas}
	\Theta_{i} \equiv \theta(\varphi_i)^{N_i}, \quad \Theta_{ij} \equiv B_\theta(\varphi_i,\varphi_j)^{N^{ij}},
\end{align}
where $N^{ij}$ denotes the least common multiple of $N_i$ and $N_j$. Using the properties of $\theta$ in Eqs.~\eqref{eq: theta quadratic} and \eqref{eq: bilinear braiding}, it can be checked that $\Theta_i$ and $\Theta_{ij}$ are well-defined -- they do not change if the elementary fluxes are modified by attachment of gauge charges. 

In what follows, it useful to express $\Theta_i$ and $\Theta_{ij}$ in terms of the group cohomological data that specifies an Abelian TQD. 
The results of Ref.~\cite{propitius} imply that $\Theta_i$ and $\Theta_{ij}$ can be expressed in terms of the parameters in $\mathcal{I}$ as:
\begin{align} \label{eq: Theta in terms of ni and nij}
    \Theta_i = e^{\frac{2 \pi i n_i }{N_i}}, \quad \Theta_{ij} = e^{\frac{2 \pi i n_{ij}}{N_{ij}}}.
\end{align}
We see that a nonzero $n_i$ implies that the elementary flux $\varphi_i$ has nontrivial statistics, and a nonzero $n_{ij}$ implies that the elementary fluxes $\varphi_i$ and $\varphi_j$ have nontrivial mutual braiding. 

To provide intuition for the quantities in Eq.~\eqref{eq: Theta in terms of ni and nij}, let us consider the TQDs corresponding to the group $G=\ZZ_2$. The group cohomology $H^3[\ZZ_2,U(1)]$ is equivalent to $\ZZ_2$, and the nontrivial element is cohomologous to a type I cocycle. This implies that there are two distinct Abelian TQDs for $G=\ZZ_2$, labeled by $n_1 = 0$ and $n_1 = 1$. The TQD corresponding to $n_1 = 0$ is in the same phase as the $\ZZ_2$ TC and has anyons $\{1,e,m,\psi\}$. The gauge charge is the boson $e$, and the elementary flux can be chosen to be $\varphi_1 = m$ or $\varphi_1= \psi$. In either case, $\Theta_1$ is:
\begin{align}
	\Theta_1 = \theta(m)^2 = \theta(\psi)^2 = 1.
\end{align}
The TQD corresponding to $n_1 = 1$, on the other hand, gives a model belonging to the DS phase with anyons $\{1,s,\bar{s}, s\bar{s}\}$. Here, $s\bar{s}$ is the gauge charge and the elementary flux is either $\varphi_1 =s$ or $\varphi_1 = \bar{s}$. In this case, $\Theta_1$ takes the value: 
\begin{align}
	\Theta_1 = \theta(s)^{2} = \theta(\bar{s})^2= -1.
\end{align}
In Appendix~\ref{app: K matrix}, we give further illustrative examples of Abelian TQDs using the $K$-matrix formalism.

An important relation between elementary fluxes and gauge charges can be derived from Eq.~\eqref{eq: Theta in terms of ni and nij}. To this end, we consider the product of $N_i$ elementary fluxes $\varphi_i$. The resulting anyon $\varphi_i^{N_i}$ carries vanishing gauge flux, as can be seen by braiding a charge $c_j$ around $\varphi_i^{N_i}$. 
Therefore, $\varphi_i^{N_i}$ must be generated by gauge charges. To probe the gauge charges carried by $\varphi_i^{N_i}$, we consider braiding an elementary flux $\varphi_j$ around $\varphi_i^{N_i}$. There are two cases:
\begin{enumerate}[label=(\roman*)]
\item If $j=i$, we have:
\begin{align}
B_\theta(\varphi_i,\varphi_i^{N_i}) = \frac{\theta(\varphi_i^{N_i+1})}{\theta(\varphi_i)\theta(\varphi_i^{N_i})}.
\end{align}
Since $\theta$ satisfies Eq.~\eqref{eq: theta quadratic}, this reduces to:
\begin{eqs}
B_\theta(\varphi_i,\varphi_i^{N_i}) = \frac{\theta(\varphi_i)^{N^2_i+2N_i+1}}{\theta(\varphi_i)\theta(\varphi_i)^{N_i^2}} = \theta(\varphi_i)^{2N_i}.
\end{eqs}
Finally, we employ Eq.~\eqref{eq: Theta in terms of ni and nij} to find:
\begin{align} \label{eq: braiding with phiNi}
B_\theta(\varphi_i, \varphi_i^{N_i}) = \Theta_i^2 = e^{2n_i\frac{2 \pi i }{N_i}}.
\end{align}
Eqs.~\eqref{eq: aharonov-bohm} and \eqref{eq: braiding with phiNi} tell us that $\varphi_i^{N_i}$ carries $2n_i$ copies of the gauge charge $c_i$. 
\item If $j \neq i$, we use the property of $B_\theta$ in Eq.~\eqref{eq: bilinear braiding}:
\begin{align}
B_\theta(\varphi_j, \varphi_i^{N_i}) = B_\theta(\varphi_j,\varphi_i)^{N_i}. 
\end{align}
This can then be written in terms of $\Theta_{ij}$ using Eq.~\eqref{eq: Thetas}:
\begin{align} \label{eq: j charges probe 1}
B_\theta(\varphi_j, \varphi_i^{N_i}) = \Theta_{ij}^{{N_i}/{N^{ij}}}.
\end{align}
Substituting the expression for $\Theta_{ij}$ in Eq.~\eqref{eq: Theta in terms of ni and nij} into the formula above, we have:
\begin{align} 
B_\theta(\varphi_j,\varphi_i^{N_i}) = e^{\frac{2 \pi i n_{ij}}{N_{ij}}\frac{N_i}{N^{ij}}} =  e^{n_{ij}\frac{2 \pi i }{N_{j}}},
\end{align}
where we have used that $N_{ij}N^{ij} = N_i N_j$.
Thus, $\varphi_i^{N_i}$ carries $n_{ij}$ copies of the gauge charge $c_j$.
\end{enumerate}
Combining (i) and (ii), we find that the gauge charges and elementary fluxes satisfy \footnote{Note that $n_{ij}$ is only defined up to integer multiples of $N_{ij}$. Without loss of generality, we can assume that each $N_i$ is a prime power, which implies that $N_{ij}$ is equal to $1$, $N_i$, or $N_j$. If $N_{ij}$ is $1$ or $N_j$, then Eq.~\eqref{eq: flux charge relation} is unchanged. If $N_{ij}=N_i$, on the other hand, then Eq.~\eqref{eq: flux charge relation} should be modified by a factor of $c_j^{t_{ij}N_j}$, for some integer $t_{ij}$. This amounts to redefining $\varphi_i$ by an integer number of $c_j$ charges. Therefore, we can always find a choice of elementary fluxes that satisfies Eq.~\eqref{eq: flux charge relation}.}:  
\begin{align} \label{eq: flux charge relation}
\varphi_i ^{N_i} = c_i^{2 n_i} \prod_{j \neq i} c_j^{n_{ij}}, \quad \forall i.
\end{align}
This relation is independent of the choice of $\varphi_i$ due to the fact that $\Theta_i$ and $\Theta_{ij}$, used in the derivation, do not depend on the choice of elementary fluxes. We also point out that Eq.~\eqref{eq: flux charge relation} implies that the group formed by the anyons of an Abelian TQD can differ from $G \times G$, if the parameters $n_i$ and $n_{ij}$ are nonzero. We refer to Appendix~\ref{app: fusion group} for an aside on the group structure of anyons in Abelian TQDs.

In summary, the anyons of an Abelian TQD corresponding to $G$ and $\mathcal{I}$ form a group ${A}_\text{TQD}$ with the presentation:
\begin{align} \label{eq: presentation for anyon group}
{A}_{\text{TQD}} \equiv  \big\langle \{c_i\}, \{\varphi_i\} | c_i^{N_i} = 1 , \, \varphi_i^{N_i} =  c_i^{2 n_i} \prod_{j \neq i} c_j^{n_{ij}}, \, \forall i \big\rangle,
\end{align}
where 
% $i$ and $j$ satisfy $1 \leq i < j \leq M$, and 
$\{\varphi_i\}_{i=1}^M$ is a set of elementary fluxes with an arbitrary $\varphi_i$ chosen for each $i$. Up to redefining $\varphi_i$ by gauge charges, the statistics of the anyons are determined by Eqs.~\eqref{eq: gauge charge properties}, \eqref{eq: aharonov-bohm}, and \eqref{eq: Theta in terms of ni and nij}. The group in Eq.~\eqref{eq: presentation for anyon group} along with the statistics and braiding of the gauge charges and elementary fluxes defines the anyon theory $\mathcal{A}_\text{TQD}$ of the Abelian TQD.

In the next section, our goal is to construct a Pauli stabilizer model for each choice of $G = \prod_{i=1}^M \ZZ_{N_i}$ and parameters  $\mathcal{I}$ such that its anyons satisfy Eqs.~\eqref{eq: gauge charge properties}, \eqref{eq: aharonov-bohm}, and \eqref{eq: Theta in terms of ni and nij} and form the group in Eq.~\eqref{eq: presentation for anyon group}. This implies that the anyonic excitations of the Pauli stabilizer Hamiltonian form the same anyon theory $\mathcal{A}_\text{TQD}$ as an Abelian TQD specified by $G$ and  $\mathcal{I}$. 

\subsection{Construction of the Pauli stabilizer models} \label{sec: TQD stabilizer code construction}
 
Having described the anyonic excitations of TQDs, we are ready to construct the TQD stabilizer models. The construction proceeds in two steps. In the first step, we identify a TC (or stack of TCs) on composite-dimensional qudits with the property that its anyonic excitations contain the anyons of a TQD as a quotient group. In the second step, we condense certain emergent bosons so that the deconfined excitations have the same properties as the anyons of a TQD. We describe the construction at the level of the anyons before providing explicit lattice models for the Pauli stabilizer models.

\subsubsection*{Anyon-level construction}

To capture the statistics and braiding of anyons in an Abelian TQD corresponding to a group $G=\prod_{i=1}^M \ZZ_{N_i}$, we start with a stack of $M$ TCs, where the $i^\text{th}$ TC is a $\ZZ_{N_i^2}$ TC. We note that, in some cases, it is possible to use a smaller group than $\ZZ_{N_i^2}$. However, $\ZZ_{N_i^2}$ is sufficient for the general construction of TQD stabilizer models, as the anyons in stacks of $\ZZ_{N_i^2}$ TCs are able to reproduce the statistics and braiding needed for the characteristic $\Theta_i$ and $\Theta_{ij}$ of the Abelian TQD corresponding to $G$.

We label the $e$ and $m$ excitations of the $i^\text{th}$ TC by $e_i$ and $m_i$, respectively. Recall that the $e$ and $m$ excitations
are bosons with the braiding relations:
\begin{align}
B_\theta(m_i,e_j) = \begin{cases} 
       e^{\frac{2 \pi i}{N_i^2}} &  \text{if } i=j \\
      1 & \text{if } i\neq j.
   \end{cases}
\end{align}
Anyons in the stack of TCs can be written as products of $e$ and $m$ excitations of the form: $\prod_{i=1}^M e^{p_i}_i m^{q_i}_i$,
for integers $p_i,q_i \in \ZZ_{N_i^2}$. The statistics and braiding of the anyons follow from the properties of the $e$ and $m$ excitations and are given by the general formulas:
\begin{align} \label{eq: general statistics}
\theta\left(\prod_{i=1}^M e^{p_i}_i m^{q_i}_i\right) &= \prod_{i=1}^M e^{\frac{2 \pi i }{N_i^2}p_i q_i} \\ \label{eq: general braiding}
B_\theta\left( \prod_{i=1}^M e^{p_i}_i m^{q_i}_i ,\prod_{i=1}^M e^{r_i}_i m^{s_i}_i \right) &= \prod_{i=1}^M e^{\frac{2 \pi i}{N_i^2}(p_i s_i + q_i r_i)}.
\end{align}
Our first objective is to identify anyons in this system that exhibit the same statistics and braiding as the anyons of an Abelian TQD corresponding to $G=\prod_{i=1}^M \ZZ_{N_i}$ and $\mathcal{I}$, where $\mathcal{I}$ specifies a product of type I and type II cocycles, as described in the previous section. 

Let us first consider the excitations $\{e_i^{N_i}\}$.  These excitations  generate the group $G$ under fusion and are bosons with trivial mutual braiding:
\begin{align}
\theta(e_i^{N_i}) = 1, \quad B_\theta(e_i^{N_i},e_j^{N_j}) = 1, \quad \forall i,j.
\end{align}
Therefore, they satisfy the same conditions as in Eq.~\eqref{eq: gauge charge properties} for gauge charges of an Abelian TQD. We suggestively define:
\begin{align} \label{eq: def chargei}
\chargei \equiv e_i^{N_i}.
\end{align}

To describe excitations that mimic the properties of elementary fluxes, we set a convention for the ordering of the factors $\ZZ_{N_i}$ of $G$. We first assume (without loss of generality) that each $N_i$ is a power of a prime. We then (implicitly) order the factors of $G$ so that $N_{i}<N_{i+1}$, for all $i$. For example, we order 
\begin{eqs} \nonumber
   \ZZ_{27}  \times \ZZ_4 \times \ZZ_5 \times \ZZ_{2}  \quad \text{as} \quad  \ZZ_2 \times  \ZZ_4 \times \ZZ_5 \times \ZZ_{27}.
\end{eqs}

Given this ordering on the factors of $G$, we consider the excitations $\{\fluxi\}$ for $1 \leq i \leq M$, where $\fluxi$ is defined as:
\begin{align} \label{eq: def fluxi}
\fluxi \equiv m_i e_i^{n_i}\prod_{j>i}e_j^{\frac{N_j}{N_i}n_{ij}}.
\end{align}
Here, the product is over all $j$ with $j>i$.
Note that due to our choice of ordering, $\frac{N_j}{N_i}n_{ij}$ is an integer. Indeed, if $N_i$ does not divide $N_j$, then the greatest common divisor of $N_i$ and $N_j$ is $N_{ij}=1$, implying that $n_{ij}$ must be zero, since it is an element of $\ZZ_{N_{ij}}$.
From the general formula in Eq.~\eqref{eq: general braiding}, the braiding of $\fluxi$ and the excitation $\chargei$ is:
\begin{align}
B_\theta(\fluxi,\chargej)  = \begin{cases} 
      e^{2 \pi i / N_i} & \text{if } i=j \\
      1 & \text{if } i\neq j.
   \end{cases}
\end{align}
This agrees with the braiding of an elementary flux and a gauge charge, as in Eq.~\eqref{eq: aharonov-bohm}.
Furthermore, the statistics and mutual braiding of $\fluxi$ tell us, for all $i,j$:
\begin{align}
\Theta_i = e^{\frac{2 \pi i n_i }{N_i}},  \quad
\Theta_{ij}  = e^{\frac{2 \pi i n_{ij} }{N_{ij}}},
\end{align}
which matches the values for $\Theta_i$ and $\Theta_{ij}$ in Eq.~\eqref{eq: Theta in terms of ni and nij}. 

Thus far, we have found excitations 
$\chargei$ and $\fluxi$
that exhibit the same statistics and braiding as the gauge charges and elementary fluxes
of an Abelian TQD. Now, we need to ensure that they also have the same fusion rules. However, using the definition of $\tilde{\varphi}_i$, the product of $N_i$ copies of $\fluxi$ is equivalent to:
\begin{eqs} \label{eq: varphi power}
{\tilde{\varphi}_i}^{N_i} = m_i^{N_i} \chargei^{n_i} \prod_{j>i} \chargej^{ n_{ij}}, \quad \forall i.
\end{eqs}  
This implies that 
$\chargei$ and $\fluxi$
 fail to satisfy the   relations of gauge charges $c_i$ and elementary fluxes $\varphi_i$ in Eq.~\eqref{eq: flux charge relation}, rewritten here:
\begin{align} \label{eq: flux charge relation 2}
\varphi_i ^{N_i} = c_i^{2 n_i} \prod_{j \neq i} c_j^{n_{ij}}, \quad \forall i.
\end{align}

This discrepancy can be resolved by condensing a particular set of bosonic anyons $\{b_i\}$, with $1 \leq i \leq M$. The bosons are chosen so that the anyons of the resulting theory obey the relations in Eq.~\eqref{eq: flux charge relation 2}. In particular, we construct an anyon theory $\mathcal{C}$ by condensing the collection of bosons $\{b_i\}$, defined by:
\begin{eqs} \label{eq: condense this}
b_i \equiv m_i^{-N_i}e_i^{N_in_i}\prod_{j<i}e_j^{N_jn_{ij}},
\end{eqs}
where the product is over $j$ with $j<i$.
The first thing to note is that the bosons have trivial mutual braiding relations:
\begin{align}
\quad B_\theta(b_i,b_j) = 1, \quad \forall i,j.
\end{align}
Therefore, they can be condensed without confining one another. Second, the bosons have trivial braiding relations with the anyons $\chargei$ and $\fluxi$:
\begin{align}
B_\theta(\chargei,b_j) = 1, \quad B_\theta(\fluxi, b_j) = 1, \quad \forall i,j. 
\end{align}
This means that the equivalence classes $[\chargei]$ and $[\fluxi]$ represent deconfined anyons in $\mathcal{C}$. Here, we have used the notation $[\,\cdot\,]$ described in Section~\ref{sec: primer} to represent equivalence classes of excitations related by fusion with condensed bosons. 

The condensation of the bosons produces relations between $[\chargei]$ and $[\fluxi]$ analogous to the relations in Eq.~\eqref{eq: flux charge relation 2}.
To see this, we express $b_i$ in terms of $\fluxi$ and $\chargei$. A straightforward calculation shows that $b_i$ is equivalent to:
\begin{align}
b_i = \fluxi^{-N_i} \times \chargei^{2n_i} \prod_{j \neq i} \chargej^{n_{ij}}.
\end{align}
Therefore, after condensation, we have:
\begin{eqs} 
[b_i] = [\fluxi^{-N_i} \times \chargei^{2n_i} \prod_{j \neq i} \chargej^{n_{ij}}] = [1], \quad \forall i.
\end{eqs}
Given that the equivalence relation is well-defined under fusion, we obtain:
\begin{eqs} \label{eq: desired flux charge relation}
[\fluxi]^{N_i} = [\chargei]^{2n_i} \prod_{j \neq i} [\chargej]^{n_{ij}}, \quad \forall i.
\end{eqs} 
Eq.~\eqref{eq: desired flux charge relation} shows us that the deconfined anyons $[\chargei]$ and $[\fluxi]$ satisfy the same relations  as the gauge charges and elementary fluxes in Eq.~\eqref{eq: flux charge relation 2}. 

At this point, we have a set of anyons $\{[\chargei]\}$ and $\{[\fluxi]\}$ with the same statistics, braiding, and fusion rules as the gauge charges and elementary fluxes of an Abelian TQD corresponding to $G$ and the parameters $\mathcal{I}$. Hence, they generate an anyon theory that is equivalent to the anyon theory $\mathcal{A}_\text{TQD}$ of the Abelian TQD. This means that the condensed theory $\mathcal{C}$ contains $\mathcal{A}_\text{TQD}$ as a subtheory.

The last step of the construction is to argue that $\mathcal{C}$ is generated by the sets of anyons $\{[\chargei]\}$ and $\{[\fluxi]\}$. This implies that $\mathcal{C}$ is equivalent to $\mathcal{A}_{\text{TQD}}$. To this end, we use the following observation (shown below): the condensation of the bosons $\{[\chargei]\}$ in $\mathcal{C}$ produces a theory without any deconfined anyons. Note that the bosons $\{\chargei\}$ and $\{b_i\}$ in the stack of $\ZZ_{N_i^2}$ TCs braid trivially with each other, so they can be condensed without confining one another. Therefore, condensing the bosons $\{[\chargei]\}$ in $\mathcal{C}$ is the same as condensing the bosons $\{\chargei\}$ followed by condensing the bosons $\{[b_i]\}$.

For the sake of contradiction, suppose that there exists an anyon $a$ that remains deconfined and nontrivial after condensing the sets of bosons $\{\chargei\}$ and $\{b_i\}$. Since $a$ is deconfined after the condensation of the $\chargei$ anyons, it must braid trivially with each $\chargei = e_i^{N_i}$. This means that $a$ takes the form:
\begin{align}
a = \prod_{i=1}^M e_i^{p_i} \prod_{i=1}^M m_i^{s_iN_i},
\end{align}
with $p_i \in \ZZ_{N_i^2}$ and $s_i \in \ZZ_{N_i}$. After condensing the bosons in $\{\chargei\}$, each $b_i$ can be identified with a product of $m_i$ excitations:
\begin{align}
[b_i] = [m_i^{-N_i}], \quad \forall i.
\end{align}
Therefore, condensing the boson labeled by $b_i$ is equivalent to condensing the boson labeled by $m_i^{-N_i}$. If $a$ is deconfined after condensing the set of bosons $\{[b_i]\}$, then it must have trivial braiding relations with $m_i^{-N_i}$, for all $i$. This constrains $a$ to be of the form:
\begin{align} \label{eq: a anyon form}
a = \prod_{i=1}^M e_i^{r_iN_i} \prod_{i=1}^M m_i^{s_iN_i},
\end{align}
for $r_i \in \ZZ_{N_i}$. The expression in Eq.~\eqref{eq: a anyon form} conflicts with the assumption that $a$ is nontrivial after the condensation of the two sets of bosons. Indeed, after condensing the bosons $\chargei$, we have:
\begin{align}
[a] = [ \prod_{i=1}^M m_i^{s_iN_i}].
\end{align}
Then, condensation of the bosons $[b_i] = [m_i^{-N_i}]$ implies that $[a]$ can be identified with the trivial anyon. 
Thus, there are no deconfined anyons after condensing the set of bosons $\{[\chargei]\}$.

The observation above tells us that $\mathcal{C}$ is modular, i.e., for every anyon $[a]$ in $\mathcal{C}$ there exists an anyon $[a']$ that braids nontrivially with $[a]$. This is because, if an anyon braided trivially with all other anyons in $\mathcal{C}$, then it would be deconfined after condensing $\{[\chargei]\}$ \footnote{Alternatively, the transparent anyon (i.e., with trivial braiding relations) could become identified with the trivial anyon after condensing $\{[\chargei]\}$. This means that the transparent anyon is equivalent to products of $[\chargei]$ bosons. This is not possible, however, because products of the anyons $[\chargei]$ braid nontrivially with $[\fluxi]$ anyons. Thus, this would contradict the fact that they are transparent.}. Since $\mathcal{C}$ is modular, we can apply the following result from Ref.~\cite{M03}: any modular anyon theory $\mathcal{A}$, that contains a modular subtheory $\mathcal{A}_1$, factorizes as:
\begin{align}
\mathcal{A} = \mathcal{A}_1 \boxtimes \mathcal{A}_2,
\end{align}
for some modular theory $\mathcal{A}_2$. The product $\boxtimes$, defined in Section~\ref{sec: primer}, implies that the anyons in $\mathcal{A}_1$ braid trivially with the anyons in $\mathcal{A}_2$.
$\mathcal{A}_\text{TQD}$ forms a modular subtheory of $\mathcal{C}$. Therefore, $\mathcal{C}$ factorizes as:
\begin{align}
\mathcal{C} = \mathcal{A}_\text{TQD} \boxtimes \mathcal{B},
\end{align}
for some modular anyon theory $\mathcal{B}$. The anyon theory $\mathcal{B}$ must be trivial, because, after condensing the $[\chargei]$ bosons, there are no deconfined anyons. If $\mathcal{B}$ was nontrivial, then the anyons in $\mathcal{B}$ would remain deconfined after condensing $\{[\chargei]\}$. We conclude that $\mathcal{C}$ is equal to $\mathcal{A}_\text{TQD}$. 
 
\subsubsection*{Lattice-level construction}
 
Having described the construction at the level of the anyons, we now turn to an explicit lattice construction of the TQD stabilizer models. The stabilizer model for the TQD specified by $G= \prod_{i=1}^M \ZZ_{N_i}$ and the set of integers $\mathcal{I}$ [Eq.~\eqref{eq: Abelian TQD parameters}] is defined on a square lattice, where
each edge hosts $M$ qudits, and the $i^\text{th}$ qudit has dimension $N_i^2$. The Pauli X and Pauli Z operators for the $i^\text{th}$ qudit at the edge $e$ are given by:
\begin{align}
X_{e,i} \equiv \sum_{j \in \ZZ_{N_i^2}} |j+1\rangle \langle j | , \quad Z_{e,i} \equiv \sum_{j \in \ZZ_{N_i^2}} \omega^j |j \rangle \langle j |.
\end{align}
When it is clear from context, we omit the edge label $e$.

The construction of the TQD stabilizer model begins with a stack of $\ZZ_{N_i^2}$ TCs. The Hamiltonian for the decoupled $\ZZ_{N_i^2}$ TCs is:
\begin{align}
H_\text{TC} \equiv \sum_{i=1}^M H_\text{TC}^{(i)},
\end{align}
where $H_\text{TC}^{(i)}$ denotes the $\ZZ_{N_i^2}$ TC in the layer $i$. Explicitly, $H_\text{TC}^{(i)}$ is the sum of vertex terms $A^\text{TC}_{v,i}$ and plaquette terms $B^\text{TC}_{p,i}$:
\begin{align}
H_\text{TC}^{(i)} \equiv -\sum_v A^\text{TC}_{v,i} - \sum_p B^\text{TC}_{p,i} + \text{h.c.},
\end{align}
with $A^\text{TC}_{v,i}$ and $B^\text{TC}_{p,i}$ represented pictorially by:
\begin{align}
A^\text{TC}_{v,i} \equiv \vcenter{\hbox{\includegraphics[scale=.23,trim={0cm 0cm 0cm 0cm},clip]{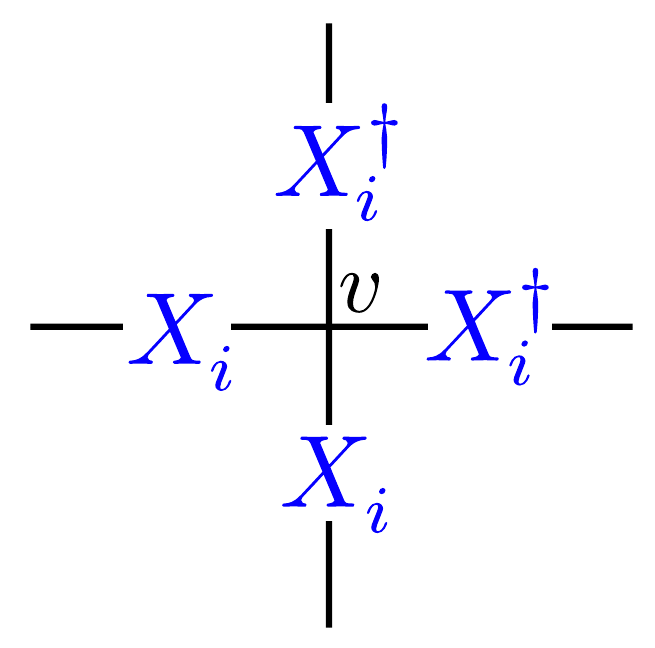}}}, \quad B^\text{TC}_{p,i} \equiv \vcenter{\hbox{\includegraphics[scale=.23,trim={0cm 0cm 0cm 0cm},clip]{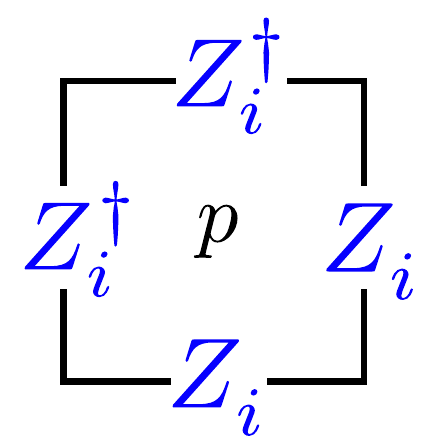}}}.
\end{align}
The associated stabilizer group for the Hamiltonian $H_\text{TC}$ is generated by the full set of vertex terms and plaquette terms of the Hamiltonian $H_\text{TC}$:
\begin{align}
\mathcal{S}_\text{TC} \equiv \langle \{A_{v,i}^\text{TC}\}_{i=1}^M, \{B_{p,i}^\text{TC}\}_{i=1}^M \rangle.
\end{align}

The TQD stabilizer model is then derived by condensing the bosons $b_i$, defined in Eq.~\eqref{eq: condense this} as:
\begin{eqs} 
b_i = m_i^{-N_i}e_i^{N_in_i}\prod_{j<i}e_j^{N_jn_{ij}}.
\end{eqs}
We take the short string operators for the bosons $b_i$ to be:
\begin{align} \label{eq: }
C_{e,i} \equiv \vcenter{\hbox{\includegraphics[scale=.23,trim={0cm 0cm 0cm 0cm},clip]{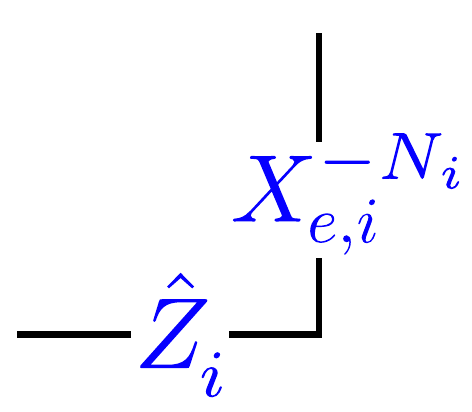}}}, \,\,\, \vcenter{\hbox{\includegraphics[scale=.23,trim={0cm 0cm 0cm 0cm},clip]{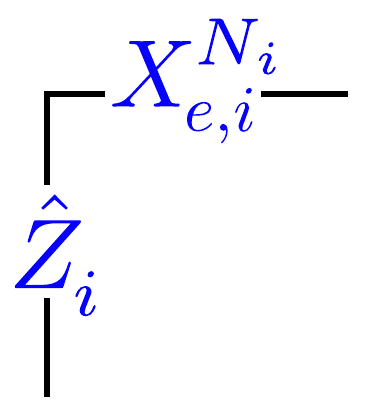}}},
\end{align}
where $\hat{Z}_i$ is shorthand for the product of Pauli Z operators:
\begin{align}
\hat{Z}_i \equiv Z_i^{N_in_i}\prod_{j<i}Z_j^{N_j n_{ij}}.
\end{align}
To build longer string operators, one would need to account for the orientation of a path along the dual lattice, but for our purposes the operators $C_{e,i}$ suffice.
The short string operators $C_{e,i}$ generate the stabilizer group $\mathcal{S}_C$:
\begin{align}
\mathcal{S}_C \equiv \langle \{C_{e,i}\}_{i=1}^M \rangle.
\end{align}

We now follow the same logic as used in Section~\ref{sec: DS construction} to condense the set of bosons $\{b_i\}$. Let us define $\mathcal{S}_\text{TC}^C$ to be the group of stabilizers in $\mathcal{S}_\text{TC}$ that commute with the elements of $\mathcal{S}_C$:
\begin{align}
\mathcal{S}^C_\text{TC} \equiv \{S \in \mathcal{S}_\text{TC} : S C  = C S, \, \forall C \in \mathcal{S}_C\}.
\end{align}
With some foresight, this is generated by the set of operators $\{A_{v,i}\}_{i=1}^M$ and $\{B_{p,i}\}_{i=1}^M$, defined graphically as:
\begin{align}
A_{v,i} \equiv \vcenter{\hbox{\includegraphics[scale=.22,trim={0cm 0cm 0cm 0cm},clip]{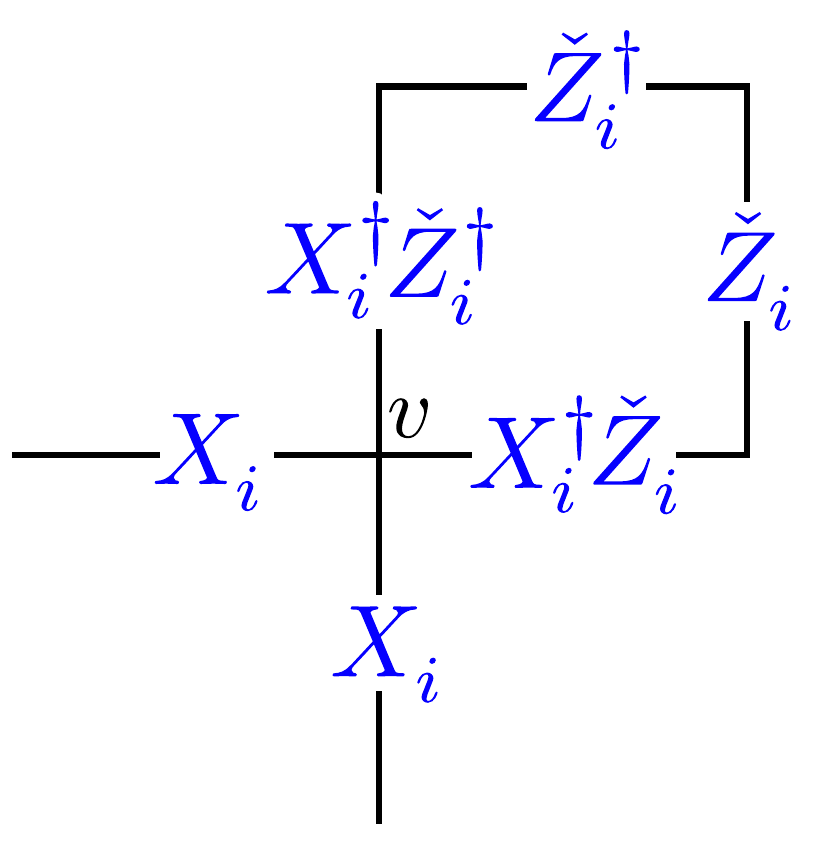}}}, \quad B_{p,i} \equiv \vcenter{\hbox{\includegraphics[scale=.22,trim={0cm 0cm 0cm 0cm},clip]{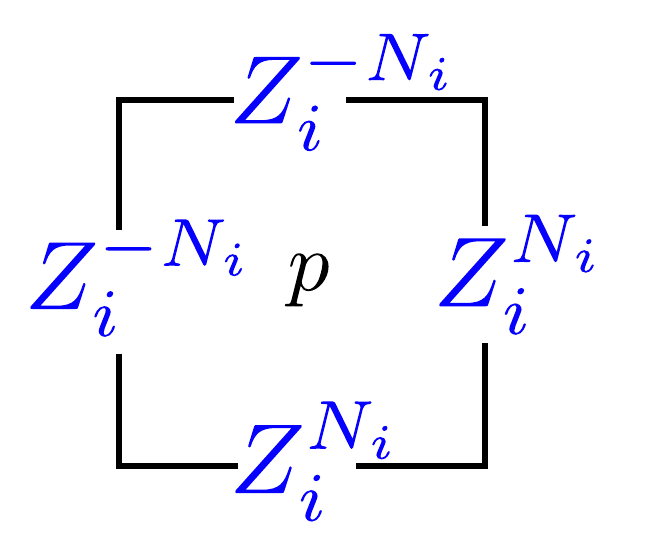}}}.
\end{align}
Here, $\check{Z}_i$ is notation for the operator:
\begin{align}
\check{Z}_i \equiv Z_i^{n_i} \prod_{j < i} Z_j^{n_{ij}}.
\end{align}
The stabilizer group of the condensed theory is then:
\begin{align}
\mathcal{S}_\text{TQD} \equiv \langle \mathcal{S}^C_\text{TC}, \mathcal{S}_C \rangle = \langle \{A_{v,i}\}_{i=1}^M, \{B_{p,i}\}_{i=1}^M, \{C_{e,i}\}_{i=1}^M \rangle.
\end{align}

The corresponding TQD stabilizer Hamiltonian $H_\text{TQD}$ is defined as:
\begin{align}
H_\text{TQD} \equiv -\sum_{v,i} A_{v,i} - \sum_{p,i} B_{p,i} - \sum_{e,i} C_{e,i} + \text{h.c.}.
\end{align}
The unit gauge charge $c_i$ and an elementary flux $\varphi_i$ can be created by the short string operators below:
\begin{align}
W_e^{c_i} &\equiv \vcenter{\hbox{\includegraphics[scale=.24,trim={0cm 0cm 0cm 0cm},clip]{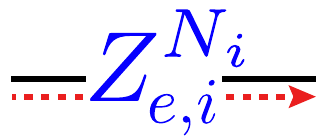}}}, \,\,\, \vcenter{\hbox{\includegraphics[scale=.24,trim={0cm 0cm 0cm 0cm},clip]{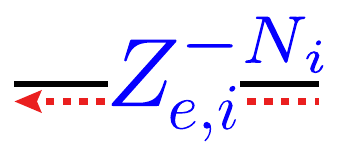}}}, \,\,\, \vcenter{\hbox{\includegraphics[scale=.24,trim={0cm 0cm 0cm 0cm},clip]{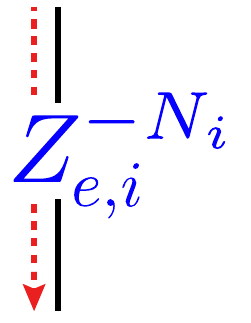}}}, \,\,\,\vcenter{\hbox{\includegraphics[scale=.24,trim={0cm 0cm 0cm 0cm},clip]{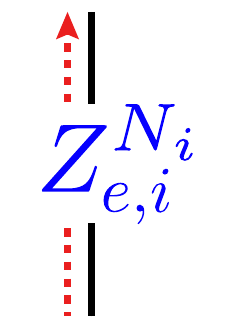}}} \\
W_e^{\varphi_i} &\equiv \vcenter{\hbox{\includegraphics[scale=.23,trim={0cm 0cm 0cm 0cm},clip]{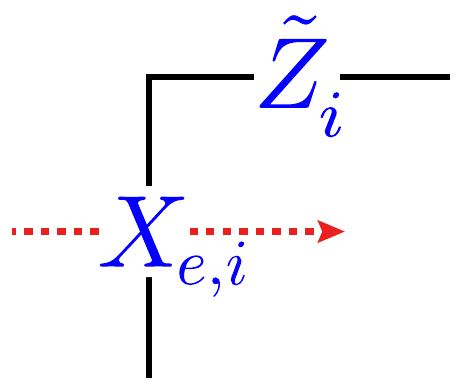}}}, \,\,\, \vcenter{\hbox{\includegraphics[scale=.23,trim={0cm 0cm 0cm 0cm},clip]{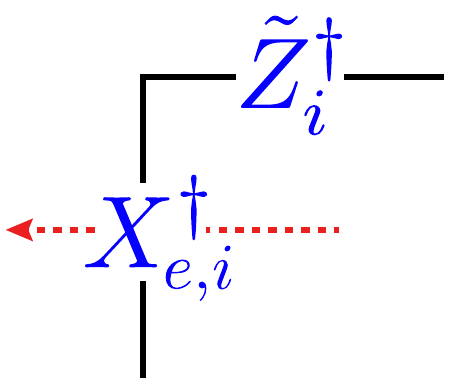}}}, \,\,\, \vcenter{\hbox{\includegraphics[scale=.23,trim={0cm 0cm 0cm 0cm},clip]{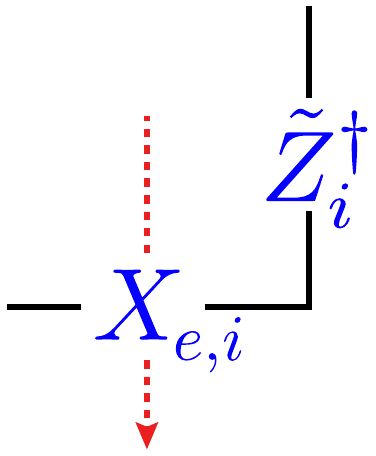}}}, \,\,\, \vcenter{\hbox{\includegraphics[scale=.23,trim={0cm 0cm 0cm 0cm},clip]{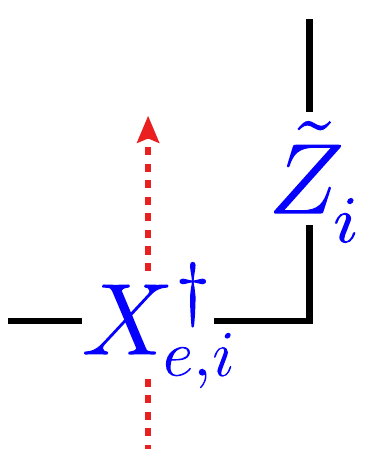}}},
\end{align}
where $\tilde{Z}_i$ is given by the product of Pauli Z operators:
\begin{align}
\tilde{Z}_i \equiv Z_i^{n_i} \prod_{j>i} Z_j^{\frac{N_j}{N_i}n_{ij}}.
\end{align}
These agree with the formulas for $\chargei$ and $\fluxi$ in Eqs.~\eqref{eq: def chargei} and \eqref{eq: def fluxi}. Longer string operators, along oriented paths $\gamma$ in the direct lattice and $\bar{\gamma}$ in the dual lattice, are given by:
\begin{align}
W_{\gamma}^{c_i} \equiv \prod_{e \in \gamma} W_e^{c_i}, \quad W_{\bar{\gamma}}^{\varphi_i} \equiv \prod_{e \in \bar{\gamma}} W_e^{\varphi_i}.
\end{align}

Lastly, as a consistency check, we compute the ground state degeneracy of $H_\text{TQD}$ on a torus. Similar to the calculation in Section~\ref{sec: DS Pauli stabilizer code}, we count the number of independent constraints on the states in the ground state subspace, defined by:
\begin{align}
\mathcal{H}_L \equiv \{|\psi \rangle : S|\psi \rangle = |\psi \rangle , \, \forall S \in \mathcal{S}_\text{TQD}\}.
\end{align}
We expect the ground state degeneracy to have order $|G|^2$, since the ground state degeneracy of a topological order on a torus is equivalent to the number of anyons in the theory \cite{kitaev2006anyons}.

To count the number of constraints on the ground state subspace, we let $N_v$ denote the number of vertices in the lattice. Then, for each layer $i$, there are $N_v$ vertex terms, $N_v$ plaquette terms, and $2N_v$ edge terms. Each of these yields an order $N_i$ constraint. The vertex term $A_{v,i}$ gives an order $N_i$ constraint, since $A_{v,i}^{N_i}$ is a product of plaquette and edge terms. On a torus, there are also global relations among the vertex stabilizers and plaquette stabilizers that need to be considered. In particular, for every $i$, the vertex terms and plaquette terms satisfy:
\begin{align}
\prod_v A_{v,i} =1, \quad \prod_p B_{p,i} =1.
\end{align}
Therefore, we only have $N_v-1$ independent vertex stabilizers and plaquette stabilizers. This gives us a total of $4N_v-2$ constraints with order $N_i$. Since there are two $N_i^2$-dimensional qudits per vertex, the dimension of the Hilbert space $\mathcal{H}_i$ in the layer $i$ is:
\begin{align}
 \text{dim}(\mathcal{H}_i) = (N_i^2)^{2N_v},
\end{align}
meaning that the dimension of the total Hilbert space $\mathcal{H}$ is:
\begin{align}
\text{dim}(\mathcal{H}) = \prod_{i=1}^M (N_i^2)^{2N_v}.
\end{align}
Given that there are $4N_v-2$ constraints with order $N_i$ for each layer $i$, the dimension of the ground state subspace is equal to:
\begin{align}
\text{dim}(\mathcal{H}_L) = \prod_{i=1}^M \frac{(N_i^2)^{2N_v}}{N_i^{4N_v-2}} = \prod_{i=1}^M N_i^2. 
\end{align}
This is equivalent to $|G|^2$, the order of the fusion group for the TQD. 

Note that the calculation above is sufficient to show that the set of operators  $\{A_{v,i}\}_{i=1}^M$ and $\{B_{p,i}\}_{i=1}^M$ generate the stabilizer group $\mathcal{S}^C_\text{TC}$. This is because, in the absence of the global relations (i.e., on a simply connected manifold), the ground state is nondegenerate. Consequently, any element of $\mathcal{S}_\text{TC}$ that commutes with the elements of $\mathcal{S}_\text{TQD}$, must already belong to $\mathcal{S}_\text{TQD}$ (see Appendix~C of Ref.~\cite{EKLH21}).

\section{Pauli stabilizer models of SPT phases} \label{sec: SPT}

As established in Sections~\ref{sec: primer} and \ref{sec: TQD stabilizer codes}, TQDs are derived by gauging the symmetry of models belonging to symmetry-protected topological (SPT) phases. Specifically, TQDs corresponding to a group $G$ and cocycle $\omega \in H^3[G,U(1)]$, are obtained by gauging the $G$ symmetry of the associated SPT models. Conversely, models of SPT phases can be constructed from TQDs by gauging $1$-form symmetries \footnote{Recall that, in two-dimensions, $1$-form symmetries are generated by operators supported along closed paths.}.
In this section, we demonstrate that gauging certain $1$-form symmetries of the TQD stabilizer models produces Pauli stabilizer models of SPT phases. To make the discussion concrete, we focus on a stabilizer model for a $\ZZ_2$ SPT phase, obtained by gauging a $1$-form symmetry of the DS stabilizer Hamiltonian. Similar arguments can be used to construct Pauli stabilizer models for SPT phase characterized by group cocycles of the form in Eq.~\eqref{eq: type I type II product}. 

Indeed, TQDs, and topologically ordered systems more generally, possess $1$-form symmetries
generated by anyon string operators along closed paths. 
Models belonging to the DS phase, for example, have a $1$-form $\ZZ_2 \times \ZZ_2$ symmetry with generators given by loops of semion $s$ and boson $s \bar{s}$ string operators. We point out that $1$-form symmetries in a topologically ordered system need not correspond to anyonic excitations, however. The DS stabilizer Hamiltonian in Section~\ref{sec: double semion stabilizer code} has a $1$-form $\ZZ_4 \times \ZZ_2$ symmetry generated by the vertex terms $A_v$ and plaquette terms $B_p$. Only the $\ZZ_2 \times \ZZ_2$ subgroup corresponds to anyon string operators. 

Given the connection between $1$-form symmetries and anyons, we can make a more precise statement about the construction of SPT models from TQDs. Namely, models for SPT phases arise from gauging the $1$-form symmetries associated to the gauge charges of TQDs. After gauging the $1$-form symmetry associated to the gauge charges, the system (i) gains a $0$-form symmetry and (ii) has trivial topological order (in the absence of symmetries) -- both of which are required for SPT phases. 
The theory obtained by gauging the symmetry has trivial topological order, since gauging $1$-form symmetries has the same effect as condensing the corresponding anyons (elaborated upon for the DS stabilizer model below). The gauge charges of the TQD form a Lagrangian subgroup (Section~\ref{sec: primer}) -- thus, condensing the gauge charges confines all of the anyons. 

We now focus on the DS stabilizer model. For the DS stabilizer model, in particular, the gauge charge is the boson $s \bar{s}$ with $\ZZ_2$ fusion rules. Therefore, we obtain a model for a $\ZZ_2$ SPT phase by gauging the $1$-form symmetry corresponding to $s \bar{s}$. For convenience, we re-write the DS stabilizer Hamiltonian $H_\text{DS}$:
\begin{align}
H_\text{DS} = -\sum_v A_v - \sum_p B_p -\sum_e C_e,
\end{align}
with $A_v$, $B_p$, and $C_e$ defined by:
\begingroup
\allowdisplaybreaks
\begin{equation} \label{eq: DS terms2}
\begin{gathered}
A_v = \vcenter{\hbox{\includegraphics[scale=.23,trim={.5cm 0cm 1.5cm 0cm},clip]{Figures/stabilizer_Av-eps-converted-to.pdf}}}, \qquad B_p = \vcenter{\hbox{\includegraphics[scale=.23,trim={0cm 0cm 0cm 0cm},clip]{Figures/stabilizer_Bp-eps-converted-to.pdf}}}, \\
 C_e = \vcenter{\hbox{\includegraphics[scale=.23,trim={0cm 0cm 0cm 0cm},clip]{Figures/stabilizer_Ceh-eps-converted-to.pdf}}}, \,\,\, \vcenter{\hbox{\includegraphics[scale=.23,trim={0cm 0cm 0cm 0cm},clip]{Figures/stabilizer_Cev-eps-converted-to.pdf}}}.
 \end{gathered}
\end{equation}
\endgroup
We also recall that the string operators for the anyon $s \bar{s}$ are generated by the short string operators:
\begin{align}
W^{s \bar{s}}_e = \vcenter{\hbox{\includegraphics[scale=.22,trim={0cm 0cm 0cm 0cm},clip]{Figures/stabilizer_Wssbarh-eps-converted-to.pdf}}}, \,\,\,\vcenter{\hbox{\includegraphics[scale=.22,trim={0cm 0cm 0cm 0cm},clip]{Figures/stabilizer_Wssbarv-eps-converted-to.pdf}}}.
\end{align}
Explicitly, the $1$-form symmetry corresponding to $s \bar{s}$ is generated by operators of the form:
\begin{align}
W_\lambda^{s \bar{s}} = \prod_{e \in \lambda} W^{s \bar{s}}_e, 
\end{align}
where $\lambda$ denotes a loop in the direct lattice. 

At this point, one could gauge the $1$-form symmetry following the usual minimal coupling prescription \cite{LG12,yoshida2015topological,kubica2018ungauging}. However, we find it illuminating to take a different approach. For Pauli stabilizer Hamiltonians, gauging a $1$-form $G$ symmetry corresponding to a boson $b$ is equivalent to condensing $b$ anyons bound to $0$-form $G$ charges. More concretely, for the DS stabilizer model, the $1$-form symmetry associated to $s \bar{s}$ can be gauged by condensing $s \bar{s}$ anyons bound to $0$-form $\ZZ_2$ charges, as described below.

\begin{figure}
\centering
    \includegraphics[width=.25\textwidth]{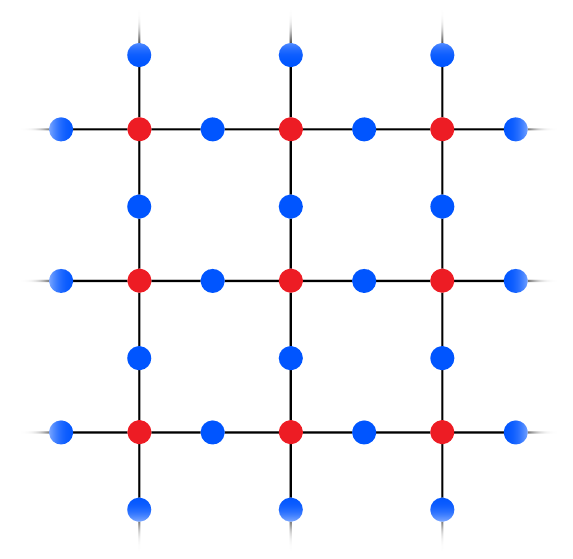}
     \caption{The Pauli stabilizer model of the $\ZZ_2$ SPT phase is defined on a square lattice with a four-dimensional qudit (blue) at each edge and a qubit (red) at each vertex.}
     \label{fig: squareSPTdof}
\end{figure}

To make this explicit, we add a qubit to each vertex of the square lattice, as in Fig.~\ref{fig: squareSPTdof}. We use $X_v$ and $Z_v$ to denote the Pauli X and Pauli Z operators at the vertex $v$. In what follows, we graphically represent $X_v$ and $Z_v$ with red operators. We define a modified Hamiltonian $\hat{H}_\text{DS}$ with ancillary qubits to be:
\begin{align}
\hat{H}_\text{DS}= H_\text{DS} - \sum_v X_v,
\end{align}
which corresponds to the stabilizer group:
\begin{align}
\hat{\mathcal{S}}_\text{DS} \equiv \langle \{A_v\}, \{B_p\}, \{X_v\} \rangle.
\end{align}
By construction, this Hamiltonian has an additional $0$-form $\ZZ_2$ symmetry generated by the product of Pauli X operators $\prod_v X_v$. The Pauli Z operator $Z_v$ creates a $0$-form $\ZZ_2$ charge at the vertex $v$. In terms of the minimal coupling prescription, the qubits on the vertices are the gauge fields. 

Next, we introduce the short string operators:
\begin{align} \label{eq: define De}
D_e \equiv \vcenter{\hbox{\includegraphics[scale=.24,trim={0cm 0cm 0cm 0cm},clip]{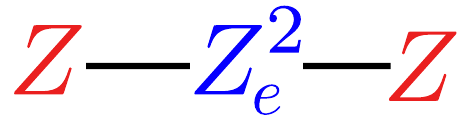}}}, \quad \vcenter{\hbox{\includegraphics[scale=.24,trim={0cm 0cm 0cm 0cm},clip]{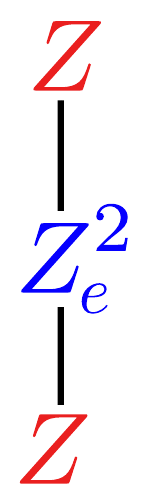}}},
\end{align}
which proliferate $s \bar{s}$ anyons bound to $0$-form $\ZZ_2$ charges. 
The claim is that gauging the $1$-form symmetry is equivalent to condensing the excitations created by the short string operators $D_e$.
From the perspective of minimal coupling, the operators in Eq.~\eqref{eq: define De} can be understood as the $1$-form Gauss's law. In accordance, products of $D_e$ along a closed path recover a $1$-form symmetry operator.

To condense the $s\bar{s}$ anyons bound to $\ZZ_2$ charges, we follow the logic described in Sections~\ref{sec: DS construction} and \ref{sec: TQD stabilizer code construction}. We start by defining the stabilizer group $\mathcal{S}_D$ generated by the set of short string operators $\{D_e\}$:
\begin{align}
\mathcal{S}_D \equiv \langle \{D_e\} \rangle.
\end{align}
We then define the stabilizer group $\hat{\mathcal{S}}_\text{DS}^D$ formed by the elements of $\hat{\mathcal{S}}_\text{DS}$ that commute with the operators in $\mathcal{S}_D$. Formally, this is the stabilizer group:
\begin{align} \label{eq: commutes with D}
\hat{\mathcal{S}}_\text{DS}^D \equiv \langle S \in \hat{\mathcal{S}}_\text{DS} : SD = DS, \, \forall D \in \mathcal{S}_D \rangle. 
\end{align}
The only generators of $\hat{\mathcal{S}}_\text{DS}$ that fail to commute with the elements of $\mathcal{S}_D$ are the vertex terms $A_v$ and $X_v$. Their products of the form $A_vX_v$, however, commute with every element of $\mathcal{S}_D$. We claim that the stabilizer group in Eq.~\eqref{eq: commutes with D} is generated by the following elements:
\begin{align}
\hat{\mathcal{S}}^D_\text{DS} = \langle \{A_vX_v\}, \{C_e\}, \{D_e\} \rangle.
\end{align}
Note that the plaquette terms $B_p$ are not included as generators, since they can be generated by products of the edge terms. The products $A_vX_v$ can be interpreted as coupling the vertex terms $A_v$ to the gauge fields. This ensures that the Hamiltonian terms are gauge invariant, i.e., commute with the $D_e$ terms.

Finally, the stabilizer group of the condensed theory is: 
\begin{align}
\mathcal{S}_\text{SPT} \equiv \langle \hat{\mathcal{S}}^D_\text{DS}, \mathcal{S}_D \rangle,
\end{align}
and the corresponding stabilizer Hamiltonian is:
\begin{align}
H_\text{SPT} \equiv -\sum_v A_vX_v - \sum_e C_e -\sum_e D_e + \text{h.c.}.
\label{eqn:SPT_Ham}
\end{align}
The Hamiltonian terms are pictured below: 
\begin{eqs}
A_vX_v &= \vcenter{\hbox{\includegraphics[scale=.23,trim={0cm 0cm 0cm 0cm},clip]{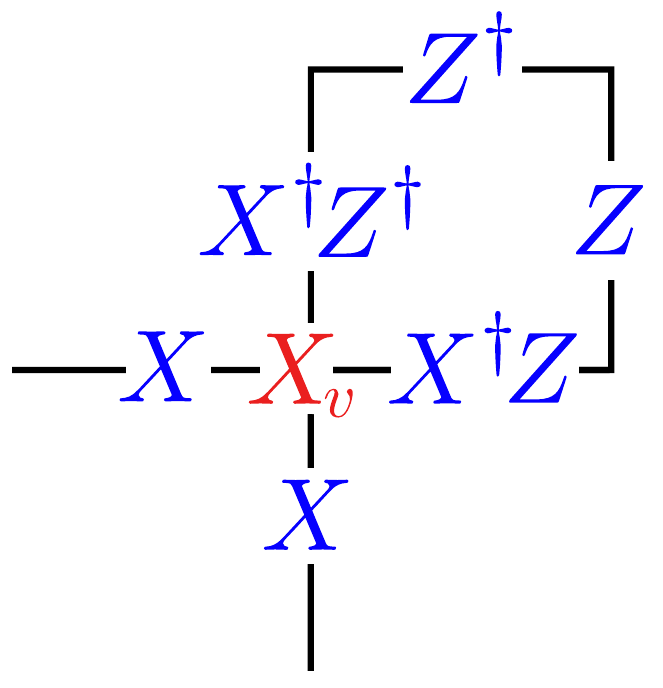}}}, \\ 
C_e &= \vcenter{\hbox{\includegraphics[scale=.23,trim={0cm 0cm 0cm 0cm},clip]{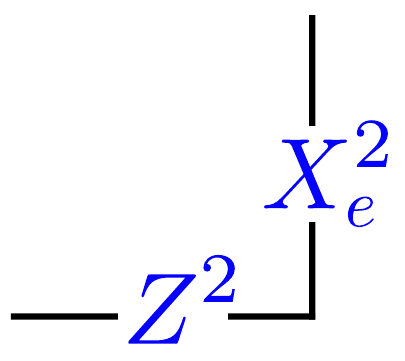}}}, \,\,\, \vcenter{\hbox{\includegraphics[scale=.23,trim={0cm 0cm 0cm 0cm},clip]{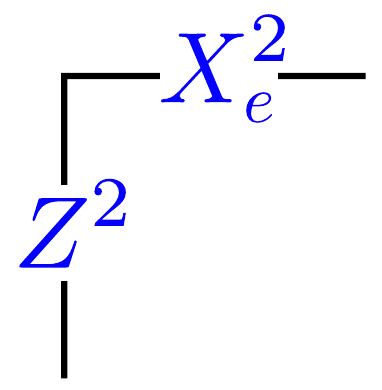}}} \\
D_e &= \vcenter{\hbox{\includegraphics[scale=.23,trim={0cm 0cm 0cm 0cm},clip]{Figures/stabilizer_Deh-eps-converted-to.pdf}}}, \,\,\, \vcenter{\hbox{\includegraphics[scale=.23,trim={0cm 0cm 0cm 0cm},clip]{Figures/stabilizer_Dev-eps-converted-to.pdf}}}.
\end{eqs}
Since the condensation procedure does not affect the $\ZZ_2$ symmetry of $\hat{H}_\text{DS}$, the Hamiltonian $H_\text{SPT}$ has a $0$-form $\ZZ_2$ symmetry generated by $\prod_v X_v$.

In what follows, we confirm that $H_\text{SPT}$ has no anyonic excitations by computing the ground state degeneracy of $H_\text{SPT}$ on a torus. We then use the methods of Ref.~\cite{EN14} to diagnose its SPT order. We find that the Hamiltonian in Eq.~\eqref{eqn:SPT_Ham} describes the $\ZZ_2$ SPT phase characterized by the nontrivial element of $H^3[\ZZ_2,U(1)]$.
 
Similar to the previous calculations of the ground state degeneracy in Sections~\ref{sec: DS Pauli stabilizer code} and \ref{sec: TQD stabilizer code construction}, we count the number of constraints in terms of the number of vertices $N_v$. Given that $A_vX_v$ squares to a product of $C_e$ and $D_e$ terms, the Hamiltonian terms give $5N_v$ order two constraints. Unlike the case of the TQD stabilizer models, there are no global relations amongst these constraints. Taking into account the qubit at each vertex and the four-dimensional qudit at each edge, the dimension of the total Hilbert space is $2^{N_v}4^{2N_v}$. After imposing the $5N_v$ order two constraints, we find that the ground state subspace $\mathcal{H}_L$ has dimension:
\begin{align}
\text{dim}(\mathcal{H}_L) = 2^{N_v}4^{2N_v}/2^{5N_v} = 1.
\end{align} 
Therefore, $H_\text{SPT}$ has a unique ground state on a torus (in fact on any manifold without boundary) and does not admit anyonic excitations. 

\begin{figure}
\centering
    \includegraphics[width=.35\textwidth]{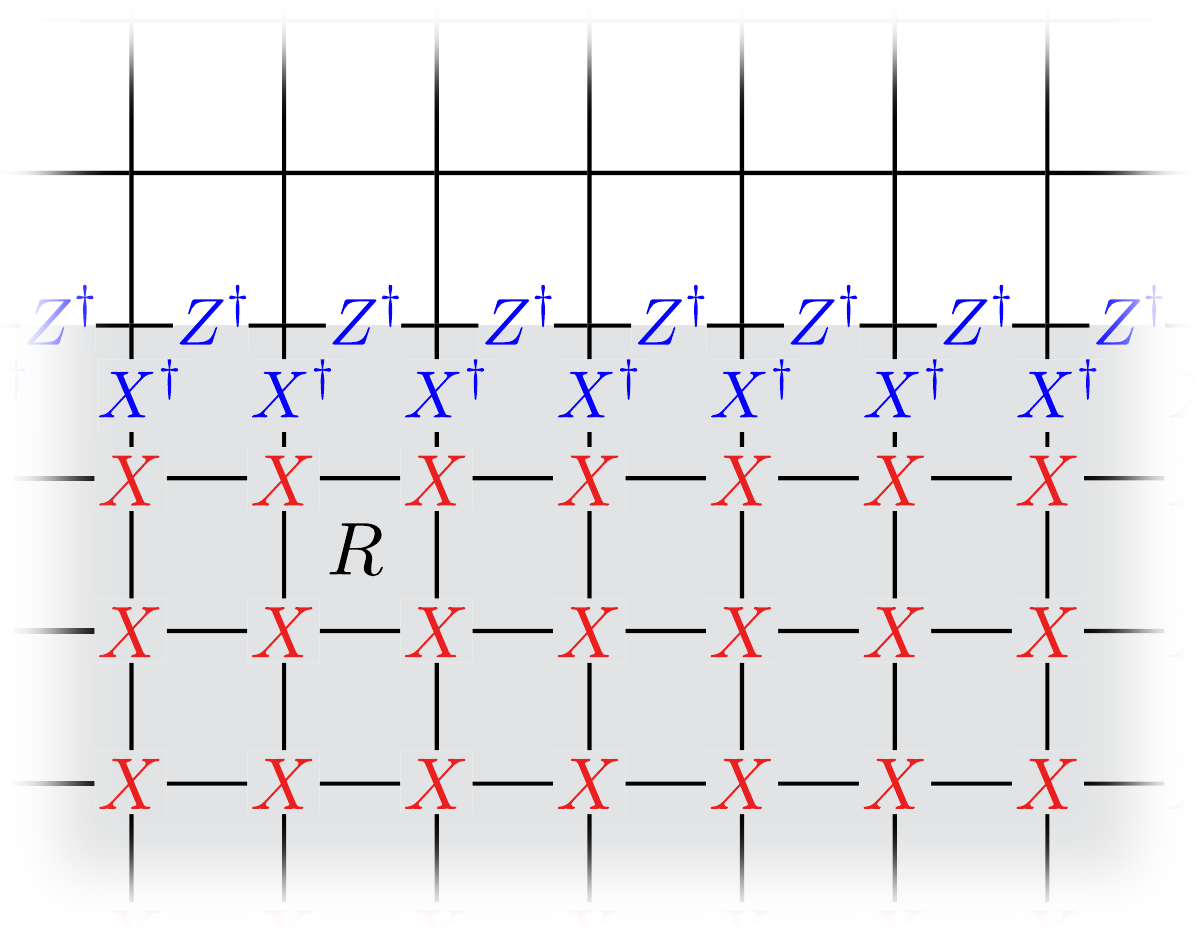}
     \caption{The action of $\tilde{P}_R(1)$ is shown above. It is equivalent to $P_R(1)$ for sites that are both inside the region $R$ (shaded grey) and away from the boundary of $R$.}
     \label{fig: stabilizersym}
\end{figure}

We now deduce the SPT phase described by $H_\text{SPT}$ by following the methods of Ref.~\cite{EN14} -- i.e., by considering an effective boundary symmetry action, we compute a group cocycle $\omega \in H^3[\ZZ_2,U(1)]$. For simplicity, we assume $H_\text{SPT}$ is defined on an infinite plane and use $R$ to denote the lower half plane. We then study the effective boundary symmetry action along the boundary of $R$. To this end, we define ${P}_R(g)$ to be the symmetry action restricted to $R$, i.e.:
\begin{align}
P_R(g) = \prod_{v \in R} X_v^g,
\end{align}
where $g$ is a $\{0,1\}$-valued element of $\ZZ_2$.
We also define $\tilde{P}_R(g)$ to be:
\begin{align}
\tilde{P}_R(g) = \prod_{\substack{A_vX_v \subset R}} (A_vX_v)^g,
\end{align}
with the product over all $A_vX_v$ whose support is contained within $R$.
$\tilde{P}_R(1)$ is portrayed in Fig~\ref{fig: stabilizersym}. Notice that away from the boundary of $R$, the action of ${P}_R(1)$ matches that of $\tilde{P}_R(1)$. As argued in Ref.~\cite{EKLH21}, their difference gives the effective boundary symmetry action $\mathcal{P}(g)$:
\begin{align}
\mathcal{P}(g) \equiv P_R(g) \tilde{P}_R(g)^\dagger.
\end{align}
$\mathcal{P}(1)$ acts as the identity away from the boundary of $R$ and near the boundary of $R$ can be graphically represented as:
\begin{align} \label{eq: ebsa}
\mathcal{P}(1) = \vcenter{\hbox{\includegraphics[scale=.35,trim={0cm 0cm 0cm 0cm},clip]{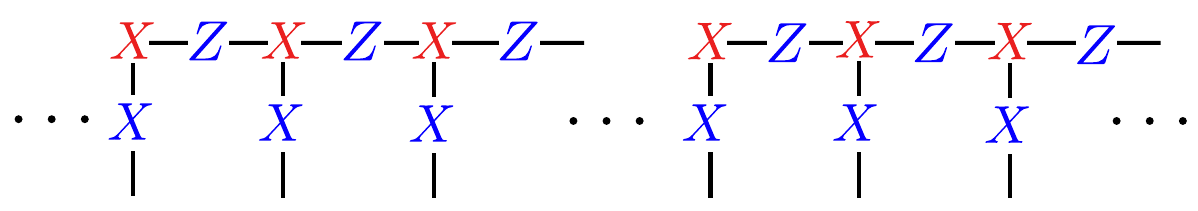}}}.
\end{align}

We emphasize that $\mathcal{P}(g)$ is a tensor product of Pauli operators. Conventional wisdom says that, if the effective boundary symmetry action is a tensor product, then the SPT model belongs to the trivial SPT phase \cite{WSW15}. However, the more accurate statement is that the SPT model belongs to the trivial SPT phase if the effective boundary symmetry action is a tensor product of linear representations of the symmetry. $\mathcal{P}(g)$ is not a tensor product of linear representations of $\ZZ_2$. $\mathcal{P}(g)$ only satisfies the $\ZZ_2$ group laws in the boundary Hilbert space $\mathcal{H}_B$, given by the set of states:
\begin{align}
 \mathcal{H}_B \equiv \{|\psi \rangle : S|\psi\rangle = |\psi \rangle, \, \forall S \in \mathcal{S}_\text{SPT} \text{ with } \text{supp}(S)\subset R\}.
\end{align}
Here, we have used $\text{supp}(S)$ to denote the support of the stabilizer $S$. (See Refs.~\cite{EN14,yoshida2015topological,EKLH21} for more details on the boundary Hilbert space of SPT models.) Indeed the effective boundary symmetry action squares to:
\begin{align}
\mathcal{P}(1)^2 = \vcenter{\hbox{\includegraphics[scale=.35,trim={0cm 0cm 0cm 0cm},clip]{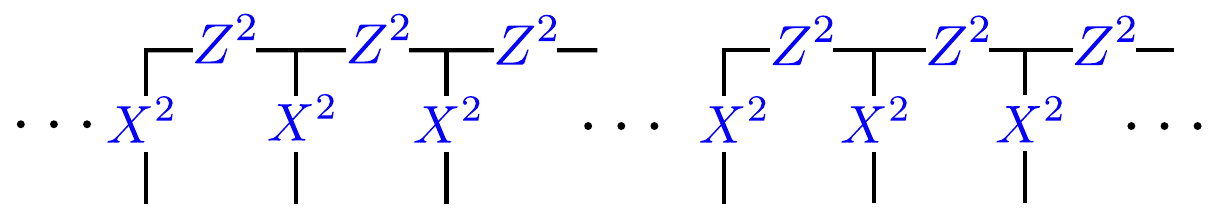}}},
\end{align}
which is a product of stabilizers whose support is contained within $R$.

The next step in diagnosing the SPT order is to truncate the effective boundary symmetry action. We truncate the effective boundary symmetry action in Eq.~\eqref{eq: ebsa} to an interval $\ell$ along the boundary of $R$ with endpoints $v_L$ and $v_R$. According to the arguments in Refs.~\cite{EKLH21} and \cite{EN14} the ambiguities at the endpoints of the truncation do not affect the characterization of the SPT phase. We choose the truncation of $\mathcal{P}(1)$ to 
take the form:
\begin{align}
\mathcal{P}_\ell(1) \equiv \vcenter{\hbox{\includegraphics[scale=.35,trim={0cm 0cm 0cm 0cm},clip]{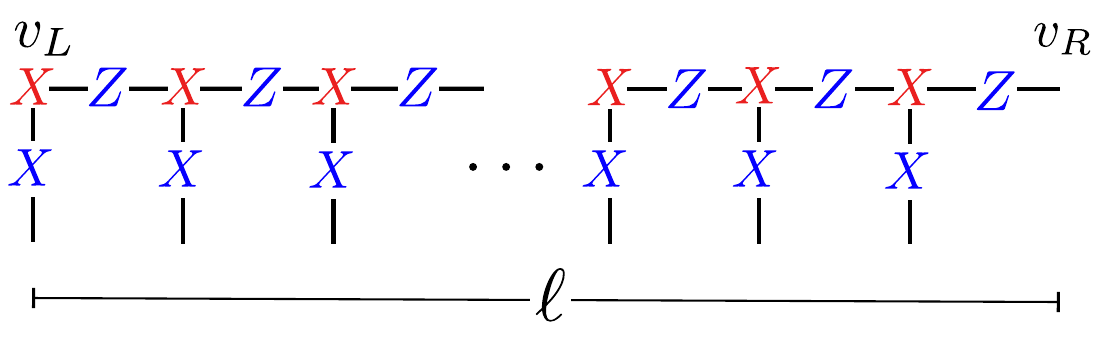}}}.
\end{align}
Furthermore, we take $\mathcal{P}_\ell(0)$ to be the identity. In the boundary Hilbert space, $\mathcal{P}_\ell(g)$ only satisfies the $\ZZ_2$ group laws up to operators at the endpoints:
\begin{align} \label{eq: eff boundary segment squared}
\mathcal{P}_\ell(1)^2 \sim \vcenter{\hbox{\includegraphics[scale=.33,trim={0cm 0cm 0cm 0cm},clip]{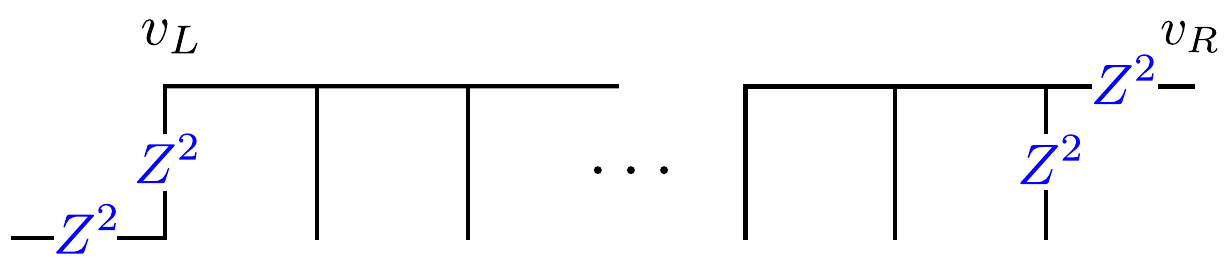}}},
\end{align}
where $\sim$ denotes that the relation holds in $\mathcal{H}_B$. We define $\Omega(g,h)$, for $g,h \in \ZZ_2$ to be the right hand side of Eq.~\eqref{eq: eff boundary segment squared} if $g=h=1$ and the identity otherwise. Then the truncated effective boundary symmetry action satisfies:
\begin{align} \label{eq: failure of group law}
\mathcal{P}_\ell(g) \mathcal{P}_\ell(h) \sim \Omega(g,h) \mathcal{P}_\ell(g+h).
\end{align}
$\Omega(g,h)$ can be decomposed as:
\begin{align} \label{eq: Omega decomposition}
\Omega(g,h) = \Omega_{v_L}(g,h) \Omega_{v_R}(g,h),
\end{align}
where $\Omega_{v_L}(g,h)$ and $\Omega_{v_R}(g,h)$ are localized near $v_L$ and $v_R$, respectively. We take $\Omega_{v_L}(g,h)$ to be:
\begin{align}
\Omega_{v_L}(g,h) \equiv \vcenter{\hbox{\includegraphics[scale=.365,trim={0cm 0cm 0cm 0cm},clip]{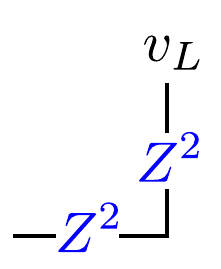}}}.
\end{align}

The last step is to consider the associativity of the truncated effective symmetry actions for $g,h,k \in \ZZ_2$:
\begin{align} \label{eq: associativity}
\mathcal{P}_\ell(g)[\mathcal{P}_\ell(h)\mathcal{P}_\ell(k)] = [\mathcal{P}_\ell(g)\mathcal{P}_\ell(h)]\mathcal{P}_\ell(k).
\end{align}
Substituting Eq.~\eqref{eq: failure of group law} into Eq.~\eqref{eq: associativity}, we obtain:
\begin{align}
[\mathcal{P}_\ell(g)\Omega(h,k)\mathcal{P}_\ell(g)^\dagger]\Omega(g,h+k) \sim \Omega(g,h) \Omega(g+h,k).
\end{align}
Given the decomposition of $\Omega(g,h)$ in Eq.~\eqref{eq: Omega decomposition}, the condition above is satisfied up to a phase at each endpoint, i.e., at the endpoint $v_L$, we have:
\begin{multline} \label{eq: omega definition SPT}
[\mathcal{P}_\ell(g)\Omega_{v_L}(h,k)\mathcal{P}_\ell(g)^\dagger]\Omega_{v_L}(g,h+k) \\
\sim \omega(g,h,k) \Omega_{v_L}(g,h) \Omega_{v_L}(g+h,k).
\end{multline}
Here, $\omega$ is the $U(1)$-valued group cocycle in $H^3[\ZZ_2,U(1)]$ that characterizes the SPT phase. From Eq.~\eqref{eq: omega definition SPT}, we see that $\omega$ is given by:
\begin{align}
\omega(g,h,k)=
\begin{cases}
-1 & \text{if } g,h,k =1, \\
1 & \text{otherwise}.
\end{cases}
\end{align}
This represents the nontrivial element of $H^3[\ZZ_2,U(1)]$, implying that $H_\text{SPT}$ describes the nontrivial $\ZZ_2$ SPT phase. We would also like to point out that the calculation of $\omega$ above, is analogous to the calculation of the $F$-symbol \cite{KL20,KL21} for the semion $s$ in the DS stabilizer model. This is because  the effective boundary symmetry action only differs from the $s$ string operator by $X_v$ operators, which do not affect the calculation. 

We have now shown that the nontrivial two-dimensional $\ZZ_2$ SPT phase can be modeled by the Pauli stabilizer Hamiltonian $H_\text{SPT}$. We emphasize that this does not conflict with the results of Ref.~\cite{EKLH21}, since the argument in Ref.~\cite{EKLH21} assumes that the $\ZZ_2$ SPT model is defined on qubits, as opposed to four-dimensional qudits. Furthermore, the construction of Pauli stabilizer models for SPT phases presented here is restricted to SPT phases characterized by group cocycles that are products of type I and type II cocycles. We are unable to construct Pauli stabilizer models for SPT phases characterized by type III cocycles, in agreement with Ref.~\cite{EKLH21}. This is also consistent with the fact that type III cocycles correspond to TQDs with non-Abelian anyons, which cannot be modeled by Pauli stabilizer Hamiltonians.

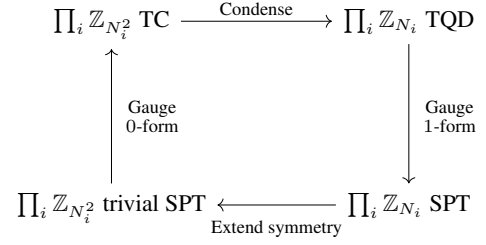
\begin{figure}
    \centering
    \[\begin{tikzcd}
	{\prod_i\mathbb{Z}_{N_i^2} \text{ TC}} && {\prod_i \mathbb{Z}_{N_i} \text{ TQD}} \\
	{\,\,\,\,\,\,\,\,\,\,\,\,\,\,\,\,\,\,\,\,\substack{\text{\scriptsize Gauge} \\ \text{\scriptsize $0$-form}}} && {\,\,\,\,\,\,\,\,\,\,\,\,\,\,\,\,\,\,\,\,\substack{\text{\scriptsize Gauge} \\ \text{\scriptsize $1$-form}}} \\
	{\prod_i\mathbb{Z}_{N_i^2} \text{ trivial SPT}} && {\prod_i \mathbb{Z}_{N_i} \text{ SPT}}
	\arrow["{\text{\scriptsize Condense}}", from=1-1, to=1-3]
	\arrow[from=1-3, to=3-3]
	\arrow[from=3-1, to=1-1]
	\arrow["{\text{\raisebox{-.3cm}{\scriptsize Extend symmetry}}}", from=3-3, to=3-1]
\end{tikzcd}\]
    \caption{An SPT model protected by a $\prod_i \mathbb{Z}_{N_i}$ symmetry (and characterized by a product of type I and type II cocycles) is trivialized by extending the symmetry to $\prod_i\mathbb{Z}_{N_i^2}$. Subsequently gauging the $0$-form symmetry produces a model in the same phase as a $\prod_i\mathbb{Z}_{N_i^2}$ TC. One can then condense bosons according to the prescription in Section~\ref{sec: TQD stabilizer codes} to obtain a $\prod_i \mathbb{Z}_{N_i}$ TQD stabilizer model and gauge the $1$-form symmetries corresponding to the gauge charges to arrive at a stabilizer model belonging to the $\prod_i \mathbb{Z}_{N_i}$ SPT phase.}
    \label{fig: commutative diagram}
\end{figure}

We note that the $\ZZ_2$ SPT model belongs to a trivial SPT phase, when considered as an SPT model protected by a $\ZZ_4$ symmetry. This is because the effective boundary symmetry action in Eq.~\eqref{eq: ebsa} is a tensor product of linear representations of $\ZZ_4$. In other words, the SPT model can be trivialized by extending the $\ZZ_2$ symmetry to a $\ZZ_4$ symmetry \cite{Prakash2018Unwinding,Juven2018Boundaries}. More generally, an SPT model protected by a $\prod_i \mathbb{Z}_{N_i}$ symmetry, which is characterized by a product of type I and type II cocycles, can be trivialized by extending the symmetry to $\prod_i\mathbb{Z}_{N_i^2}$. This yields the commutative diagram shown in Fig.~\ref{fig: commutative diagram}, which further motivates the construction of the $\prod_i \mathbb{Z}_{N_i}$ TQD stabilizer model from a $\prod_i\mathbb{Z}_{N_i^2}$ TC.

\section{Discussion}

We have constructed a Pauli stabilizer model for every twisted quantum double (TQD) with Abelian anyons. Our strategy was to condense bosonic anyons in decoupled toric codes (TCs) so that the remaining deconfined excitations are those of the Abelian TQD. As an example, we constructed a Pauli stabilizer model of the double semion (DS) topological order in Section~\ref{sec: double semion stabilizer code}. We explicitly verified that the DS stabilizer model belongs to the same phase as the DS string-net model by finding a finite-depth quantum circuit (with ancillary degrees of freedom) that maps between the ground state subspaces. In addition, we described how Pauli stabilizer models of SPT phases can be obtained from the TQD stabilizer models. We made the construction explicit for the nontrivial $\ZZ_2$ SPT phase of Ref.~\cite{levin2012braiding}.

Our work builds on the classification of topological Pauli stabilizer codes initiated by Refs.~\cite{bombin2014structure,BDCP12,Haah2018a}. We conjecture that our models give a complete classification of topological Pauli stabilizer codes up to finite-depth Clifford circuits with ancilla. For four-dimensional qudits, for example, this would imply that every Pauli stabilizer model is locally equivalent to decoupled copies of $\mathbb{Z}_4$ TCs, $\mathbb{Z}_2$ TCs, DS stabilizer codes, and six-semion stabilizer codes (defined in Appendix~\ref{app: K matrix}).
However, the proof of such a classification is challenging, as it involves working with polynomial rings over finite rings as opposed to polynomial rings over finite fields in the case of prime-dimensional qudits. More specifically, the current approach to classification requires a rigorous statement about the existence of a bosonic anyon. The argument in Ref.~\cite{HFH18} uses Hasse-Minkowski theorem to prove that there is a gapped one-dimensional boundary, implying that there is a nontrivial Lagrangian subgroup for a topologically ordered system (see Section~\ref{sec: primer}). This does not directly apply to $\mathbb{Z}_4$ and hence, new techniques need to be developed for this case. Furthermore, the current classification requires a proof that all local excitations of a Pauli stabilizer Hamiltonian are fully mobile in two dimensions. We expect that a generalization of Hilbert's syzygy theorem is needed for systems of composite-dimensional qudits. We leave the details to forthcoming works. 

Another important future objective is to understand the quantum error correcting properties of the TQD stabilizer codes. As a first step, it would be interesting to compute the optimal error thresholds using the statistical mechanical mappings of Refs.~\cite{dennis2002memory,Bombin2010subsystem,Andrist2015error,Chubb2021statistical}. In such mappings, a spin is associated with each local stabilizer generator and coupling strengths are determined by the so-called Nishimori conditions~\cite{Nishimori_1981}. This yields a disordered statistical mechanical Hamiltonian, where the disorder realization depends on the configuration of Pauli errors on the code space. Subsequently, the phase diagram for the model can be studied using Monte Carlo methods, and the optimal error threshold of a topological Pauli stabilizer code can be read off from the critical point along the Nishimori line. One could also consider other methods for estimating the error thresholds, such as those of Refs.~\cite{AnwarFast2014,HutterHDRG2015,WatsonFast2015}, which were used to compute the thresholds of $\ZZ_p$ TCs with prime $p$. Beyond studying the error correcting properties, it would also be interesting to explore whether there are fault-tolerant operations that are more natural to implement on TQD stabilizer codes as compared to TCs.  

The Pauli stabilizer models presented here may be of interest beyond their potential for quantum error correction. In particular, they are a substantial simplification from their string-net counterparts and may be amenable to simulation on many-body qudit platforms. Moreover, as mentioned in Section~\ref{sec: DS construction}, the ground state of the DS stabilizer Hamiltonian can be prepared by making two-body measurements of a $\ZZ_4$ TC ground state. This gives an efficient construction of the DS ground state starting from the $\ZZ_4$ TC ground state. One could make further measurements to obtain the ground state of the $\ZZ_2$ SPT model of Section~\ref{sec: SPT}. Anyon condensation in TCs also allows for the construction of symmetry-enriched topological phases, so similar methods as Section~\ref{sec: SPT} could be used to create simulable models of these phases. It may be insightful to compare the condensation approach employed here to other means of preparing topological states, such as in Ref.~\cite{liu2021methods}.

Finally, while TQDs capture all Abelian anyon theories that admit gapped boundaries \cite{KS11,KKOSS21}, there are of course, two-dimensional Abelian anyon theories that possess obstructions to gapped boundaries (e.g., a nonzero chiral central charge). It is expected that such Abelian anyon theories cannot be described by Pauli stabilizer models in two spatial dimensions \cite{kitaev2006anyons,KapustinThermal2020}. However, Abelian anyon theories without gapped boundaries have been realized on the boundary of commuting projector Hamiltonians in three spatial dimensions \cite{walker2012}. Refs.~\cite{Haah2021QCA} and \cite{CDESTW21} have taken this one step further and constructed three-dimensional Pauli stabilizer models that host chiral Abelian anyon theories (and ungappable Abelian anyon theories, more generally) on their two-dimensional surface. Up to stacking with two-dimensional TQD stabilizer models and condensing bosons, these provide a three-dimensional Pauli stabilizer model for every two-dimensional Abelian anyon theory, wherein the anyon theory is realized on the boundary. We also note that it may be interesting to consider fault-tolerant quantum computation in these models along the lines of Refs.~\cite{Raussendorf2006oneway} and \cite{roberts20203fermion}.

Abelian anyon theories without gapped boundaries have also been identified in the context of topological subsystem codes, such as the three-fermion subsystem code in Refs.~\cite{Bombin2009fermions,Bombin2010subsystem,Suchara2011subsystem,roberts20203fermion}. Roughly speaking, topological subsystem codes correspond to a parameter space of frustrated Hamiltonians with common conserved quantities, which are taken to be the stabilizers of the subsystem code \cite{Bombin2010subsystem}. In the honeycomb model of Ref.~\cite{kitaev2006anyons}, for example, loops of fermionic string operator are preserved throughout the phase diagram and define the stabilizer group \footnote{Note that, although the model of Ref.~\cite{kitaev2006anyons} can be interpreted as a topological subsystem code, it does not encode any logical qubits. See Ref.~\cite{Hastings2021dynamically}, however.}. With this, one can assign an anyon theory to a topological subsystem code based on the conserved quantities. The basic idea is that the stabilizers of the subsystem code generate a $1$-form symmetry associated to an Abelian anyon theory. Preliminary work suggests that we can leverage our TQD stabilizer codes to build a topological subsystem code for every two-dimensional Abelian topological order, regardless of whether the theory admits gapped boundaries. Such models may provide natural candidates for systems that host non-Abelian anyons, similar to the honeycomb model of Ref.~\cite{kitaev2006anyons}. 

\vspace{0.05in}
\noindent{\it Acknowledgements -- } TDE thanks Lukasz Fidkowski and Meng Cheng for valuable insights on the classification of Abelian topological orders. TDE would also like to acknowledge Timothy H. Hsieh and Zi-Wen Liu for inspirational discussions about the double semion phase. A.D. thanks Bowen Yang for useful discussions on related problems. This work was supported by the JQI fellowship at the University of Maryland (YC), the Simons Foundation through the collaboration on Ultra-Quantum Matter (651438, AD; 651444, WS), NSERC (NT), the It from Qubit collaboration (DJW), and by the Institute for Quantum Information and Matter, an NSF Physics Frontiers Center (PHY-1733907, AD, WS).

\appendix

\section{Relation to string-net model ground states} \label{app: string-net ground states}

In Section~\ref{sec: relation to nonPauli}, we mapped the ground state subspace of the DS stabilizer Hamiltonian $H_\text{DS}$ to that of the DS string-net model $H_\text{DS}^\text{s-n}$ indirectly -- by mapping $H_\text{DS}$ to $H_\text{DS}^\text{s-n}$ using ancillary degrees of freedom, a finite-depth quantum circuit, and ground state preserving changes to the intermediate Hamiltonians. In this Appendix, we make the mapping of the ground states explicit. To do so, we employ notation from simplicial cohomology (reviewed in the next section). 

\subsection{Simplicial cohomology notation} \label{app: cohomology notation}

In the discussion below, we make use of concepts from simplicial cohomology to construct a finite-depth quantum circuit that maps the ground state subspace of the DS stabilizer Hamiltonian to the ground state subspace of the DS string-net model.
We begin by summarizing the terminology used in the construction and provide intuition for some of the formulas. More details and useful examples of simplicial cohomology used in the context of quantum many-body systems can be found in Ref.~\cite{CT21}.

\begin{figure}
\centering
\subfloat[\label{fig: branchingtriangular}]{\includegraphics[width=.35\textwidth]{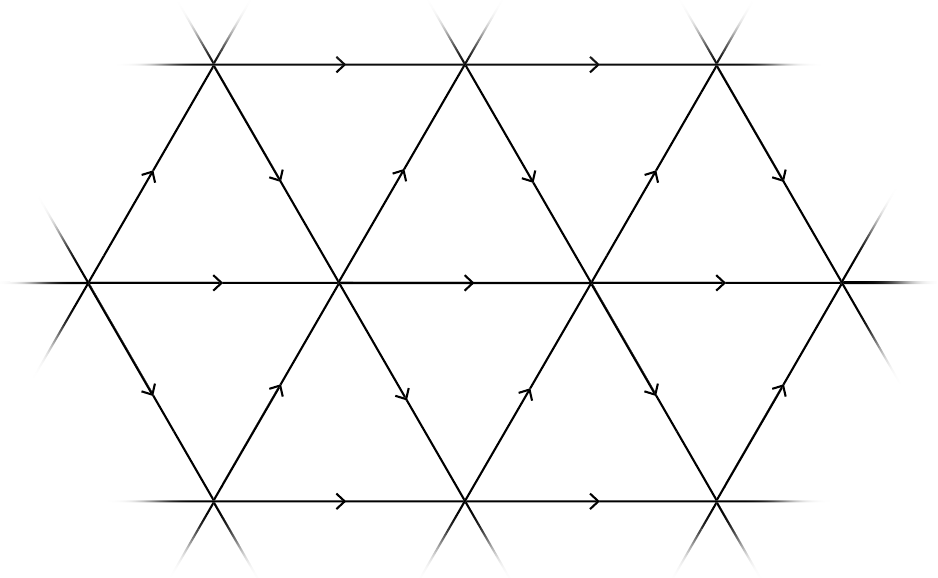}} \\
\subfloat[\label{fig: branchingorder}]{\includegraphics[width=.22\textwidth]{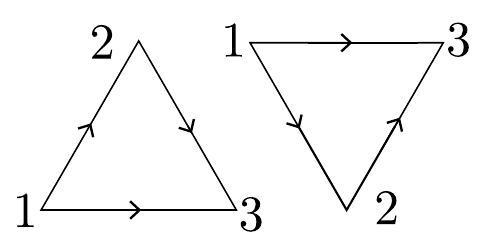}}
     \caption{(a) We consider triangulations with a branching structure -- each edge is assigned an orientation such that there no cycles around any of the triangles. (b) The vertices of each triangle can be ordered according to the number of incident edges.}
\end{figure}

In what follows, we consider two-dimensional triangulated manifolds $\mathcal{M}$ equipped with a branching structure, i.e., an assignment of an orientation to each edge of the lattice with the property that there are no cycles around any face (see Fig.~\ref{fig: branchingtriangular}, for example). The branching structure determines an ordering of the vertices of a triangle, as shown in Fig.~\ref{fig: branchingorder}. We refer to the vertices $v$, edges $e$, and faces $f$ as $0$-simplices, $1$-simplices, and $2$-simplices, respectively. We label a $p$-simplex $\sigma_p$ by its vertices $\langle 1 \ldots p+1 \rangle$, where the vertices are ordered according to the branching structure. For example, a face can be labeled as $\langle 123 \rangle$ with edges  $\langle 23 \rangle$, $\langle 13 \rangle$, and $\langle 12 \rangle$ and vertices $\langle1\rangle$, $\langle2\rangle$, and $\langle3\rangle$.

A $p$-chain on $\mathcal{M}$ is a formal sum of $p$-simplices with coefficients in an Abelian group $A$. An arbitrary $p$-chain $s_p$ takes the form:
\begin{align}
s_p = \sum_p \alpha_p \sigma_p, \text{ with } \alpha_p \in A.
\end{align}
Explicitly, arbitrary $1$-chains $s_1$ and $2$-chains $s_2$ can be written in the form:
\begin{align} \label{eq: arbitrary chains}
s_1 = \sum_{\langle 12 \rangle} \alpha_{\langle 12 \rangle} \langle 12 \rangle, \quad
s_2 = \sum_{\langle 123 \rangle} \alpha_{\langle 123 \rangle} \langle 123 \rangle.
\end{align}
The $p$-chains over $A$ form a group denoted by $C_p[\mathcal{M},A]$. For two-dimensional manifolds, the group $C_p[\mathcal{M},A]$ is taken to be the trivial group for $p < 0$ and $p>2$.

The boundary operator $\partial$ is a linear map from the group of $(p+1)$-chains to the group of $p$-chains:
\begin{align}
\partial: C_{p+1}[\mathcal{M},A] \to C_p[\mathcal{M},A].
\end{align}
The action of the boundary operator on the arbitrary chains in Eq.~\eqref{eq: arbitrary chains} is given by:
\begin{eqs}
\partial s_1 &= \sum_{\langle 12 \rangle} \alpha_{\langle 12 \rangle} (\langle 2 \rangle - \langle 1 \rangle), \\
\partial s_2 
&= \sum_{\langle 123 \rangle} \alpha_{\langle 123 \rangle} (\langle 23 \rangle - \langle 13 \rangle + \langle 12 \rangle).
\end{eqs}
While $p$-chains and the boundary operator $\partial$ do not appear in the main text, they are essential to defining cochains and the coboundary operator, described below.

A $p$-cochain $\bm{a}$ is a linear map from the group of $p$-chains to $A$:
\begin{align}
\bm{a}: C_p[\mathcal{M},A] \to A.
\end{align}
The Abelian group formed by $p$-cochains is denoted by $C^p[\mathcal{M},A]$. Letting $A = \ZZ_N$, for some $N$, we use $\bm{v}$ and $\bm{e}$ to represent the cochains:
\begin{align}
\bm{v}(v') = 
\begin{cases}
1 & \text{ if } v' = v, \\
0 & \text{ otherwise},
\end{cases} \quad 
\bm{e}(e') = 
\begin{cases}
1 & \text{ if } e' = e, \\
0 & \text{ otherwise}.
\end{cases}
\end{align}
Any $0$-cochain $\bm{a}$ or $1$-cochain $\bm{c}$ can then be expressed as a linear combination of the cochains $\bm{v}$ and $\bm{e}$:
\begin{align}
\bm{a} = \sum_v \alpha_v \bm{v}, \quad \bm{c} = \sum_e \alpha_e \bm{e},
\end{align}
for some $\alpha_v,\alpha_e \in \ZZ_N$.

The coboundary operator $\delta$ is a linear map from the group of $p$-cochains to the group of $(p+1)$-cochains:
\begin{align}
\delta :  C^p[\mathcal{M},A] \to C^{p+1}[\mathcal{M},A].
\end{align}
Given a $p$-cochain $\bm{a}$, the coboundary of $\bm{a}$ is defined by:
\begin{align}
\delta \bm{a} (s) = \bm{a} (\partial s),
\end{align}
where $s$ is an arbitrary $(p+1)$-chain. For example, taking $A$ to be $\ZZ_2$, the coboundary of $\bm{v}$ satisfies:
\begin{align} \label{eq: coboundary v}
\delta \bm{v} (e) = 
\begin{cases} 
1 & \text{ if } \bm{v}(\partial e) = 1, \\
0 & \text{ otherwise}.
\end{cases}
\end{align}
This says that $\delta \bm{v}(e)$ is $1$ for every edge connected to $v$. More generally, let $\bm{b}$ be the $\ZZ_2$-valued $0$-cochain:
\begin{align}
\bm{b} = \sum_v \beta_v \bm{v},
\end{align}
for some $\beta_v \in \ZZ_2$.
Then the coboundary of $\bm{b}$ is equal to:
\begin{align}
\delta \bm{b} = \sum_v \beta_v \delta \bm{v},
\end{align}
which, given Eq.~\eqref{eq: coboundary v}, evaluates to $1$ on edges along closed paths in the dual lattice (see Fig.~\ref{fig: DSloops}). This shows that the states $|\delta \bm{b} \rangle$ in Appendix~\ref{app: ground state mapping}  correspond to configurations of loops in the dual lattice.

Lastly, we introduce the cup product $\cup$. The cup product maps a $p$-cochain and a $q$-cochain to a $(p+q)$-cochain:
\begin{align}
\cup: C^p[\mathcal{M},A] \times C^q[\mathcal{M},A] \to C^{p+q}[\mathcal{M},A].
\end{align}
The cup product of the $p$-cochain $\bm{a}_p$ and the $q$-cochain $\bm{a}_q$ evaluated on arbitrary $(p+q)$-simplex $\sigma_{p+q} = \langle 1 \ldots p+q+1 \rangle$ is:
\begin{align}
\bm{a}_p \cup \bm{a}_q (\sigma_{p+q}) = \bm{a}_p (\langle 1 \ldots p+1 \rangle) \bm{a}_q(\langle p+1 \ldots p+q+1 \rangle).
\end{align}
Here, $\langle 1 \ldots p+1 \rangle$ is the $p$-simplex formed by the first $p+1$ vertices of $\sigma_{p+q}$ and $\langle p+1 \ldots p+q+1 \rangle$ is the $q$-simplex formed by the last $q+1$ vertices of $\sigma_{p+q}$.

As an example of the cup product, we consider the $\ZZ_2$-valued cochain $\bm{e}' \cup \bm{e}$ from Eq.~\eqref{eq: Ce cohomology}. $\bm{e}'$ and $\bm{e}$ are $1$-cochains, so the cup product is a $2$-cochain. Evaluated on a face $\langle 123 \rangle$, we have:
\begin{align}
\bm{e}' \cup \bm{e}(\langle 123 \rangle) = \bm{e}'(\langle 12 \rangle) \bm{e}(\langle 23 \rangle).
\end{align}
This is nonzero if and only if $e' = \langle 12 \rangle$ and $e = \langle 23 \rangle$. Another interesting example is the cup product $\bm{v} \cup \delta \bm{v}$, which appears in Eq.~\eqref{eq: Av simplification}. $\bm{v} \cup \delta \bm{v}$ is a $\ZZ_2$-valued $2$-cochain satisfying:
\begin{align}
\bm{v} \cup \delta \bm{v} (\langle 12 \rangle) = \bm{v}(\langle 1 \rangle) \delta \bm{v}(\langle 12 \rangle),
\end{align}
for any edge $\langle 12 \rangle$. This is nonzero if and only if $v = \langle 1 \rangle$ and $\langle 12 \rangle$ is connected to $v$. In other words, $\bm{v} \cup \delta \bm{v}$ evaluates to $1$ on an edge $e$ if and only if $e$ is oriented outwards from $v$.

\subsection{Ground state mapping} \label{app: ground state mapping}

Having introduced notation from simplicial cohomology, we are now ready to show explicitly that the ground states of the DS stabilizer model can be mapped to the ground states of the DS string-net model using a finite-depth quantum circuit. To simplify the discussion, we start with a DS stabilizer model defined on a triangular lattice equipped with a branching structure (Fig.~\ref{fig: branchingtriangular}). Although the argument below assumes that the manifold is simply-connected (so that the ground state of the DS stabilizer model is unique) it can be generalized to arbitrary two-dimensional orientable manifolds straightforwardly.

We consider a DS stabilizer Hamiltonian of the form:
\begin{align}
H_\text{DS} = -\sum_v A_v - \sum_f B_f - \sum_e C_e + \text{h.c.},
\end{align}
with the terms given graphically as:
% \begin{eqs}
% A_v &= \vcenter{\hbox{\includegraphics[scale=.23,trim={0cm 0cm 0cm 0cm},clip]{Figures/stabilizer_triAv-eps-converted-to.pdf}}},  \\
% B_f &= \vcenter{\hbox{\includegraphics[scale=.23,trim={0cm 0cm 0cm 0cm},clip]{Figures/stabilizer_triBf1-eps-converted-to.pdf}}}, \, \, \, \vcenter{\hbox{\includegraphics[scale=.23,trim={0cm 0cm 0cm 0cm},clip]{Figures/stabilizer_triBf2-eps-converted-to.pdf}}},\\
% C_e &= \vcenter{\hbox{\includegraphics[scale=.23,trim={0cm 0cm 0cm 0cm},clip]{Figures/stabilizer_triCe1-eps-converted-to.pdf}}}, \,\,\, \vcenter{\hbox{\includegraphics[scale=.23,trim={0cm 0cm 0cm 0cm},clip]{Figures/stabilizer_triCe2-eps-converted-to.pdf}}}, \,\,\, \vcenter{\hbox{\includegraphics[scale=.23,trim={0cm 0cm 0cm 0cm},clip]{Figures/stabilizer_triCe3-eps-converted-to.pdf}}}.
% \end{eqs}
\begin{eqs} \nonumber
A_v = \vcenter{\hbox{\includegraphics[scale=.23,trim={0cm 0cm 0cm 0cm},clip]{Figures/stabilizer_triAv-eps-converted-to.pdf}}},
\end{eqs}
\begin{align}
B_f &= \vcenter{\hbox{\includegraphics[scale=.23,trim={0cm 0cm 0cm 0cm},clip]{Figures/stabilizer_triBf1-eps-converted-to.pdf}}}, \, \, \, \vcenter{\hbox{\includegraphics[scale=.23,trim={0cm 0cm 0cm 0cm},clip]{Figures/stabilizer_triBf2-eps-converted-to.pdf}}},\\ \nonumber
C_e &= \vcenter{\hbox{\includegraphics[scale=.23,trim={0cm 0cm 0cm 0cm},clip]{Figures/stabilizer_triCe1-eps-converted-to.pdf}}}, \,\,\, \vcenter{\hbox{\includegraphics[scale=.23,trim={0cm 0cm 0cm 0cm},clip]{Figures/stabilizer_triCe2-eps-converted-to.pdf}}}, \,\,\, \vcenter{\hbox{\includegraphics[scale=.23,trim={0cm 0cm 0cm 0cm},clip]{Figures/stabilizer_triCe3-eps-converted-to.pdf}}}.
\end{align}
Note that the form of $C_e$ depends on the orientation of the edge $e$. It is convenient to write $C_e$ uniformly, using a cup product. In particular, $C_e$ can be written as:
\begin{align} \label{eq: Ce cohomology}
C_e = X^2_e \prod_{e',f} (Z^2_{e'})^{\bm{e'} \cup \bm{e}(f)},
\end{align}
for any edge $e$.
Here, $\bm{e}$ denotes the $\ZZ_2$-valued $1$-cochain that evaluates to $1$ on the edge $e$ and zero otherwise.

The ground state $|\psi_\text{DS}\rangle$ of the DS stabilizer Hamiltonian can be obtained by projecting the computational zero state to the ground state subspace. Letting $N_e$ denote the number of edges and $|0\rangle^{\otimes N_e}$ be the computational zero state, the ground state is \footnote{If the manifold is not simply connected, one can also consider ground states with nontrivial holonomy by including projectors built from string operators along homologically nontrivial cycles.}:
\begin{align}
|\psi_\text{DS}\rangle = \prod_e (1+C_e) \prod_v (1 + A_v + A_v^2+A_v^3) |0\rangle^{\otimes N_e}.
\end{align} 
For simplicity, here and throughout this section, we ignore the normalization of the state. 
Note that it is not necessary to include projectors for the face terms $B_f$, since $|0\rangle^{\otimes N_e}$ is already in the $+1$ eigenspace of the face terms. After expanding the $C_e$ and $A_v$ projectors, they may be written as:
\begin{align} \label{eq: Ce projector cohomology}
\prod_e (1+C_e) &= \sum_{\bm{c}_e} \prod_{e} C_{e}^{\bm{c}(e)} \\
\prod_v (1+A_v+A^2_v+A^3_v) &= \sum_{\bm{a}} \prod_{v} A_{v}^{\bm{a}(v)},
\end{align}  
where the sums on the right-hand side run over cochains $\bm{c} \in C^1[\mathcal{M},\ZZ_2]$ and $\bm{a} \in C^0[\mathcal{M},\ZZ_4]$.
The $A_v$ projector can be further simplified by acting with the Pauli Z operators on the computational zero state. More specifically, for any vertex $v$, $A_v$ applied to $|0\rangle^{\otimes N_e}$ is equal to:
 \begin{align} \label{eq: Av simplification}
 A_v |0\rangle^{\otimes N_e} = \prod_e (X^2_e)^{\bm{v} \cup \delta \bm{v}(e)}X_e^{\delta\bm{v}(e)}|0\rangle^{\otimes N_e},
 \end{align}
where $\delta$ is the coboundary operator and $\bm{v}$ is the $\ZZ_2$-valued $0$-cochain that evaluates to $1$ on $v$ and $0$ otherwise \footnote{One could of course consider a $\ZZ_4$-valued cochain $\bm{v}$, but we choose $\bm{v}$ to be $\ZZ_2$-valued to simplify the algebra later in the calculation.}. Pictorially, the operator on the right-hand side of Eq.~\eqref{eq: Av simplification} is:
\begin{align}
\prod_e (X^2_e)^{\bm{v} \cup \delta \bm{v}(e)}X_e^{\delta\bm{v}(e)} = \vcenter{\hbox{\includegraphics[scale=.23,trim={0cm 0cm 0cm 0cm},clip]{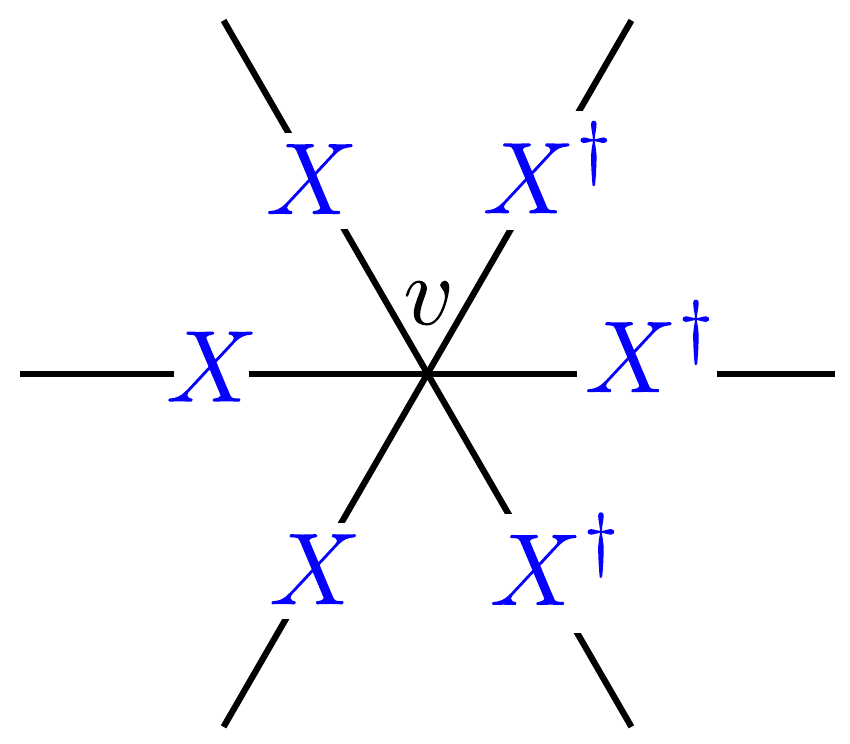}}}
\end{align}
We now have that the ground state of the DS stabilizer Hamiltonian is equal to:
 \begin{eqs} \label{eq: DS gs simplified}
|\psi_\text{DS}\rangle = &\Big[\sum_{\bm{c}} \prod_{e}(X^2_e)^{\bm{c}(e)} \prod_{f,e'} (Z^2_{e'})^{\bm{e'} \cup \bm{c}(f)}\Big] \\
 \times &\Big[\sum_{\bm{a}} \prod_e (X^2_e)^{\bm{a} \cup \delta \bm{a}(e)}X_e^{\delta\bm{a}(e)} \Big] |0\rangle^{N_e},
 \end{eqs}
where we have substituted $C_e$ in Eq.~\eqref{eq: Ce cohomology} into the expression for the projector in Eq.~\eqref{eq: Ce projector cohomology}.
 
To compare $|\psi_\text{DS}\rangle$ to the ground state of the DS string-net model, we map $|\psi_\text{DS}\rangle$ to a system of qubits. We introduce a pair of qubits to each edge $e$ and label the qubits by $A$ and $B$. As in Section~\ref{sec: relation to nonPauli}, the operator algebra for the pair of qubits at edge $e$ is generated by the Pauli X and Pauli Z operators: $X^A_e, \, Z^A_e, \, X^B_e, \, Z^B_e$.
The state on four-dimensional qudits is mapped to the system of qubits by applying the finite-depth quantum circuit $\mathcal{U}_{2,4}$, defined by:
\begin{align}\label{eq: qudit to qubits2}
Z_e \longleftrightarrow S_e^AZ^B_e, \quad X_e \longleftrightarrow X^A_e  CX^{AB}_e.
\end{align}
This unitary maps the computational zero state on qudits to the computational zero state on qubits:
\begin{align}
 \mathcal{U}_{2,4} |0\rangle^{\otimes N_e} = |0,0\rangle^{\otimes N_e},
\end{align}
where the two entries of the state on the right-hand side correspond to the $A$ and $B$ sites, and the qubits on the left-hand side and the qudits on the right-hand side have been suppressed.

The ground state in Eq.~\eqref{eq: DS gs simplified} is mapped to qubits by conjugating the projectors by $\mathcal{U}_{2,4}$. This gives us:
 \begin{eqs} 
\mathcal{U}_{2,4} |\psi_\text{DS}\rangle = &\Big[\sum_{\bm{c}} \prod_{e}(X^B_e)^{\bm{c}(e)} \prod_{f,e'} (Z^A_{e'})^{\bm{e'} \cup \bm{c}(f)}\Big] \\
 \times &\Big[\sum_{\bm{a}} \prod_e (X^B_e)^{\bm{a} \cup \delta \bm{a}(e)}(X^A_e)^{\delta\bm{a}(e)} \Big] |0,0\rangle^{\otimes N_e}.
 \end{eqs}
Note that the factors of $CX_e^{AB}$ act trivially on the computational zero state $|0,0\rangle^{\otimes N_e}$. Since $\bm{a}$ only appears in the exponent of order two operators, it can be replaced by a $\ZZ_2$-valued $1$-cochain $\bm{b} \in C^1[\mathcal{M},\ZZ_2]$. The ground state is then equivalent to:
\begin{eqs} \label{eq: DS projected on qubits}
    \mathcal{U}_{2,4} |\psi_\text{DS}\rangle = &\Big[\sum_{\bm{c}} \prod_{e}(X^B_e)^{\bm{c}(e)} \prod_{f,e'} (Z^A_{e'})^{\bm{e'} \cup \bm{c}(f)}\Big] \\
    \times &\Big[\sum_{\bm{b}} \prod_e (X^B_e)^{\bm{b} \cup \delta \bm{b}(e)}(X^A_e)^{\delta\bm{b}(e)} \Big] |0,0\rangle^{\otimes N_e}.
\end{eqs}
 
To evaluate the expression in Eq.~\eqref{eq: DS projected on qubits} further, we introduce cohomological notation for the computational basis states. Let $\{a_e^A\}$ and $\{a^B_e\}$ label the $\ZZ_2$ values of the computational basis state:
\begin{align}
|\{a^A_e\},\{a^B_e\}\rangle \equiv \bigotimes_e |a^A_e, a^B_e \rangle.
\end{align} 
We define the $\ZZ_2$-valued $1$-cochains $\bm{a}^A$ and $\bm{a}^B$ by the conditions:
\begin{align}
\bm{a}^A(e) =  a^A_e, \quad
\bm{a}^B(e) =  a^B_e.
\end{align}
The computational basis state $|\{a^A_e\},\{a^B_e\}\rangle$ can then be labeled by the corresponding $1$-cochains, i.e.:
\begin{align}
|\bm{a}^A,\bm{a}^B \rangle \equiv |\{a^A_e\},\{a^B_e\}\rangle.
\end{align}

Returning to the expression for $\mathcal{U}_{2,4} |\psi_\text{DS}\rangle$ in Eq.~\eqref{eq: DS projected on qubits}, we apply the Pauli operators to the computational zero state. Then, using the cohomological notation for the basis states, we find:
\begin{eqs}
\mathcal{U}_{2,4} |\psi_\text{DS}\rangle = \sum_{\bm{b},\bm{c}} \prod_{f} (-1)^{\delta \bm{b} \cup \bm{c}(f)} |\delta \bm{b}, \bm{b} \cup \delta \bm{b} + \bm{c} \rangle.
\end{eqs}
This can be reduced to a more familiar form by shifting the sum over $\bm{c}$, so that:
\begin{align}
\bm{c} \rightarrow \bm{c} + \bm{b} \cup \delta \bm{b}.
\end{align}
Making this substitution, the state $\mathcal{U}_{2,4} |\psi_\text{DS}\rangle$ becomes:
\begin{eqs} \label{eq: DS state cohomological reduced}
\mathcal{U}_{2,4} |\psi_\text{DS}\rangle= \sum_{\bm{b},\bm{c}} \prod_{f} (-1)^{\delta \bm{b} \cup \bm{b} \cup \delta \bm{b}(f) + \delta \bm{b} \cup \bm{c}(f)} |\delta \bm{b}, \bm{c} \rangle.
\end{eqs}

The sign $(-1)^{\delta \bm{b} \cup \bm{c}(f)}$ in Eq.~\eqref{eq: DS state cohomological reduced} can be produced by applying a control-$Z$ gate between the $A$ site on an edge $\langle 12 \rangle$ and the $B$ site on an edge $\langle 23 \rangle$ of the face $f=\langle 123 \rangle$, where the ordering of the vertices is determined by the branching structure (see Appendix~\ref{app: cohomology notation}). Letting $CZ^{AB}_{ 12 , 23 }$ denote this control-$Z$ gate, we define $\mathcal{U}_{AB}$ to be the following finite-depth quantum circuit: 
\begin{align} \label{eq: U_AB}
\mathcal{U}_{AB} \equiv \prod_{\langle 123 \rangle} CZ^{AB}_{ 12 , 23 }.
\end{align}
Applying $\mathcal{U}_{AB}$ to $\mathcal{U}_{2,4} |\psi_\text{DS}\rangle$ yields:
\begin{align}
\mathcal{U}_{AB}\mathcal{U}_{2,4} |\psi_\text{DS}\rangle = \sum_{\bm{b},\bm{c}} \prod_{f} (-1)^{\delta \bm{b} \cup \bm{b} \cup \delta \bm{b}(f)} |\delta \bm{b}, \bm{c}.\rangle
\end{align}
With this, the $A$ and $B$ sites have been disentangled. 
Ignoring the product state on the $B$ sites, we are left with:
\begin{align} \label{eq: gauged Z2 spt state}
\mathcal{U}_{AB}\mathcal{U}_{2,4} |\psi_\text{DS}\rangle = \sum_{\bm{b}} \prod_{f} (-1)^{\delta \bm{b} \cup \bm{b} \cup \delta \bm{b}(f)} | \delta \bm{b} \rangle.
\end{align}

The state in Eq.~\eqref{eq: gauged Z2 spt state} is mapped to the ground state of the DS string-net model by applying another finite-depth quantum circuit of control-$Z$ gates. We let $CZ^{AA}_{12 , 23}$ denote the control-$Z$ operator between the $A$ site on edge $\langle 12 \rangle$ and the $A$ site on edge $\langle 23 \rangle$. Then, we define $\mathcal{U}_{AA}$ to be the finite-depth quantum circuit:
\begin{align}
\mathcal{U}_{AA} = \prod_{f \in F_\text{up}} CZ^{AA}_{\langle 12 \rangle \langle 23 \rangle},
\end{align}
where the product is over all upward pointing triangles. Acting with $\mathcal{U}_{AA}$ on $\mathcal{U}_{AB}\mathcal{U}_{2,4} |\psi_\text{DS}\rangle$ gives the state:
\begin{align}
\mathcal{U}_{AA} \mathcal{U}_{AB}\mathcal{U}_{2,4} |\psi_\text{DS}\rangle = \sum_{\bm{b}} \Psi(\bm{b}) | \delta \bm{b} \rangle,
\end{align}
where $\Psi(\bm{b})$ is the amplitude:
\begin{align}
\Psi(\bm{b}) \equiv \prod_{f} (-1)^{\delta \bm{b} \cup \bm{b} \cup \delta \bm{b}(f)} \prod_{f \in F_\text{up}} (-1)^{\delta \bm{b} \cup \delta \bm{b}(f)}.
\end{align}

\begin{figure}
\centering
    \includegraphics[width=.4\textwidth]{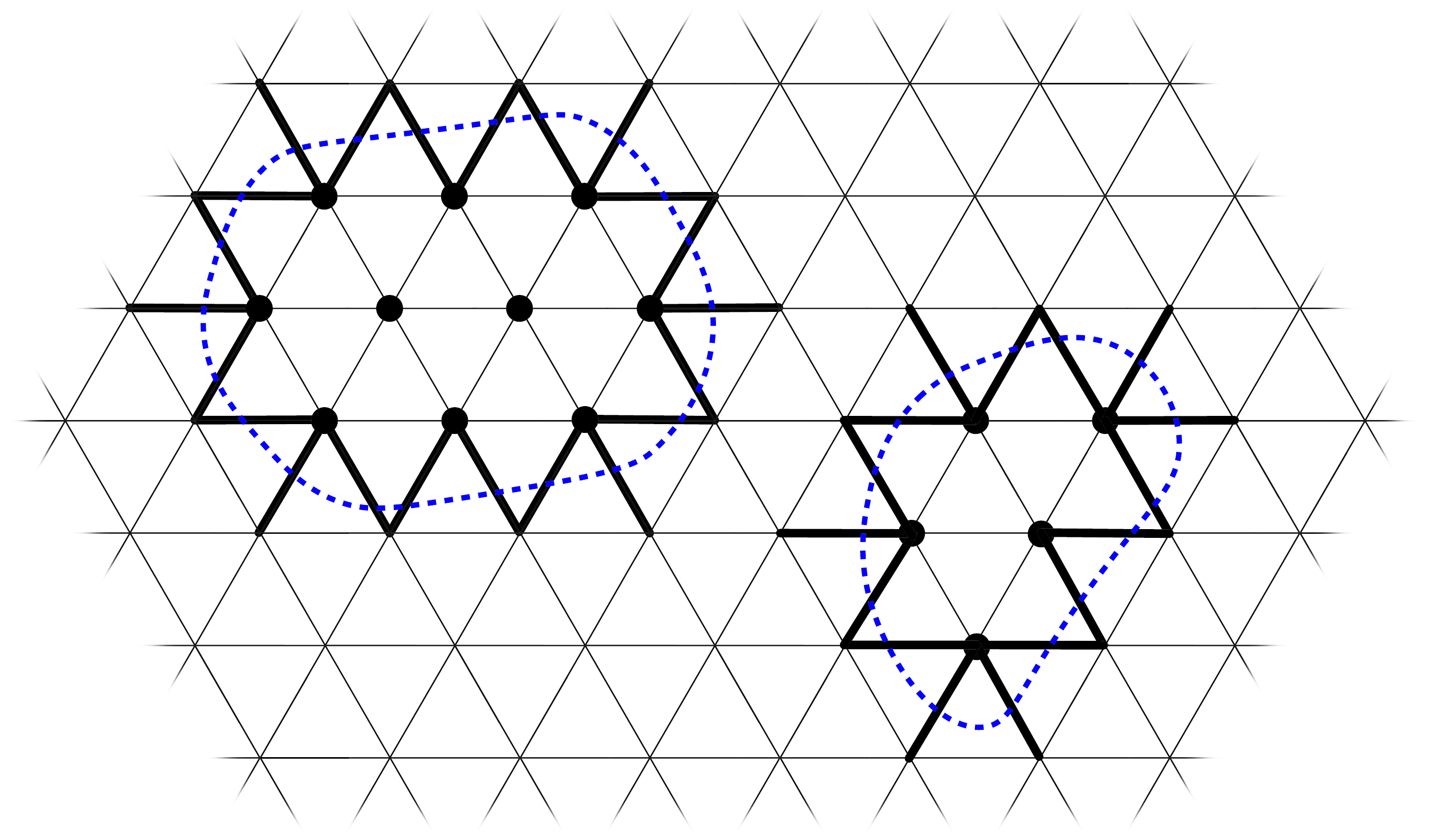}
     \caption{The state $|\delta \bm{b} \rangle$ corresponds to a configuration of loops on the dual lattice (dashed blue). The loops are defined by the edges $e$ for which $\delta \bm{b}(e) = 1$ (thickened edges). The loops bound the vertices $v$ for which $\bm{b}(v) = 1$ (black dots).}
     \label{fig: DSloops}
\end{figure}

In the next section and Refs.~\cite{CT21} and \cite{FHHT20}, it is argued that $\Psi(\bm{b})$ is equivalent to:
\begin{align}
\Psi(\bm{b}) = (-1)^{N_\text{loops}(\delta \bm{b})}.
\end{align}
Here, $N_\text{loops}(\delta \bm{b})$ is the number of loops (on the dual lattice) in the configuration corresponding to $\delta \bm{b}$ (see Fig.~\ref{fig: DSloops}). In other words, $\mathcal{U}_{AA} \mathcal{U}_{AB}\mathcal{U}_{2,4} |\psi_\text{DS}\rangle$ is equal to:
\begin{align}
\mathcal{U}_{AA} \mathcal{U}_{AB}\mathcal{U}_{2,4} |\psi_\text{DS}\rangle = \sum_{\bm{b}} (-1)^{N_\text{loops}(\delta \bm{b})} |\delta \bm{b} \rangle,
\end{align}
which is precisely the ground state of the DS string-net model on the hexagonal dual lattice. Since $\mathcal{U}_{AA} \mathcal{U}_{AB}\mathcal{U}_{2,4}$ is a finite-depth quantum circuit, the DS stabilizer Hamiltonian must belong to the same phase as the DS string-net model \cite{chen2010local}. 
In summary, $\mathcal{U}_{2,4}$ maps from four-dimensional qudits to pairs of qubits, $\mathcal{U}_{AB}$ disentangles the $B$ site qubits from the $A$ site qubits, and $\mathcal{U}_{AA}$ fixes the amplitude to be that of the DS string-net model ground state.

\subsection{String-net model ground state amplitudes} \label{app: string net gs}

In the argument above, we claimed that the amplitudes of the state $\mathcal{U}_{AA} \mathcal{U}_{AB}\mathcal{U}_{2,4} |\psi_\text{DS}\rangle$ are precisely those of the ground state of the DS string-net model. 
Let us complete the argument that $\mathcal{U}_{AA} \mathcal{U}_{AB}\mathcal{U}_{2,4} |\psi_\text{DS}\rangle$ is equivalent to:
\begin{align}
\mathcal{U}_{AA} \mathcal{U}_{AB}\mathcal{U}_{2,4} |\psi_\text{DS}\rangle = \sum_{\bm{b}} (-1)^{N_\text{loops}(\delta \bm{b})} | \delta \bm{b} \rangle.
\end{align}

To see this, we consider mapping the state to a symmetry-protected topological (SPT) state by gauging the $1$-form symmetry. This can be implemented by an operator duality, (the two-dimensional Kramers-Wannier duality) which maps between a system of qubits on the edges of the lattice and a system of qubits on the vertices \cite{levin2012braiding,yoshida2015topological,kubica2018ungauging}. For a set of $\ZZ_2$ values $\{b_v\}$, we label the computational basis state $|\{b_v\} \rangle$ by the $0$-cochain $\bm{b}$. We let $X_v$ and $Z_v$ denote the Pauli X and Pauli Z operators on the vertex $v$. With this, the explicit mapping of operators is:
\begin{align} \label{eq: KW duality}
\prod_{e \ni v} X_e \longleftrightarrow X_v, \quad Z_{\langle vv' \rangle} \longleftrightarrow Z_{v}Z_{v'},
\end{align}
where the product is over edges connected to $v$, and $\la vv' \ra$ is an arbitrary edge.
Under the duality, the state $|\delta \bm{b} \rangle$ is mapped to $|\bm{b} \ra$. Furthermore, it can be checked that $\mathcal{U}_{AA} \mathcal{U}_{AB}\mathcal{U}_{2,4} |\psi_\text{DS}\rangle$ is mapped to:
\begin{eqs}
|\psi_\text{SPT} \rangle \equiv \sum_{\bm{b}} \Psi(\bm{b}) |\bm{b} \ra.
\end{eqs}
Given the identification of $0$-cochains $\bm{b}$ with computational basis states $|\{b_v\}\ra$, this can alternatively be written as:
\begin{align}
|\psi_\text{SPT} \rangle = \sum_{\{b_v\}} \Psi(\{b_v\}) |\{b_v\} \ra.
\end{align}

We now simplify the state $|\psi_\text{SPT}\rangle$ by describing a geometric interpretation for the amplitude $ \Psi(\{b_v\})$, following Ref.~\cite{TV18}. On the triangular lattice with the branching structure shown in Fig.~\ref{fig: triDSdof}, $\Psi(\{b_v\})$ is equivalent to:
\begin{align} \label{eq: spt amplitude}
\Psi(\{b_v\}) =
 \prod_{\langle uvw \rangle} (-1)^{b_u b_v b_w} \prod_{\la vw \ra} (-1)^{b_vb_{w}} \prod_v (-1)^{b_v}.
\end{align}
To arrive at the expression above, we used the explicit formula for the cup product in Appendix~\ref{app: cohomology notation}. The amplitude in Eq.~\eqref{eq: spt amplitude} can be further reduced to the form:
\begin{align}
\Psi(\bm{b}) = (-1)^{V_1(\bm{b}) - E_1(\bm{b}) + F_1(\bm{b})},
\end{align}
where $V_1(\bm{b})$, $E_1(\bm{b})$, and $F_1(\bm{b})$ are the number of vertices, edges, and faces, respectively, contained entirely within the domains formed by the $b_v=1$ vertices in the configuration $\{b_v\}$. We let $\Sigma_1(\bm{b})$ denote the domains formed by the vertices with $b_v=1$. Then, $ \Psi(\bm{b})$ can be expressed in terms of the Euler characteristic $\chi\boldsymbol{(}\Sigma_1(\bm{b})\boldsymbol{)}$:
\begin{align}
\Psi(\bm{b}) = (-1)^{\chi \boldsymbol{(}\Sigma_1(\bm{b})\boldsymbol{)}}.
\end{align}
The Euler characteristic of an orientable surface $\Sigma$ is:
\begin{align}
\chi(\Sigma) = 2 - 2g(\Sigma) - n(\Sigma),
\end{align}
where $g(\Sigma)$ is the genus of $\Sigma$ and $n(\Sigma)$ is the number of boundary components of $\Sigma$. Therefore, letting $N_\text{dw}(\bm{b})$ be the number of domain walls in the configuration associated with $\bm{b}$, the amplitude $\Psi(\bm{b})$ is equivalent to:
\begin{align}
\Psi(\bm{b}) = (-1)^{N_\text{dw}(\bm{b})}.
\end{align}
This gives us:
\begin{align}
|\psi_\text{SPT}\ra = \sum_{\bm{b}} (-1)^{N_\text{dw}(\bm{b})} |\bm{b} \ra.
\end{align}

The final step of the argument is to gauge the $0$-form symmetry of the SPT state. This should return us to the state $\mathcal{U}_{AA} \mathcal{U}_{AB}\mathcal{U}_{2,4} |\psi_\text{DS}\rangle$. The $0$-form symmetry is gauged by applying the duality in Eq.~\eqref{eq: KW duality}, leaving us with:
\begin{align}
\mathcal{U}_{AA} \mathcal{U}_{AB}\mathcal{U}_{2,4} |\psi_\text{DS}\rangle = \sum_{\bm{b}} (-1)^{N_\text{dw}(\bm{b})} |\delta \bm{b} \ra.
\end{align}
Since $N_\text{dw}(\bm{b})$ is the same as the number of loops $N_\text{loops}(\delta \bm{b})$ in the configuration corresponding to $\delta \bm{b}$, we have the desired result:
\begin{align}
\mathcal{U}_{AA} \mathcal{U}_{AB}\mathcal{U}_{2,4} |\psi_\text{DS}\rangle =  \sum_{\bm{b}} (-1)^{N_\text{loops}(\delta \bm{b})} |\delta \bm{b} \ra.
\end{align}

\section{$K$-matrix formulation of twisted quantum doubles} \label{app: K matrix}

For completeness, we describe Abelian TQDs and the construction of the associated stabilizer models in terms of the $K$-matrix formalism. The $K$-matrix formalism is based on an integer symmetric matrix $K$, whose matrix elements encode the couplings between $U(1)$ Chern-Simons theories. We refer to Refs.~\cite{WZ92,Lu2012integer,Lu2016Classification} for further details on the connection to Chern-Simons theories. What is important for our discussion is that the universal properties of the Abelian anyon theory can be deduced entirely from the matrix $K$. 

We restrict our focus to $K$-matrices for Abelian TQDs. To this end, we take $G$ to be the finite Abelian group of the form $G=\prod_{i=1}^M \ZZ_{N_i}$ and recall that the cocycle of an Abelian TQD can be labeled by a set of integers $\mathcal{I}=\{ n_i \}_{i=1}^M \cup \{ n_{ij} \}_{i,j=1}^M$, as described in Section~\ref{sec: TQD anyons}. We define the $K$ matrix associated to the TQD to be the $2M \times 2M$ matrix:
% \begin{align}
%     K_{TQD} &= \begin{pmatrix}
%  \boldsymbol{0} & \boldsymbol N\\
%   \boldsymbol{N} & -(\boldsymbol{A}+\boldsymbol{A}^T)\\
%     \end{pmatrix}.
%     \label{eq:KmatrixTQD}
% \end{align}
\begin{align}
    K_{TQD} &= \begin{pmatrix}
 \boldsymbol{0} & \boldsymbol N\\
  \boldsymbol{N} & -\boldsymbol{S}_\mathcal{I}\\
    \end{pmatrix},
    \label{eq:KmatrixTQD}
\end{align}
% where $\boldsymbol N =\text{diag}(N_1, \ldots, N_M)$ and
where $\boldsymbol N$ is the $M\times M$ diagonal matrix:
\begin{align}
    \boldsymbol N =\begin{pmatrix}
 N_1 &  & &  & \\
    & N_2 & & & \\
    & & N_{3}& & \\
 & & &\ddots & \\
 &&&&N_{M}
    \end{pmatrix},
\end{align}
and $\boldsymbol{S}_\mathcal{I}$ is the $M \times M$ symmetric matrix:
\begin{align}
    \boldsymbol{S}_\mathcal{I} =\begin{pmatrix}
 2n_1 & n_{12} & n_{13}& \cdots & n_{1M}\\
  n_{12} & 2n_2  &n_{23}& \cdots & n_{2M}\\
  n_{13}& n_{23} & 2n_{3}& \cdots & n_{3M}\\
 \vdots & \vdots & \vdots &\ddots & \vdots\\
 n_{1M} &n_{2M}&n_{3M}&&2n_{M}
    \end{pmatrix}.
\end{align}
% \begin{align}
%     \boldsymbol A =\begin{pmatrix}
%  n_1 & n_{12} & n_{13}& \cdots & n_{1M}\\
%   & n_2  &n_{23}& \cdots & n_{2M}\\
%   && n_{3}& \cdots & n_{3M}\\
%  & & &\ddots & \vdots\\
%  &&&&n_{M}
%     \end{pmatrix}
% \end{align}
We note that the $ij$ matrix element of $\boldsymbol{S}_\mathcal{I}$ is precisely $n_{ij}$.

% There is some freedom in the choice of $K_\text{TQD}$, which we elaborate on later in the section. For now, we note that t
This choice of $K_\text{TQD}$ has the property that the gauge charges $c_i$ and (some choice of) elementary fluxes $\varphi_i$ correspond to unit vectors $l_{c_i}$ and $l_{\varphi_i}$ of length $2M$. In particular, $l_{c_i}$ and $l_{\varphi_i}$ are the unit vectors with a $1$ in the $(M+i)^\text{th}$ or $i^\text{th}$ entry, respectively. %For example, $l_{c_1}$ and $l_{\varphi_1}$ are:
%\begin{align}
%l_{c_1}^T = (0 \,\, 1 \,\, 0   \cdots  0), \quad l_{\varphi_1}^T = (1 \,\, 0 \,\, 0 \cdots 0).
%\end{align}
Further, since every anyon $a$ in the TQD can be generated by $c_i$ and $\varphi_i$:
\begin{align}
a = \prod_{i=1}^M \varphi_i^{p_i}\prod_{j=1}^M c_j^{q_j},
\end{align}
we can assign an integer vector $l_a$ to each anyon $a$ such that the $i^\text{th}$ entry is $p_i$ and the $(M+i)^\text{th}$ entry is $q_i$:
\begin{align}
l_a^T = (p_1 \ p_2 \ \cdots p_M \  q_1 \  q_{2} \cdots q_{M}).
\end{align}
For any two anyons $a$ and $a'$, the vectors $l_a$ and $l_{a'}$ satisfy:
\begin{align} \label{eq: fusion respected}
l_{a \times a'} = l_a + l_{a'}.
\end{align}

There is some ambiguity in the assignment of integer vectors to anyons within the $K$-matrix formalism. To see this, 
notice that the columns of $K$ encode the relations of the gauge charges and elementary fluxes. For example, the first column corresponds to the relation
$
 c_1^{N_1}=1, 
$
while the $(M+1)^\text{th}$ column tells us:
\begin{align}
\varphi_1^{N_1} c_1^{-2n_1} \prod_{i \neq 1} c_i^{-n_{1i}} = 1.
\end{align}
Therefore, we may freely redefine the vectors $l_a$ by integer multiples of the columns of $K$, i.e.:
\begin{align} \label{eq: vector identification}
    l_a \sim l_a + \sum_j m_j \text{ col}_j (K)
\end{align}
where $m_j$ are integers and $\text{col}_j (K)$ is the $j^\text{th}$ column of $K$.

The exchange statistics and braiding relations for anyons $a$ and $a'$ can be extracted from the inverse of $K$ using the formulas:
\begin{align}
\theta(a) &= e^{\pi i l_a^T K^{-1} l_a}, &
B_\theta(a,a') &= e^{2 \pi i l_a^T K^{-1} l_{a'}}.
\end{align}
In terms of the quadratic form $q$ and associated bilinear function $b_q$ (both defined in Section~\ref{sec: primer}), we have:
\begin{align}  \label{eq: statistics and braiding from K}
 q(a) &= \frac{1}{2} l_a^TK^{-1}l_a, &
 b_q(a,a')  &=  l_a^TK^{-1}l_{a'}
\end{align}
The relations in Eq.~\eqref{eq: statistics and braiding from K} can be checked by computing the inverse of the $K_\text{TQD}$, which we find to be
\begin{align}
    K_{TQD}^{-1} &= \begin{pmatrix}
 \boldsymbol{N}^{-1}\boldsymbol{S}_\mathcal{I}\boldsymbol{N}^{-1} & \boldsymbol{N}^{-1}\\
 \boldsymbol{N}^{-1}&\boldsymbol{0}
    \end{pmatrix}.
\end{align}
From which, we obtain:
\begin{align}
\begin{gathered}
    q(\varphi_i) = \frac{n_i}{N_i^2}, \quad q(c_i) = 0, \\
    b_q(c_i,\varphi_j) = \delta_{ij} \frac{1}{N_i}, \quad b_q(\varphi_i,\varphi_j) =  \frac{n_{ij}}{N_iN_j}.
\end{gathered}
\end{align}
These agree with the statistics and braiding relations of the gauge charges and elementary fluxes in Eq.~\eqref{eq: Theta in terms of ni and nij}.

We now see that all of the characteristic properties of the TQD can be determined from the matrix $K_\text{TQD}$. In particular, the fusion rules of the anyons are given by Eqs.~\eqref{eq: fusion respected} and \eqref{eq: vector identification}, and the statistics and braiding can be computed from Eq.~\eqref{eq: statistics and braiding from K}. 

Before turning to examples, we clarify some freedom in the definition of a $K$-matrix $K$. The freedom comes from choosing a different set of generators for the anyons. For example, for $K_\text{TQD}$, the generators are implicitly chosen to be the gauge charges and a set of elementary fluxes. A different set of generators corresponds to transforming the vectors $l_a$ as $Wl_a$, where $W$ is an integer matrix belonging to $GL(2M,\ZZ)$. To preserve the statistics and braiding, $K^{-1}$ must transform as:
\begin{align}
K^{-1} \to (W^T)^{-1}K^{-1}W^{-1},
\end{align}
which means that $K$ is mapped according to:
\begin{align} \label{eq: K transformation}
K \to WKW^T.
\end{align}
Therefore, any two $K$-matrices related by a transformation as in Eq.~\eqref{eq: K transformation} describe the same anyon theory.

% \vspace{.2cm}
% \begin{center}
%     \textbf{\small Examples}
% \end{center}
% \vspace{.2cm}

\subsubsection*{Examples}

We are now prepared to consider examples, starting with the $\ZZ_N$ TC. In this case, $K$ is the $2 \times 2$ matrix:
\begin{align}
K=\begin{pmatrix}
 0 & N \\
N & 0  
    \end{pmatrix}.
\end{align}
The anyons of the $\ZZ_N$ TC are generated by $e$ and $m$, which correspond to the unit vectors:
\begin{align}
    l_e = \begin{pmatrix}0\\1\end{pmatrix}, \quad l_{m} = \begin{pmatrix}1\\0\end{pmatrix}.
\end{align}
Both $e$ and $m$ have order $N$ under fusion, which can be checked using Eqs.~\eqref{eq: fusion respected} and \eqref{eq: vector identification}. Furthermore, according to Eq.~\eqref{eq: statistics and braiding from K}, $e$ and $m$ are bosons and exhibit the expected Aharonov-Bohm effect:
\begin{align}
q(e)=0, \quad q(m)=0, \quad b_q(e,m)= \frac{1}{N}.
\end{align}

As our next example, we consider the $K$-matrix formalism for the DS phase. For the DS topological order we have $G=\ZZ_2$ and $n_1 = 1$. This gives us the matrix:
\begin{align}
    K =  \begin{pmatrix}
 0 & 2 \\
2 & -2  
    \end{pmatrix}.
\end{align}
We find that this matches the definitions of the anyons of the double semion model by associating:
\begin{align}
    l_{s\bar{s}} &= \begin{pmatrix}0\\1\end{pmatrix}, &  l_s &= \begin{pmatrix}1\\0\end{pmatrix}, & l_{\bar{s}} &= \begin{pmatrix}1\\1\end{pmatrix}.
\end{align}
We confirm that the exchange statistics are given by:
\begin{align}
    q(s\bar{s}) = 0, \quad  q(s) = \frac{1}{4}, \quad q(\bar{s}) = -\frac{1}{4}.
\end{align}

Let us now explore examples where the fusion group of the anyons differs from $G \times G$. First, we consider the TQD with $G=\ZZ_3$ and $n_1=1$. This corresponds to the matrix:
\begin{align}
    K =  \begin{pmatrix}
 0 & 3 \\
3 & -2  
    \end{pmatrix}.
\end{align}
Interestingly, the anyons in this theory have $\ZZ_9$ fusion rules. The generating anyon of order nine can be chosen to be the elementary flux $l_{\varphi_1}^T= (1 \ 0) $, with $q(\varphi_1)=1/9$. The excitation $\varphi_1^3$ is a boson and corresponds to the charge $c_1$ of order 3. This can be seen by noticing that:
\begin{align}
    l_{\varphi_1^3} + \text{col}_1(K) + \text{col}_2(K) = l_{c_1}.
\end{align}

Next, we consider the TQD with $G= \ZZ_2 \times \ZZ_2$ and $n_{12}=1$. Notably, this TQD is characterized by a type II cocycle. Given the general form for $K$, we find:
\begin{align}
    K =  \begin{pmatrix}
 0 & 0 & 2&0 \\
0 & 0 &0&2 \\
2 & 0 &0&-1 \\
0 & 2 &-1&0 \\
    \end{pmatrix}.
\end{align}
The elementary fluxes $l_{\varphi_1}^T =(1 \ 0 \ 0 \ 0) $ and $l_{\varphi_2}^T =(0 \ 1 \ 0 \ 0) $ are bosons with the property that they square to each others corresponding gauge charge. More precisely, they satisfy:
\begin{eqs}
   l_{\varphi_1^2}^T &=(2 \ 0 \ 0 \ 0)  \sim (0 \ 0 \ 0 \ 1) =  l_{c_2}^T,\\
l_{\varphi_2^2}^T &=(0 \ 2 \ 0 \ 0)  \sim (0 \ 0 \ 1 \ 0) =  l_{c_1}^T.
\end{eqs}
Therefore, the anyons have $\ZZ_4 \times \ZZ_4$ fusion rules.
Furthermore, the elementary fluxes have nontrivial mutual braiding:
\begin{eqs}
l_{\varphi_1}^TK^{-1}l_{\varphi_2} = b_q(\varphi_1,\varphi_2) = \frac{1}{4}.
\end{eqs}

In fact, this TQD is equivalent to the $\ZZ_4$ TC by the identification $\varphi_1 = e$ and $\varphi_2=m$. To see this explicitly, we can perform a basis transformation to relabel the anyons as those of the $\ZZ_4$ TC. To do so, we define the matrix $W$, given by:
\begin{align}
    W = \begin{pmatrix}
 1 & 0 & 0&0 \\
0 & 1 &0&0 \\
0 & 2 &0&-1 \\
2 & 0 &1&0 \\
    \end{pmatrix} \in GL(4,\ZZ).
\end{align}
Following Eq.~\eqref{eq: K transformation}, $K$ is transformed as:
\begin{align}
    W^TK W =\begin{pmatrix}
 0 & 4\\
4 & 0 
    \end{pmatrix} \oplus \begin{pmatrix}
 0 & 1\\
1 & 0 
    \end{pmatrix}.
\end{align}
Thus, the theory describes a decoupled $\ZZ_4$ TC and $\ZZ_1$ TC, where the latter factor of $\ZZ_1$ can be ignored, as it has no anyonic excitations. With this, we have mapped the anyon content of the TQD to that of the $\ZZ_4$ TC. Similarly, if $n_{12}=1$ and either $n_1=1$ or $n_2=1$, but not both, then the TQD is again equivalent to the $\ZZ_4$ TC.

As a final example, let us consider the TQD with $G=\ZZ_2 \times \ZZ_2$ and $n_1=n_2=n_{12}=1$. In this case, the $K$ matrix is:
\begin{align}
    K =  \begin{pmatrix}
 0 & 0 & 2&0 \\
0 & 0 &0&2 \\
2 & 0 &-2&-1 \\
0 & 2 &-1&-2 \\
    \end{pmatrix}.
\end{align}
Similar to the previous example, this anyon theory has $\ZZ_4 \times \ZZ_4$ fusion rules. However, in contrast, the elementary fluxes are semions. All together, this theory has six semions, six anti-semions, and four bosons. This differs from other $\ZZ_2 \times \ZZ_2$ TQDs, which all have at least one fermionic excitation. We refer to this theory as the six-semion theory, since it is closely related to the three-fermion theory, according to the classification of Abelian anyon theories in Ref.~\cite{WW20}. Specifically, they both belong to the family of anyon theories denoted by $F_{2^r}$ in Ref.~\cite{WW20}, with $r=1$ and $r=2$ corresponding to the three-fermion theory and six-semion theory, respectively. For the anyon theories in $F_{2^r}$, the fusion group is $\ZZ_{2^r} \times \ZZ_{2^r}$ and the quadratic form is given by:
\begin{align}
    q\boldsymbol{(}(a_1,a_2)\boldsymbol{)} = \frac{a_1^2+a_2^2 + a_1a_2}{2^r},
\end{align}
where $(a_1,a_2)$ is an element of $\ZZ_{2^r} \times \ZZ_{2^r}$. The chiral central charge is $4 \text{ mod }8$ if $r$ is odd and $0 \text{ mod }8$ if $r$ is even.

% \vspace{-.6cm}
\subsubsection*{Construction of TQDs from boson condensation}
% \subsection{Condensation construction}

The general construction of TQDs starting from TCs, as described in Sec. \ref{sec: TQD stabilizer code construction}, can be concisely stated in terms of $K$-matrices. To see this, we begin with a stack of TCs, given by the following $2M \times 2M$ $K$-matrix:
\begin{align}
    K_{TC} &= \begin{pmatrix}
\boldsymbol{0} &  \boldsymbol{N}^{2}\\
\boldsymbol{N}^{2} & \boldsymbol{0}
    \end{pmatrix}.
\end{align}

The next step in the construction is to condense the set of bosons in Eq.~\eqref{eq: condense this}, rewritten here for convenience:
\begin{eqs}
b_i = m_i^{-N_i}e_i^{N_in_i}\prod_{j<i}e_j^{N_jn_{ij}}.
\end{eqs}
We can verify that the $b_i$ are bosons with trivial mutual braiding by noting that:
\begin{eqs}
q(b_i) = \frac{1}{2}l_{b_i}^TK^{-1}_{TC}l_{b_i} &= 0 \text{ mod }1, \quad \forall i, \\
b_q(b_i,b_j) = l_{b_i}^TK^{-1}_{TC}l_{b_j} &= 0 \text{ mod }1, \quad \forall i,j.
\label{eq:bosonscommute}
\end{eqs}
% \begin{align}
% \frac{1}{2}[Q^TK^{-1}Q]_{ii} &= -n_i\\
% [Q^TK^{-1}Q]_{ij} &= -n_{ij}
% \end{align}
The vectors of the bosons can be compiled into a $2M \times M$ matrix $Q$, where the vectors $l_{b_i}$ form the columns of $Q$:
\begin{align}
    Q= \begin{pmatrix}
-\boldsymbol{N}\\
\boldsymbol{N}\boldsymbol{U}_\mathcal{I}
    \end{pmatrix}.
\end{align}
% \begin{align}
%     Q= \begin{pmatrix}
% -\boldsymbol{N}\\
% \boldsymbol{NA}
%     \end{pmatrix}.
% \end{align}
Here, $\boldsymbol{U}_\mathcal{I}$ is the $M \times M$ upper triangular matrix:
\begin{align}
    \boldsymbol{U}_\mathcal{I} =\begin{pmatrix}
 n_1 & n_{12} & n_{13}& \cdots & n_{1M}\\
  & n_2  &n_{23}& \cdots & n_{2M}\\
  && n_{3}& \cdots & n_{3M}\\
 & & &\ddots & \vdots\\
 &&&&n_{M}
    \end{pmatrix}.
\end{align}
Indeed, Eq.~\eqref{eq:bosonscommute} is satisfied by noting that
\begin{align}
 Q^TK_{TC}^{-1}Q = - (\boldsymbol{U}_\mathcal{I} + \boldsymbol{U}_\mathcal{I}^T) = - \boldsymbol{S}_\mathcal{I}.
\end{align}
% \begin{align}
%  Q^TK_{TC}^{-1}Q = - (\boldsymbol{A}+\boldsymbol{A}^T) \equiv 0 \text{ mod }1.
% \end{align}
% The bosons we condense in Eq.\eqref{eq: condense this} are vectors which form $M$ column vectors of the following matrix
% \begin{align}
%     Q= \begin{pmatrix}
%  -N_1 &0 & 0   &\cdots\\
% n_1N_1 & n_{12}N_1 & n_{13}N_1  &\cdots \\
%  0 &  -N_2& 0  &\cdots\\
% 0 & n_2N_2 & n_{23}N_3 &\cdots\\
%  0 &  0 & -N_{3}  &\cdots\\
%  0 &  0 &  n_3N_3&\cdots\\
%   \vdots & \vdots & \vdots&\ddots\\
%     \end{pmatrix}.
% \end{align}

The deconfined anyons, i.e., those that braid trivially with the bosons $b_i$, are generated by the columns of the following matrix:
\begin{align} \label{eq: L def}
    L= \begin{pmatrix}
\boldsymbol{I}_{M\times M} & \boldsymbol{0}\\
\boldsymbol{N}\boldsymbol{U}_\mathcal{I}^T\boldsymbol{N}^{-1}& \boldsymbol{N}
    \end{pmatrix},
\end{align}
% \begin{align}
%     L= \begin{pmatrix}
% \mathbbm 1 & \boldsymbol{0}\\
% \boldsymbol{NA}^T\boldsymbol{N}^{-1}& \boldsymbol{N}
%     \end{pmatrix}.
% \end{align}
where $\boldsymbol{I}_{M \times M}$ is the $M \times M$ identity matrix.
Note that in Eq.~\eqref{eq: L def}, the first $M$ columns correspond to the elementary fluxes of Eq.~\eqref{eq: def fluxi}, and the following $M$ columns correspond to the gauge charges of Eq.~\eqref{eq: def chargei}. We see that
\begin{align}
    L^TK_{TC}^{-1}Q \equiv \begin{pmatrix}
 \boldsymbol{0}\\
 -\boldsymbol{I}_{M \times M}
    \end{pmatrix} \text{ mod }1,
\end{align}
confirming that the deconfined anyons braid trivially with the condensed bosons. 

Note that the exchange statistics and braiding relations of the deconfined anyons are captured by the matrix $L^TK_{TC}^{-1}L$. Therefore, after condensation, the corresponding $K$-matrix is given by $L^{-1}K_{TC}{(L^{-1})}^T$. The matrix $L^{-1}K_{TC}{(L^{-1})}^T$ is precisely the $K$-matrix $K_{TQD}$, given in Eq.~\eqref{eq:KmatrixTQD}.

\section{Anyon fusion groups of twisted quantum doubles} \label{app: fusion group}

Here, we describe the group generated by the anyons of an Abelian TQD. To keep the discussion general, we consider an Abelian TQD associated to a group $G=\prod_{i=1}^M \ZZ_{N_i}$ and a cocycle specified by the set $\mathcal{I}$, as in Section~\ref{sec: TQD anyons}. We then derive the group structure by considering the fusion of elementary fluxes.

To start, we choose an elementary flux $\varphi_i$ for every $i$ in $\{1,\ldots, M\}$. After choosing the elementary fluxes, each anyon can be expressed as a unique product of elementary fluxes and gauge charges. That is, the anyons can be written in the form: 
\begin{align}
\prod_{i=1}^M \varphi_i^{g_i}\prod_{j=1}^M c_j^{k_j},
\end{align}
for some $g_i, k_i \in \ZZ_{N_i}$. The fusion of two anyons composed of elementary fluxes gives:
\begin{eqs} \label{eq: flux product}
\prod_{i=1}^M \varphi_i^{g_i} \times \prod_{i=1}^M \varphi_i^{h_i} = \prod_{i=1}^M \varphi_i^{[g_i+h_i]_{N_i}} \prod_{i=1}^M c_i^{2n_i \frac{1}{N_i} (g_i + h_i - [g_i + h_i]_{N_i})} 
 \\ \times \prod_{i=1}^M \prod_{j \neq i} c_j^{n_{ij} \frac{1}{N_i}(g_i + h_i - [g_i+h_i]_{N_i})},
\end{eqs}
where the gauge charges on the right-hand side follow from the relation in Eq.~\eqref{eq: flux charge relation} and $h_i$ is an element of $\ZZ_{N_i}$. Hence, the elementary fluxes fail to satisfy the $G$ group laws by products of gauge charges. This suggests that the group formed by the anyons is a central extension of $G$ by $G$.  

To make this explicit, we notice that an arbitrary anyon $\prod_{i=1}^M \varphi_i^{g_i}\prod_{j=1}^M c_j^{k_j}$ can be labeled by a pair $(g,k)$ in $G \times G$, where the $i^\text{th}$ components of $g$ and $k$ are $g_i$ and $k_i$, respectively. With this, the product in Eq.~\eqref{eq: flux product} becomes:
\begin{align}
(g,0) \times (h, 0) = (g+h,\lambda(g,h) ).
\end{align}
Here, $\lambda(g,h)$ is an element of $G$ whose $i^\text{th}$ component $\lambda(g,h)_i$ is:
\begin{eqs}
\lambda(g,h)_i &= 2n_i\frac{1}{N_i} (g_i+h_i-[g_i+h_i]_{N_i}) \\ &+ \sum_{j \neq i} n_{ij} \frac{1}{N_j}(g_j+h_j-[g_j+h_j]_{N_j}).
\end{eqs}
More generally, the product of $(g,k)$ and $(h,\ell)$ is:
\begin{eqs}
(g,k) \times (h,\ell) = (g+h, k+\ell+\lambda(g,h)).
\end{eqs}
The function $\lambda : G \times G \to G$ defines a $2$-cocycle in $H^2[G,G]$ and specifies the central extension of $G$ by $G$ that corresponds to the group formed by the anyons of the Abelian TQD under fusion.

\vspace{1cm}
\bibliography{bib}
\end{document}